\documentclass[useAMS,usenatbib]{mnras}
\usepackage{graphicx,amsmath,url,natbib}

\usepackage{multirow}
\usepackage[flushleft]{threeparttable}

\newcommand{\gtrsim}{\,\raisebox{-0.4ex}{$\stackrel{>}{\scriptstyle\sim}$}\,}
\newcommand{\lesssim}{\,\raisebox{-0.4ex}{$\stackrel{<}{\scriptstyle\sim}$}\,}
\newcommand{\ergs}{\mbox{ erg s}^{-1}}
\newcommand{\cc}{\mbox{ cm}^{-3}}
\newcommand{\kms}{\mbox{ km s}^{-1}}
\newcommand{\kpc}{\mbox{ kpc}}
\newcommand{\pc}{\mbox{ pc}}

\usepackage{xparse}
\usepackage{soul}
\usepackage[usenames,dvipsnames]{xcolor}

\makeatletter
\NewDocumentCommand{\sayw}{O{Green}O{Black}+m}
    {%
        \begingroup
        \setulcolor{#1}%
        \setul{-0.6ex}{1.4pt}%
        \def\SOUL@uleverysyllable{%
            \rlap{%
                \color{#2}\the\SOUL@syllable
                \SOUL@setkern\SOUL@charkern}%
            \SOUL@ulunderline{%
                \phantom{\the\SOUL@syllable}}%
        }%
        \ul{#3}%
        \endgroup
    }
\makeatother

\begin{document}

\title[Feedback on gas disks]{Relativistic jet feedback III: feedback on gas disks}
\author[D. Mukherjee et al.]{%
Dipanjan Mukherjee$^{1,2}$\thanks{E-mail: dipanjan.mukherjee@unito.it},
Geoffrey V. Bicknell,$^{2}$
Alexander Y. Wagner$^{3}$,
\newauthor
Ralph S. Sutherland$^{1}$,
and Joseph Silk$^{4,5,6}$
\\
$^{1}$Dipartimento di Fisica Generale, Universita degli Studi di Torino , Via Pietro Giuria 1, 10125 Torino, Italy\\
$^2$Australian National University, Research School of Astronomy and Astrophysics, Cotter Rd., Weston, ACT 2611, Australia \\
$^3$Center for Computational Sciences, University of Tsukuba, 1-1-1 Tennodai, Tsukuba, Ibaraki, 305-8577\\
$^4$Institut d’Astrophysique, Sorbonne Universit`e  et CNRS, UMR 7095, 98bis Bd Arago,
F-75014 Paris, France\\
$^5$Department of Physics and Astronomy, The Johns Hopkins University, Baltimore, MD 21218, USA \\
$6$BIPAC, University of Oxford,1 Keble Road, Oxford OX1 3RH, UK
}

\date{\today}
\pagerange{\pageref{firstpage}--\pageref{lastpage}} 
\pubyear{2015}
\maketitle

\begin{abstract}
We study the interactions of a relativistic jet with a dense turbulent gaseous disk of radius $\sim 2$ kpc.  We have performed a suite of simulations with different mean density, jet power and orientation. Our results show that: A) the relativistic jet couples strongly with the gas in the inner kpc, creating a cavity and launching outflows. B) The high pressure bubble inflated by the jet and its back-flow compresses the disk at the outer edges, driving inflows within the disk. C) Jets inclined towards the disk affect the disk more and launch sub-relativistic, wide-angle outflows along the minor axis. D) Shocks driven directly by the jet and the jet-driven energy bubble raise the velocity dispersion throughout the disk by several times its initial value. E) Compression by the jet-driven shocks can enhance the star formation rate in the disk, especially in a ring-like geometry close to the axis. However, enhanced turbulent dispersion in the disk also leads to quenching of star formation. Whether positive or negative feedback dominates depends on jet power, ISM density, jet orientation with respect to the disc, and the time-scale under consideration. Qualitatively, our simulations compare favourably with  kinematic and morphological signatures of several observed galaxies such as NGC~1052, NGC~3079, 3C~326 and 3C~293.
\end{abstract}


\section{Introduction}\label{sec:intro}
Outflows from Active Galactic Nuclei (AGN) have long been identified to be one of the dominant mechanisms that influences stellar mass assembly in massive galaxies \citep{silk98a,dimatteo05a,bower06a,croton06a}. Feedback from the central super massive black hole is thought to occur primarily in two ways. In the \emph{quasar} mode \citep[or the \emph{establishment} mode, ][]{wagner16a}, outflows are driven by luminous AGN accreting at near Eddington rates, and disperse star forming gas from the galaxy. Such winds are primarily considered to be wide-angled, driven by the radiation pressure from the luminous AGN \citep[e.g. ][]{proga00a,ostriker10a,bieri17b}. In the \emph{radio} mode \citep[or \emph{maintenance} mode, ][]{harrison13a} it is envisaged that large scale jets from the central black hole heat the IGM, thereby preventing cooling flows and mass build up of galaxies \citep{fabian12a}. However, it is still unclear how the central AGN affects the gas in the galaxy, and over what time scales \citep{schawinski09a,schawinski15a}. On-going star formation in AGN host galaxies \citep{harrison12a} also complicates the simple interpretation presented above, motivating the need for more detailed investigation of how the central black hole may affect the ISM of the galaxy.

Observational evidence for both forms of feedback is plentiful. Powerful radio jets extending to $\sim 100$ kpc and heating the ambient gas, especially in clusters of galaxies \citep{fabian12a}, are an archetypal  \emph{maintenance} mode scenario. AGN powered galactic outflows have been detected in several systems \citep{nesvadba08a,rupke11a,tombesi15a,nesvadba16a,collet16a}. Suppressed star formation in systems harbouring an AGN driven outflow \citep{nesvadba10a,wylezalek16a}  are an indicator of \emph{establishment} mode feedback. The role of jets from from supermassive black holes has largely been considered in the context of \emph{maintenance} mode feedback \citep{gaspari12a,gaspari12b,yang16a}. It is however not well understood whether galaxies with radio jets are also directly affected by the jet itself, before they evolve into large scale structures.

Recent observational studies \citep{nesvadba10a,nesvadba11b,alatalo15a} have shown that star formation may be significantly regulated or suppressed by turbulence and outflows induced by the radio jet. However, recent theoretical 
\citep{gaibler12a,bieri16a,fragile17a} and observational \citep{lacy17a,salome15a,bicknell00a} papers have proposed a jet induced positive feedback scenario where jet-driven shocks may induce star formation in galaxies. Hence, the manner in which a relativistic jet, often well-collimated at radio wavelengths, affects the large scale interstellar medium (ISM) of the galaxy is not well-understood. This is an open question, requiring both theoretical and observational investigation.

In earlier papers \citep{sutherland07a,wagner11a,wagner12a,mukherjee16a} we have simulated the interaction of a relativistic jet with a turbulent ISM. We have shown that the primary driver of feedback in these cases is the evolution of an energy bubble inflated by the jet. In the initial phases when the jet is confined within the galaxy's ISM, the energy bubble progresses nearly spherically, strongly interacting with the galaxy's ISM, creating multi-phase outflows. We have previously considered a spherical distribution of gas, which is appropriate for an elliptical system formed by a merger and triggering an AGN \citep{gatti16a}.

However, in the high redshift universe ($z \sim 2$), a significant fraction of the galaxies have gaseous disks \citep{glazebrook13a,wisnioski15}. If an AGN is triggered in such galaxies, the jet (or nuclear wind) encounters a disk-like morphology of the ISM. Observations of feedback from radio-loud galaxies at high redshifts \citep{collet16a,nesvadba16a} confirm this scenario. Since a disk-like geometry provides lower column depth along the path of the jet, it is not clear whether jet feedback can have a significant effect on the galaxy's ISM. 

Previously, \citet{sutherland07a,gaibler11a,gaibler12a,dugan17a,cielo17a} carried out simulations of jets and/or winds from a central black hole affecting a gas disk in a galaxy.  These results suggest that a jet can indeed affect the gas distribution as well as the star formation kilo-parsecs away from the jet axis. However these papers do not address the effects of different jet power, ISM density and jet orientation on the jet-ISM interaction in a systematic manner. Sometimes, essential physics such as relativistic effects, external gravity, or atomic cooling were ignored.

Adhering to the framework followed in our previous papers \citep{wagner11a,wagner12a,mukherjee16a}, we present here simulations of a relativistic jet interacting with a multi-phase turbulent disk. Our primary motivation is to understand how the jet and its associated energy bubble couple with the disk on extended scales ($\gtrsim 1$ kpc), the dependence on jet orientation with respect to the disk, the effect on the disk kinematics and turbulence, and the implications for the star formation rate (positive or negative). In our simulations we use the PLUTO hydrodynamic code and employ a relativistic Riemann solver, atomic cooling, external gravity, and establish a rotating  disk with a turbulent velocity dispersion.  

We structure the paper as follows. In Sec.~\ref{sec.setup} we describe the numerical set up of the simulation. We discuss the simulation code and the method used to set up the turbulent disk and the relativistic jet. In Sec.~\ref{sec.results} we present the results of the simulations. We discuss the impact of jets launched both perpendicular to the disk (Sec.~\ref{sec.vertical}) and inclined towards it (Sec.~\ref{sec.inclined}). In Sec.~\ref{sec.sfr} we discuss the possible impact of the jet feedback on the star formation rate. Finally in Sec.~\ref{sec.discuss} we summarize the results and discuss their implications. We have found qualitative similarities between the results our simulations and the observed kinematic and morphological structure of several galaxies such as 3C~326, 3C~293, NGC~1052 and NGC~3079, which we discuss in that section.

\section{Simulation set up}\label{sec.setup}
\begin{table}
\centering
\caption{ Parameters of the ambient gas and gravitational potential common to all simulations. }
\label{tab.params}
\begin{tabular}{| l | l | l |}
\hline
\multicolumn{2}{c|}{Parameters} 	& Value   \\
\hline
Baryonic core radius  & $r_B$		& 1 kpc			\\
Baryonic velocity dispersion  & $\sigma _B$	& 250 km s$^{-1}$	\\
Ratio of DM to Baryonic & $\lambda$		& 5	\\
core radius 	&& 		\\
Ratio of DM to Baryonic & $\kappa$		& 2			\\
velocity dispersion && \\
Halo Temperature  & $T_h$		& $10^7$ K		\\
Halo density at r=0  & $n_{h0}$		& $0.5$ cm$^{-3}$ 	\\
Turbulent velocity dispersion & $\sigma _t$	& 250 km s$^{-1}$	\\
of warm clouds && \\
\hline
\end{tabular} \\
\begin{tablenotes}
{\footnotesize
\item (a) For the assumed gravitational potential, total Baryonic (stellar) mass within a radius of 5 kpc and 10 kpc are: $M_B(r<5 \kpc)= 9.52\times10^{10} M_\odot$ and $M_B(r<10 \kpc)= 1.04\times10^{11} M_\odot$ respectively. 
\item (b) The total mass of dark matter  within a radius of 5 kpc and 10 kpc are: $M_{DM}(r<5 \kpc)= 2.56\times10^{11} M_\odot$ and $M_{DM}(r<10 \kpc)= 9.1\times10^{11} M_\odot$ respectively.	
}
\end{tablenotes}
\flushleft
\end{table}
\begin{table}
\centering
\caption{Simulations}\label{tab.sims}
\begin{tabular}{| c | c | c | c | c | c | c |}
\hline
	Simulation	   & Power 		& $n_{w0}$    		&$\theta _{\rm inc}$$^b$    	& $\chi$$^c$	& $\Gamma $$^d$ & Gas Mass	\\
Label	   & (ergs$^{-1}$)	& (cm$^{-3}$)$^a$ 	& 			&		&& ($10^9 M_{\odot}$) \\
\hline
A  & $10^{45}$			& 100 			& 0		&10	&5  & 2.05  \\
B  & $10^{45}$			& 200 			& 0		&10	&5  & 5.71  \\
C  & $10^{45}$			& 200 			& 20		&10	&5  & 5.71  \\
D  & $10^{45}$			& 200   		& 45		&10 	&5  & 5.71  \\
E  & $10^{45}$			& 200 			& 70		&10	&5  & 5.71  \\
\hline
F  & $10^{46}$			& 100 			& 0 		&80 	&10 & 2.05  \\
G  & $10^{46}$			& 200 			& 0		&80	&10 & 5.71  \\
H  & $10^{46}$			& 400 			& 0		&80 	&10 & 20.68 \\
\hline
\end{tabular} 
\flushleft
	$^a$ The density of the dense gas at $r=z=0$.\\
	$^b$ Inclination of the jet axis with respect to the $Z$ axis. \\
	$^c$ The ration of rest mass energy to relativistic enthalpy given by eq.~\ref{eq.chi}. \\
	$^d$ The Lorentz factor. $\Gamma=10$ corresponds to jet injection velocity of $0.995c$, and $\Gamma=5$ corresponds to $0.979c$.
\end{table}

\subsection{Code and numerical Schemes}
We have performed 3D  hydrodynamic simulations of a relativistic jet interacting with a turbulent gas disk before breaking out into the galactic halo. The simulations have been carried out using the relativistic hydrodynamic module of the publicly available PLUTO code \citep{mignone07}. The numerical techniques employed are identical to \citet{mukherjee16a}. We use the HLLC Riemann solver to solve the relativistic hydrodynamic equations \citep{mignone05a}, with a piecewise parabolic reconstruction scheme \citep[PPM method, ][]{colella84a}. A 3rd order Runge-Kutta integration scheme is used to perform the time integration. 

The simulations were performed on a Cartesian ($X$--$Y$--$Z$) domain of physical dimensions: $4\kpc \times 4 \kpc \times 8 \kpc$, with a grid of  $672 \times 672 \times 784$ cells. The $X$ and $Y$ grids are uniform with resolution $\sim 6$ pc. The $Z$ grid is uniform over the central region ($\sim \pm 1.6 \kpc$, 544 grid points), with the same resolution as the other two dimensions. Beyond $\sim \pm 1.6 \kpc$, a geometrically stretched grid was applied over 120 cells with a stretching ratio of $r_s\sim 1.017$\footnote{Resolution of successive cells in the $Z$ direction are: $\Delta z^{k+1} = r_s \times \Delta z^k$}. Thus the central region with the gas disk is well-resolved with a uniform resolution of $\sim 6 \pc^3$.

Energy losses due to non-equilibrium atomic cooling ($[\rho/\mu(T)]^2 \Lambda (T)$, where $\Lambda$ is the cooling function in $\ergs \cc$) are included using the results from the MAPPINGS V code \citep{sutherland17a}. We tabulate the cooling function $\Lambda$ as a function of $p/\rho$, and, during the simulation, we interpolate the cooling rate at every computational cell. The mean molecular weight, $\mu(T)$, also obtained from MAPPINGS V, monotonically varies from $\mu \sim 1.24$ at low temperatures to $\mu \sim 0.6$ when the gas is fully ionized. 
For temperatures beyond $\sim 10^{9}$ K, the cooling function is extended assuming Bremsstrahlung emission, similar to \citet{krause07a}. A lower temperature threshold of $\sim 1000$ K was enforced, below which cooling is turned off. Molecular cooling is not considered in this work. In the dense cores of the clouds, atomic cooling is strong enough to quickly drive the gas to the cooling floor. Thus, neglecting molecular cooling does not affect the dynamics of the gas in our simulations.

\subsection{Two phase initial density}\label{sec.twophase}
The initialization of the two phase ISM closely follows that of Paper I with some differences. An external gravitational potential ($\phi(r)$) is set up following the double isothermal prescription of \citet[e.g.][]{sutherland07a,mukherjee16a} which accounts for contribution from baryonic (stellar) and dark components. For the assumed parameters (listed in Table~\ref{tab.params}), the total baryonic (stellar) mass contained within $\sim 10 \kpc$ is $\sim 1.1\times10^{11} M_\odot$. A hot isothermal halo ($T\sim 10^7$K) in hydrostatic equilibrium is established. The density of the halo gas is determined by the following parameters: a central density  $n_{h0}$ ($R=0$), $R$ being the spherical radius, mean molecular weight $\mu \sim 0.6$ and atomic mass unit $m_a=1.6605\times10^{-24} \mbox{ g cm}^{-3}$. The density is given by
\begin{equation}
	n_h(R)=n_{h0}\exp\left[\frac{-\phi({R})}{k_B T/(\mu m_a)}\right],
\end{equation}
where  $k_B$ is the Boltzmann constant.
\begin{figure}
	\centering
	\includegraphics[width = 6.5cm, keepaspectratio] 
	{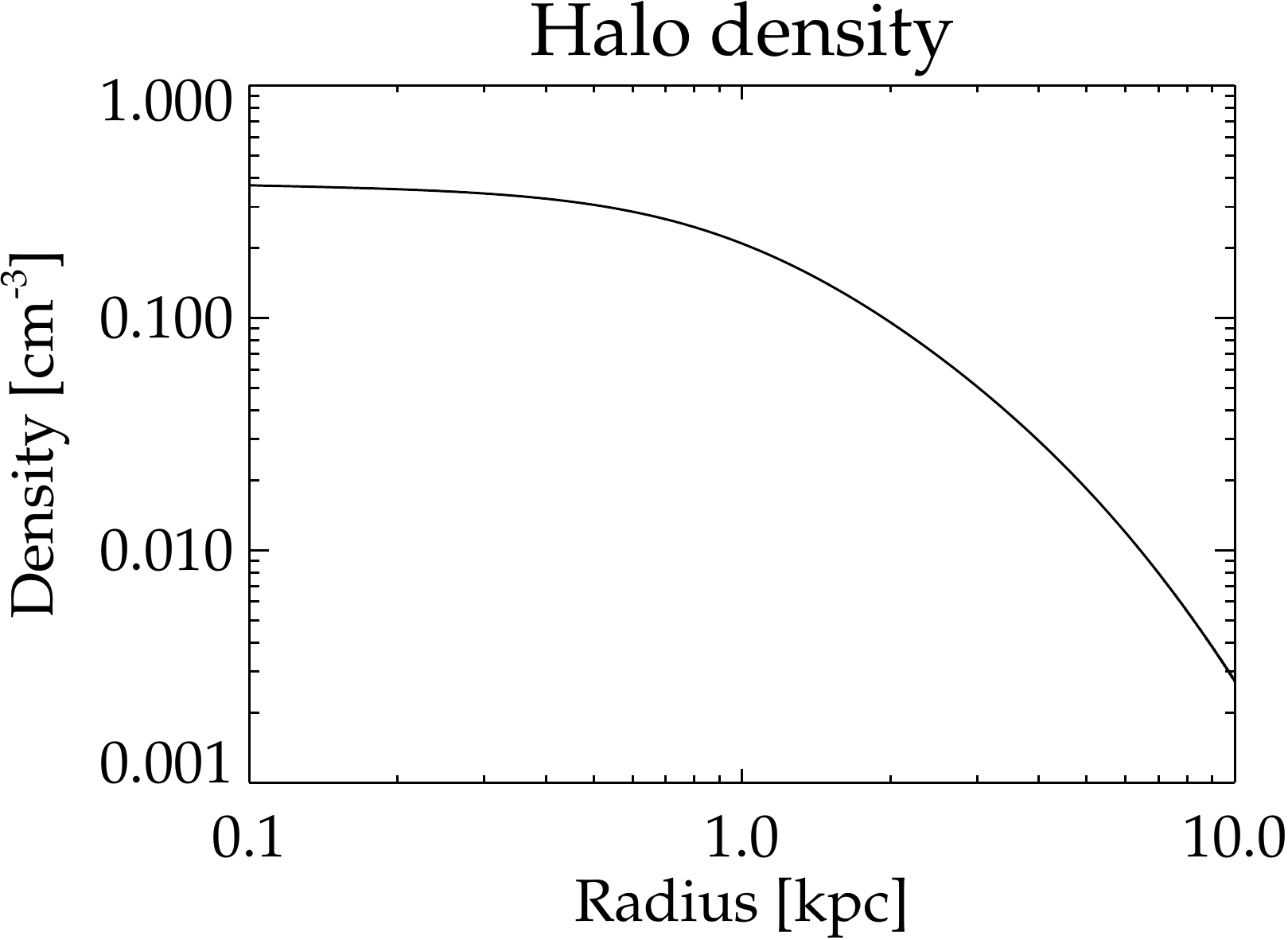}
	\caption{\small Variation of the density ($n_h(R)$) of the hot ambient medium as a function of spherical radius ($R$).}
	\label{fig.halorho}
\end{figure}
Figure~\ref{fig.halorho} shows the variation of the halo density with radius for the simulations presented here. Beyond the core radius the halo density decreases ($n_h \sim 0.2$ at $r\sim 1 \kpc$) rapidly as a power law.
\begin{figure*}
	\centering
	\includegraphics[width = 6.5cm, keepaspectratio] 
	{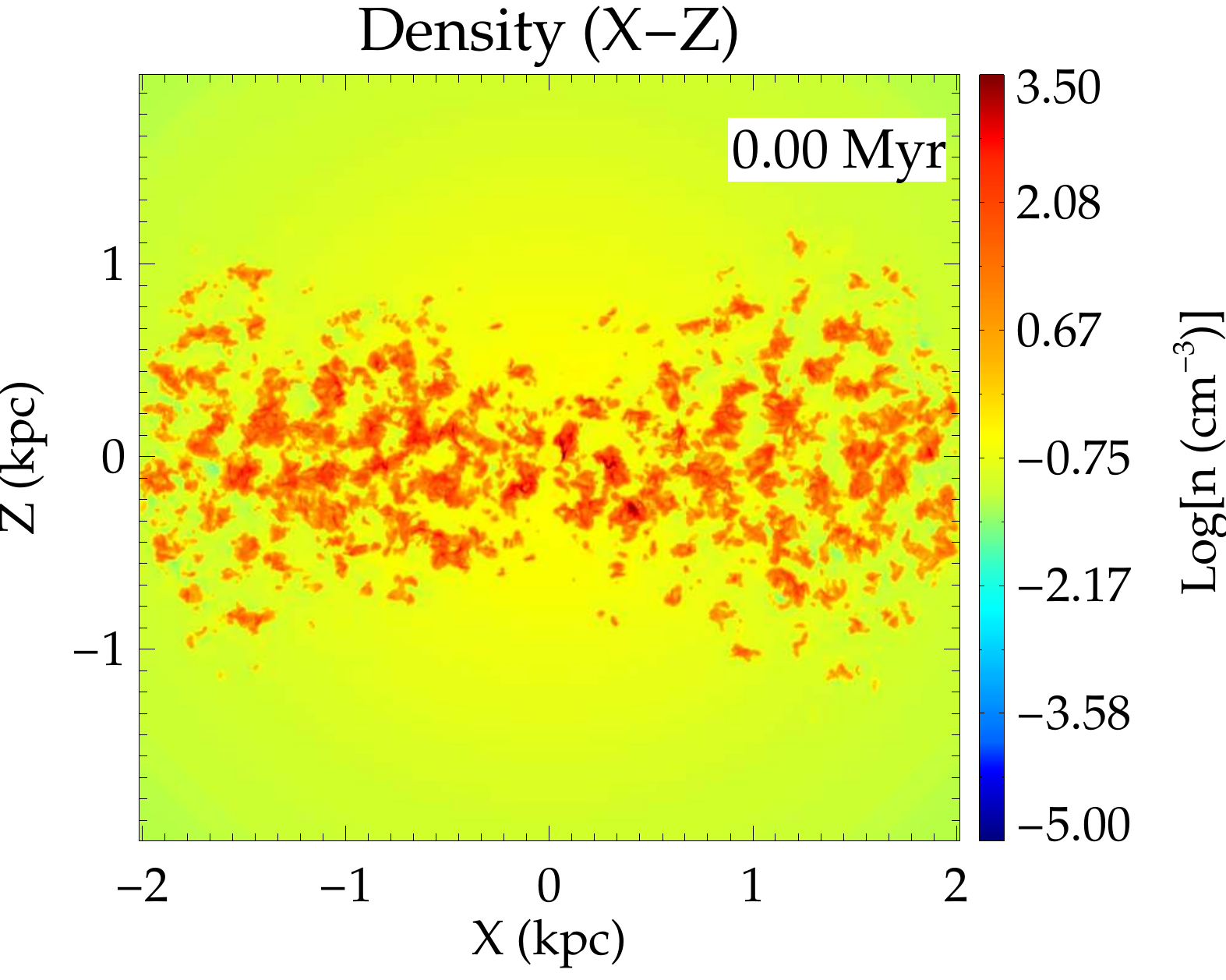}
	\includegraphics[width = 6.5cm, keepaspectratio] 
	{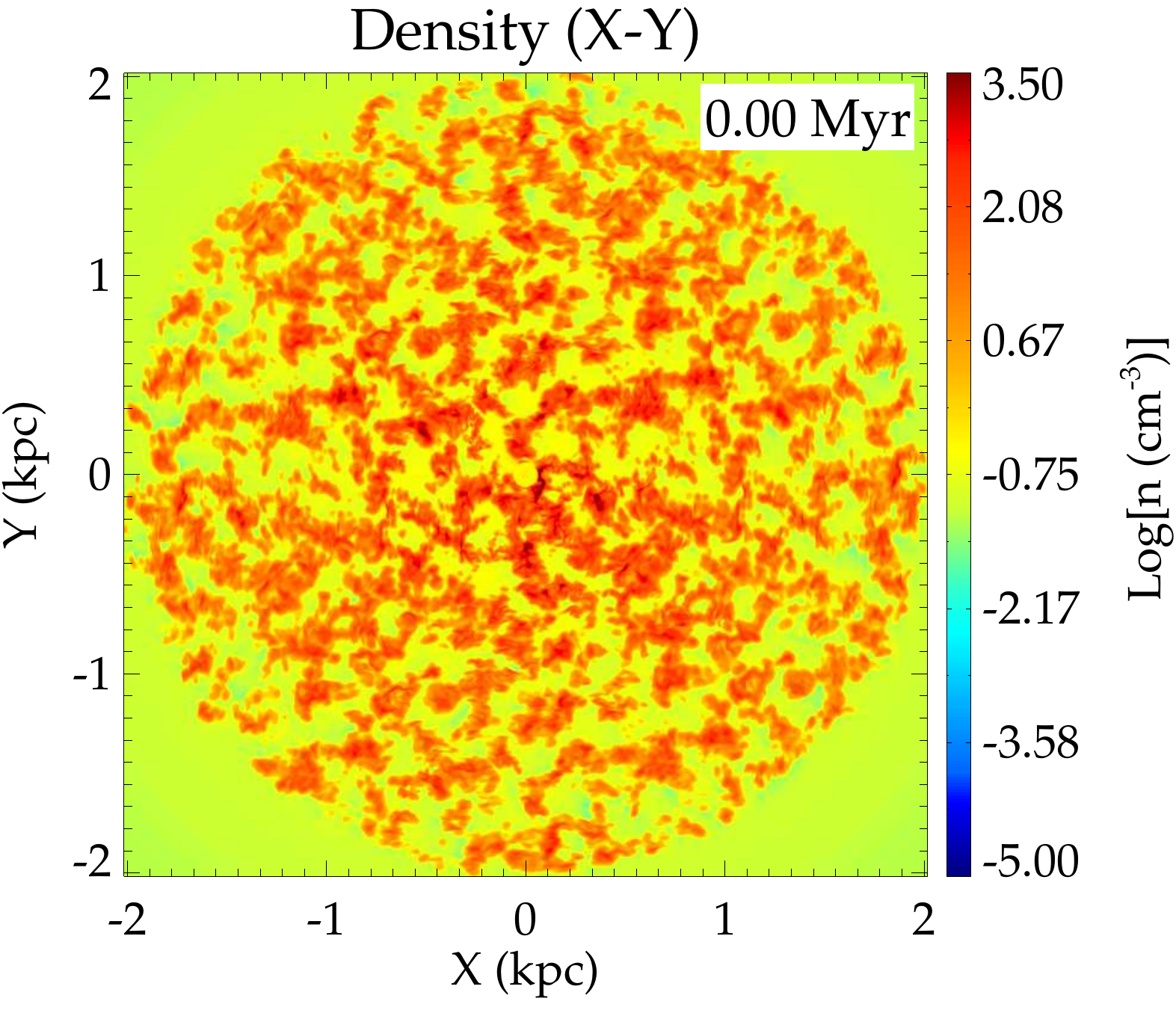}
	\caption{\small Left: Density ($\log \left(n [\cc]\right)$) of the turbulent disk in the $X-Z$ plane after brief settling. Right: Density in the $X-Y$ plane. The jet is injected into this settled turbulent disk structure. }
	\label{fig.nojet}
\end{figure*}
We embed a turbulent gas disk of radius $\sim 2 \kpc$ into the halo. The density of the turbulent disk is described by a distribution of fractal clouds. The clouds follow a single point log-normal distribution and a two point power-law correlation with a Kolmogorov spectrum \citep{sutherland07a,wagner11a,mukherjee16a}. The fractal density is created using our  publicly available pyFC routines \footnote{\url{https://pypi.python.org/pypi/pyFC}.}; the cloud distribution is then interpolated into the PLUTO computational domain. The fractal cubes generated are characterized by the following parameters: physical size $1.8 \kpc \times 1.8 \kpc 1.8 \kpc ^3$ on a grid of dimension $350 \times 350 \times 350$ points, maximum wavelength $\lambda _{\bf max} = 1.8/6= 0.3 \kpc$\footnote{This corresponds to setting $k_{\rm min}=6$ as an input parameter to the pyFC cube generator}, with a mean $\mu=1$ and dispersion of $\sqrt{5}$ describing the log-normal distribution. The average size of the largest clouds is, thus, $\lambda _{\rm max}/2 =150 \pc$.   

The fractal density distribution is then apodized with a mean warm phase density profile given by eq.~\ref{eq.disk} below. The parameters defining the number density of the warm gas $n_{\rm w} (r,z)$ as a function of cylindrical coordinates $r$ and $z$, are the central density $n_{\rm w}(0,0)$, and the total dispersion $\sigma_{\rm t}$. We neglect the contribution of the thermal component ($kT/(\mu m)$) to the turbulent dispersion, which for a cold disk at $T=10^3$ K is only a few $\kms$, much smaller than the imposed velocity dispersion. The mean warm density is:
\begin{equation}
\frac{n_w(r,z)}{n_{w0}}\!\! =  \exp\left \lbrace-\frac{1}{\sigma _t^2}  \left[ \phi(r,z) - \epsilon^2 \phi(r,0) -(1-\epsilon^2) \phi(0,0) \right]  \right\rbrace ,\label{eq.disk}
\end{equation}

\citep{sutherland07a}. In the above we assume the azimuthal velocity component to be a fraction $\epsilon$ of the Keplerian velocity:
\begin{equation}
v_{\phi}=\epsilon \left[\frac{r \partial \phi(r,z)}{\partial r}\right]^{1/2}.
\end{equation}
 For this work we assume $\epsilon=0.93$ for all simulations. The fractal cloud is set to be in pressure equilibrium with the ambient gas. Thus initially the temperature of a cloud cell is  $T_{\rm cloud}=n_hT_h/(n_w)$.  For a cloud temperature greater than a critical value ($T_{\rm cloud} > 3\times10^4$ K), the fractal density is replaced with halo gas, and this places a lower bound on the cloud density. Beyond the critical temperature the clouds are considered to be thermally unstable \citep{sutherland07a}. 

In addition to the rotation, we impose a turbulent velocity on the distribution of warm clouds. Each Cartesian component of the velocity is drawn from an independent random Gaussian distribution with dispersion $\sim 100 \kms$. The turbulent velocity field is also created using the pyFC routine on a uniform grid of size same as the density field. The velocity field is initialised with a Kolmogorov power spectrum (pre-apodization) and $\lambda_{\rm max}=0.6 \kpc$, twice that of the density field. The  $\lambda_{\rm max}=0.6 \kpc$ for the velocity field is chosen to be larger than that of the density distribution to ensure that the clouds do not immediately blow up at the start of the simulation due to the imposed turbulent velocity. The turbulent velocity field thus created is then interpolated on to the computational domain, and apodized with the medium. Computational cells not deemed as the warm medium ($T>T_{\rm cloud}$), are replaced by halo gas, with the velocity set to zero.  

The fractal disk is then evolved without injecting the jet. The clouds interact and settle into a turbulent disk, similar to the results on cloud settling discussed in Paper I. The density after settling is shown in Fig.~\ref{fig.nojet}. The settled density thus forms a clumpy turbulent disk with average cloud sizes of $\sim 100$ pc, similar to clumpy star forming regions observed in high redshift galaxies \citep{jones10a,livermore15a}. The power sepctrum of the settled disk is close to the Kolmogorov slope (see more in Appendix~\ref{append.powspec}). During the settling, the velocity dispersion decreases, typical of decaying turbulence without a driving force (also discussed in Paper I). The effective dispersion of the settled disk which the jet encounters is $\sim 80-100 \kms$,  similar to the observed dispersion in high redshift galaxies \citep{forster09,wisnioski15}.

\subsection{Jet parameters}
\begin{figure}
	\centering
	\includegraphics[width = 5cm, keepaspectratio] 
	{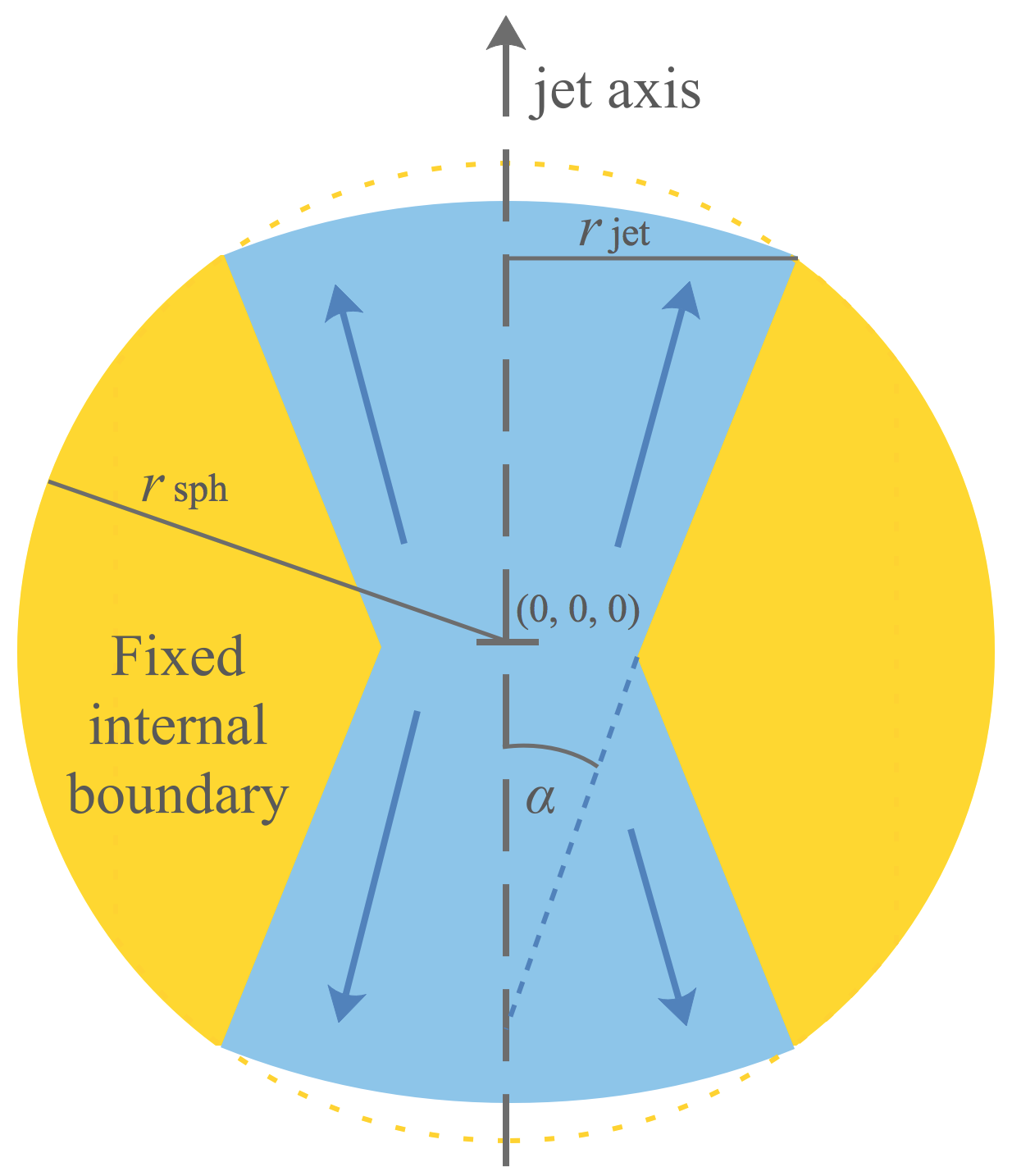}
	\caption{\small The geometry of the jet injection through an internal spherical PLUTO boundary. $r_{\rm jet}=30$ pc is the jet radius. $r_{\rm sph}=60$ pc is the radius of the internal spherical region. The origin of the outflow is the centre of the grid at $(0,0,0)$. The outflow is directed radially in the shape of a cone whose virtual apex is below $(0,0,0)$. with a half-angle of $\alpha=10^\circ$.}
	\label{fig.icecream}
\end{figure}
The jets are injected as a bi-conical outflow with an opening angle of $20^\circ$, as shown in the schematic in Fig.~\ref{fig.icecream}.  The jet radius at the injection zone, is 30 pc. This ensures at least 10 computational cells across the diameter of the jet. The injection occurs from a spherical inner boundary inside the computational domain. The computational cells inside the inner boundary are set to have values characteristic of the halo gas and are not evolved with time. The radius of the spherical inner domain is 60 pc. For simulations A, B, F, G and H (see Table~\ref{tab.sims}) the jet axis is aligned with the $Z$ axis. For simulation C, D and E, the jet axis is inclined from the $Z$ axis by an angle $\theta_{\rm inc}$.
The simulation parameters are listed in Table~\ref{tab.sims}.

The following parameters characterize the relativistic jets injected in the simulation domain. The Lorentz factor of the jet is $\Gamma=1/\sqrt{1-\beta^2}$, $\beta=v/c$ being the jet velocity in units of the speed of light. We assume an ideal equation of state \citep{mignone07a}, where the specific enthalpy ($h$) is related to the pressure and density by
\begin{equation}
	\rho h=\rho c^2+\frac{\gamma _{\rm ad}}{\gamma _{\rm ad}-1} p
\end{equation}
The adiabatic index is assumed to be $\gamma _{\rm ad}=5/3$, which is appropriate for the thermal gas in the turbulent disk into which the energy from the jets is deposited \citep[as discussed in][]{mukherjee16a}. We define the parameter $\chi$ for a jet with adiabatic index $\gamma _{\rm ad}$ and jet pressure $p_{\rm jet}$ as
\begin{equation}
	\chi = \frac{\rho c^2}{\rho h -\rho c^2}=\left(\frac{\gamma _{\rm ad}-1}{\gamma _{\rm ad}} \right) \frac{\rho c^2}{p_{\rm jet}}\label{eq.chi}
\end{equation}
 The kinetic power injected by the jet across a surface $A_{\rm jet}$  is given by
\begin{align}
    P_{\rm j} &=\left(\Gamma ^2 \rho h - \Gamma \rho c^2 \right) c \beta A_{\rm jet} \nonumber \\
              &=\frac{\gamma _{\rm ad}}{\gamma _{\rm ad} -1} c p_{\rm jet} \Gamma ^2 \beta A_{\rm jet} \left(1 + \frac{\Gamma -1}{\Gamma} \chi \right). \label{eq.pow}
\end{align}	
For a given value of jet power ($P_{\rm j}$), Lorentz factor ($\Gamma$) and $\chi$ (as in Table~\ref{tab.sims}), the pressure inside the jet is obtained from eq.~\ref{eq.pow}. The density inside the jet can then be computed from eq.~\ref{eq.chi}.

\section{Results}\label{sec.results}
We have performed a suite of simulations with different jet powers and orientations with respect to the disk (see Table.~\ref{tab.sims}). In the following section we discuss the results of the simulations.

\subsection{Vertical jets}\label{sec.vertical}
	\begin{figure*}
	\centering
	\includegraphics[width = 6.5cm, keepaspectratio] 
	{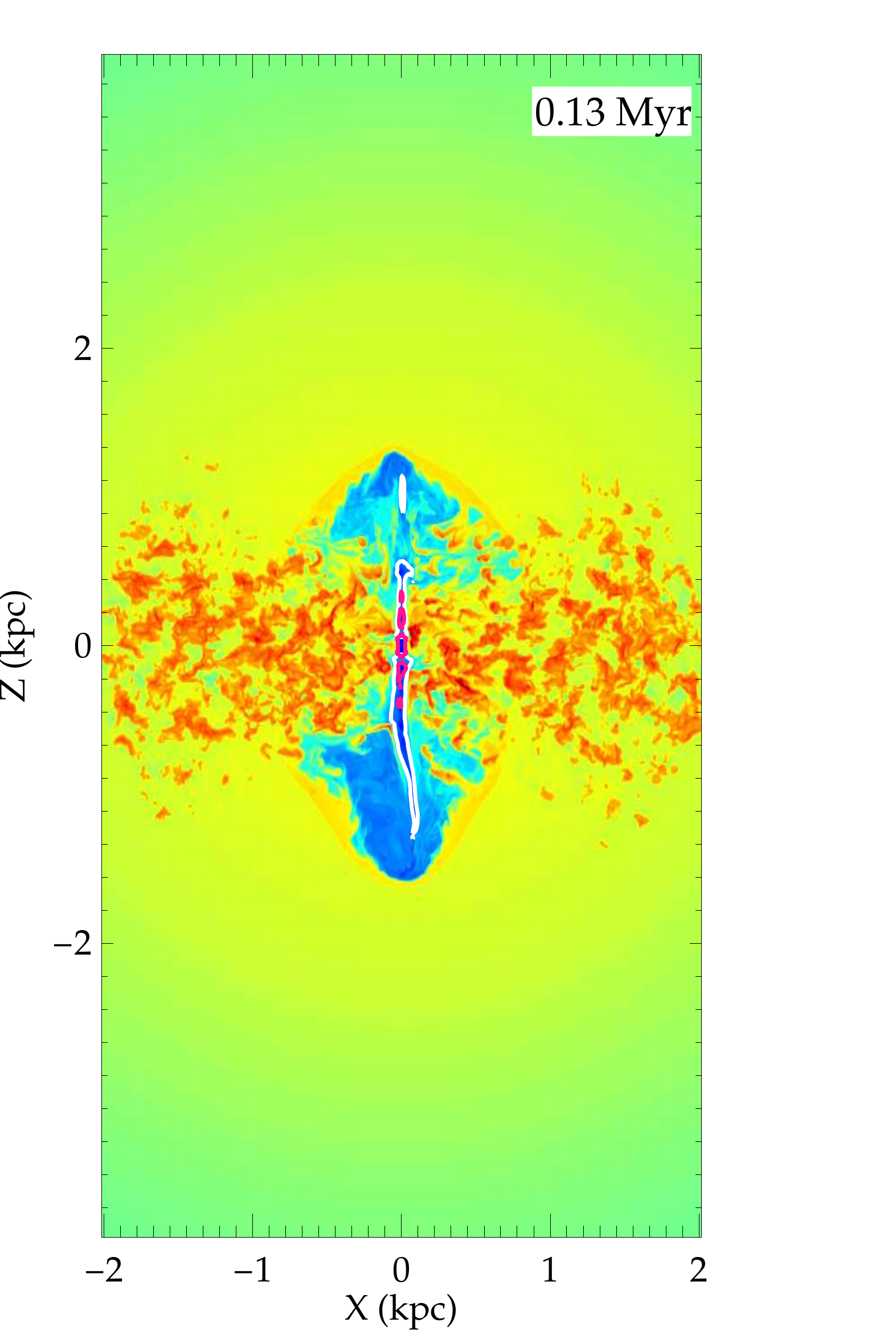}\vspace{-0cm}\hspace{-2cm}
	\includegraphics[width = 6.5cm, keepaspectratio] 
	{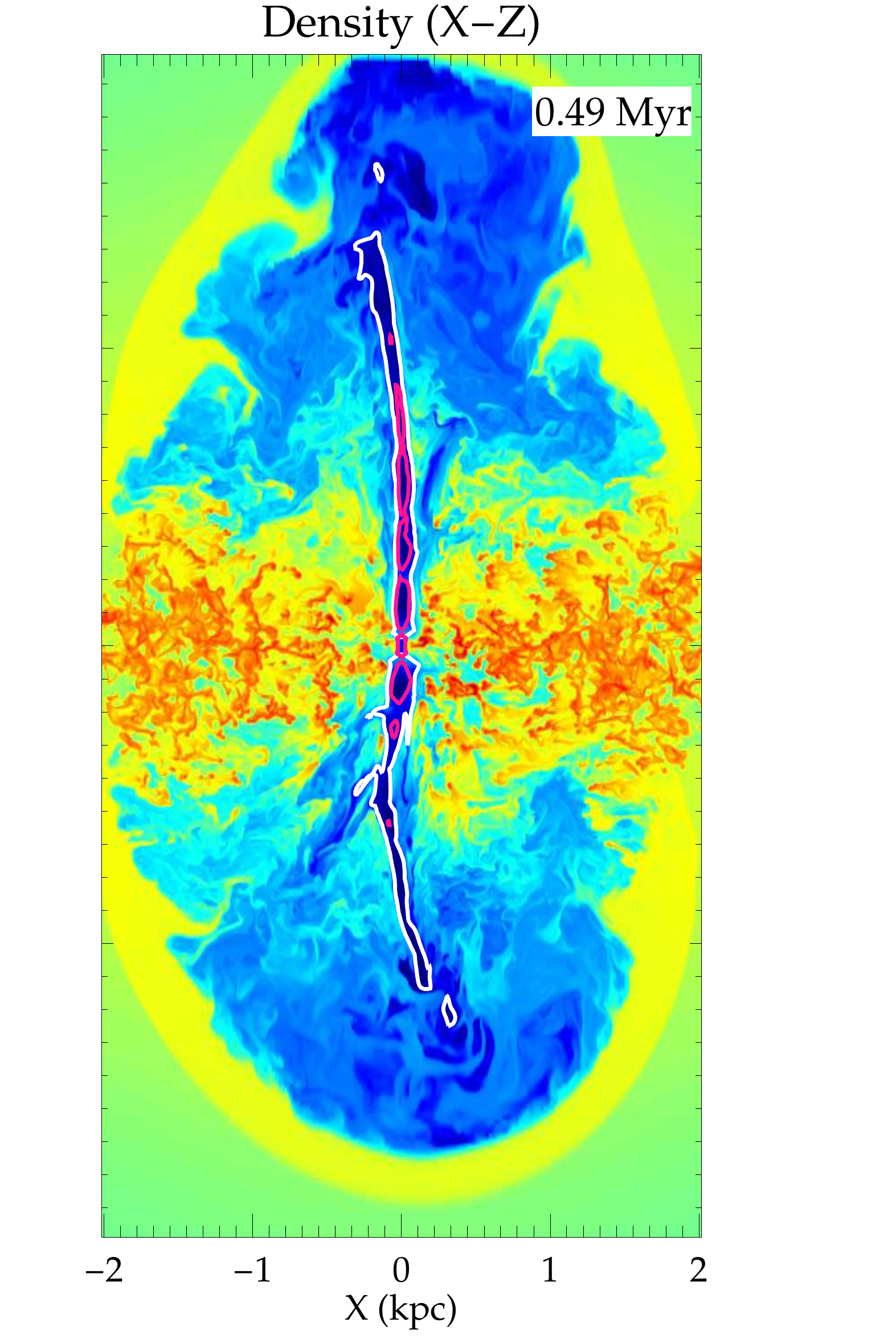}\vspace{-0.cm}\hspace{-2cm}
	\includegraphics[width = 6.5cm, keepaspectratio] 
	{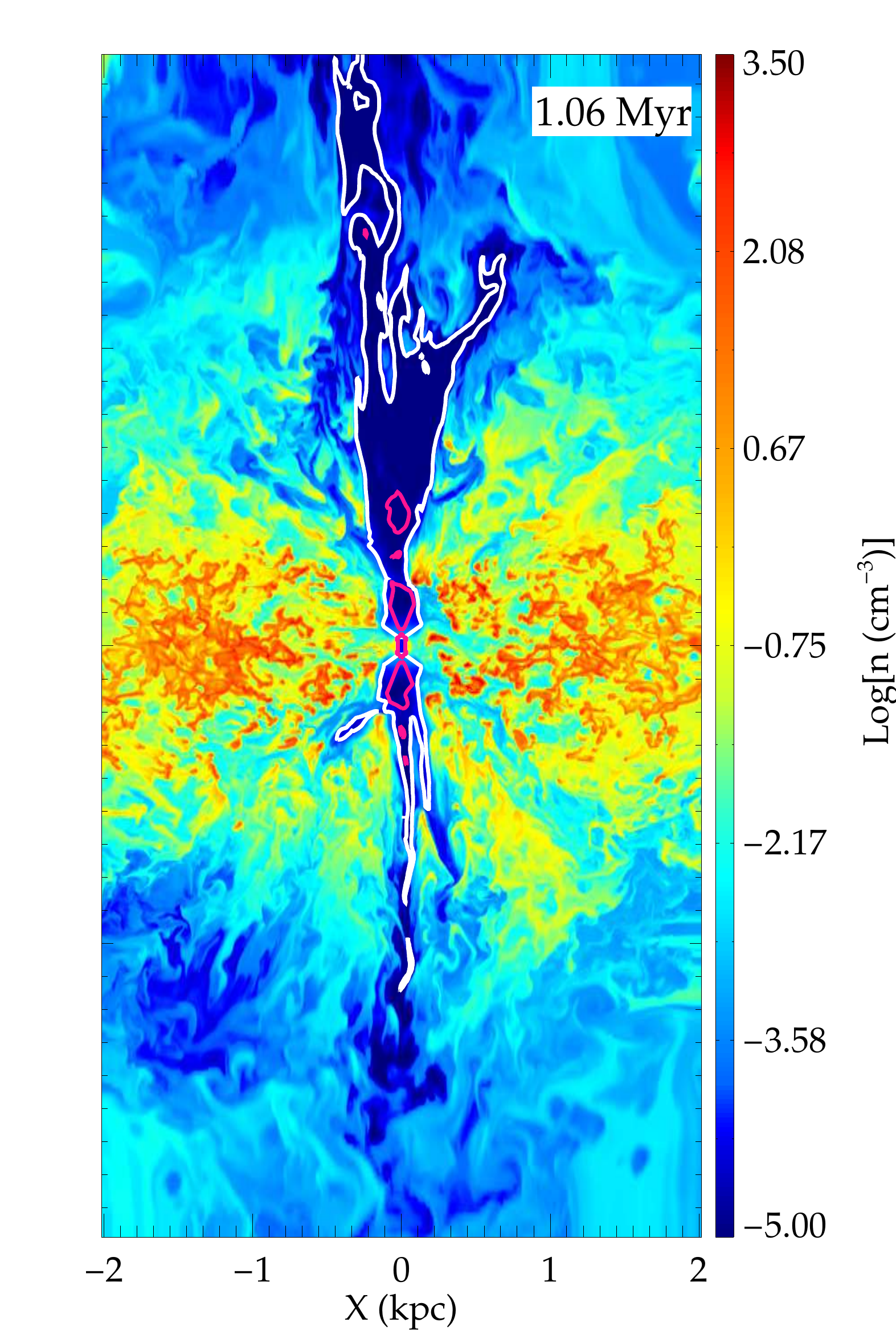}\vspace{-0cm}
	\includegraphics[width = 6.5cm, keepaspectratio] 
	{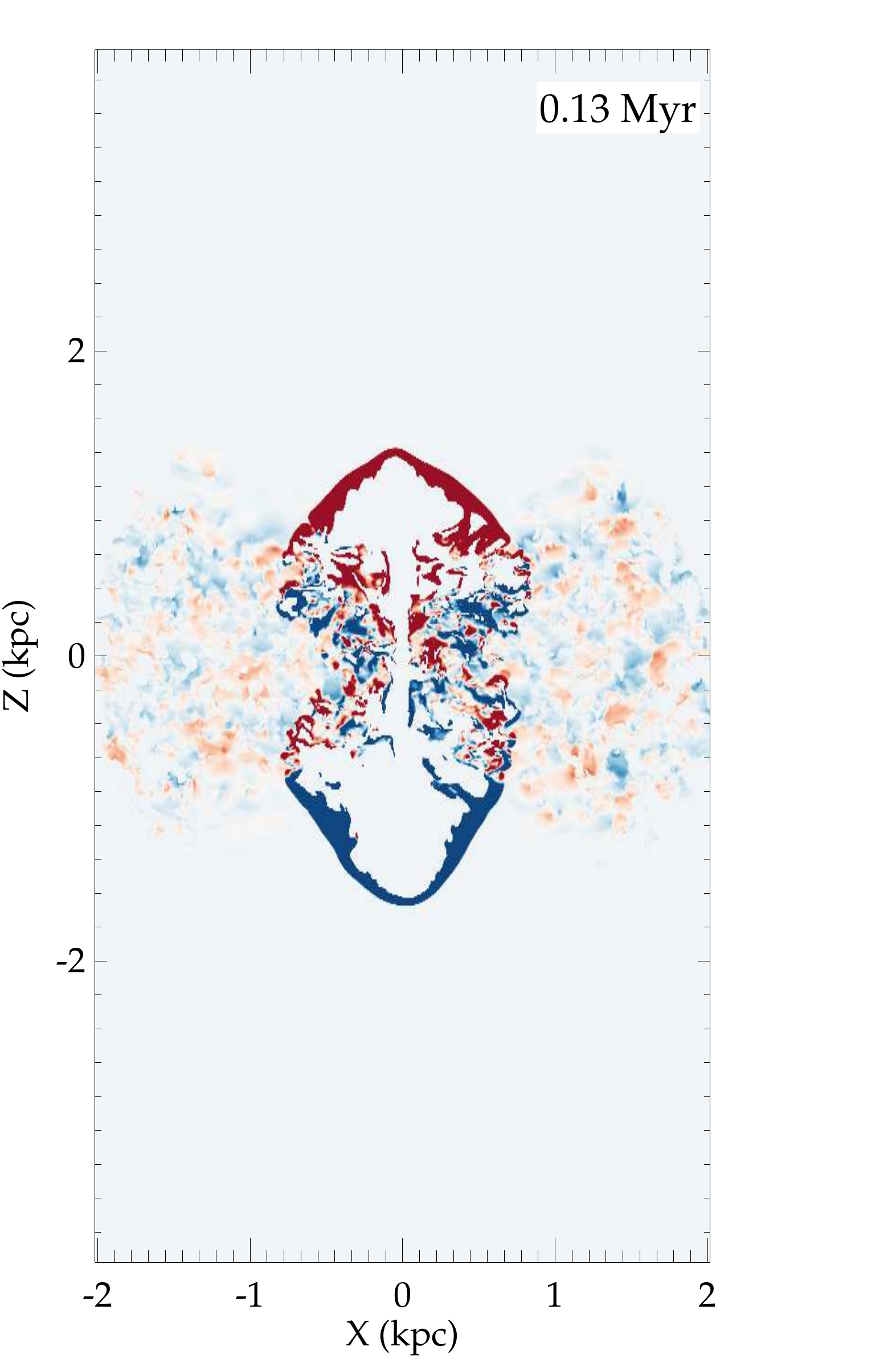}\vspace{-0.cm}\hspace{-2cm}
	\includegraphics[width = 6.5cm, keepaspectratio] 
	{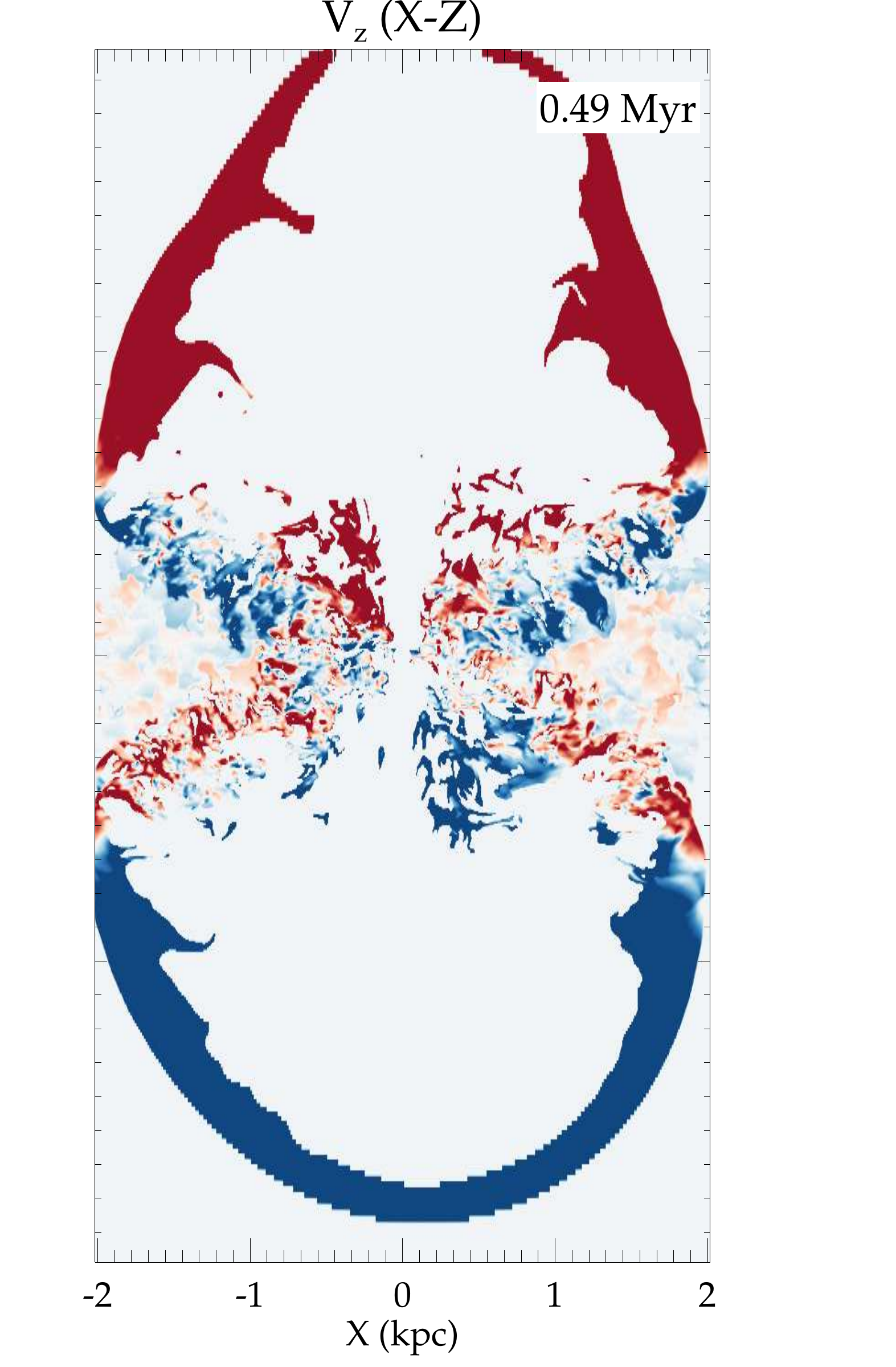}\vspace{-0.cm}\hspace{-2cm}
	\includegraphics[width = 6.5cm, keepaspectratio] 
	{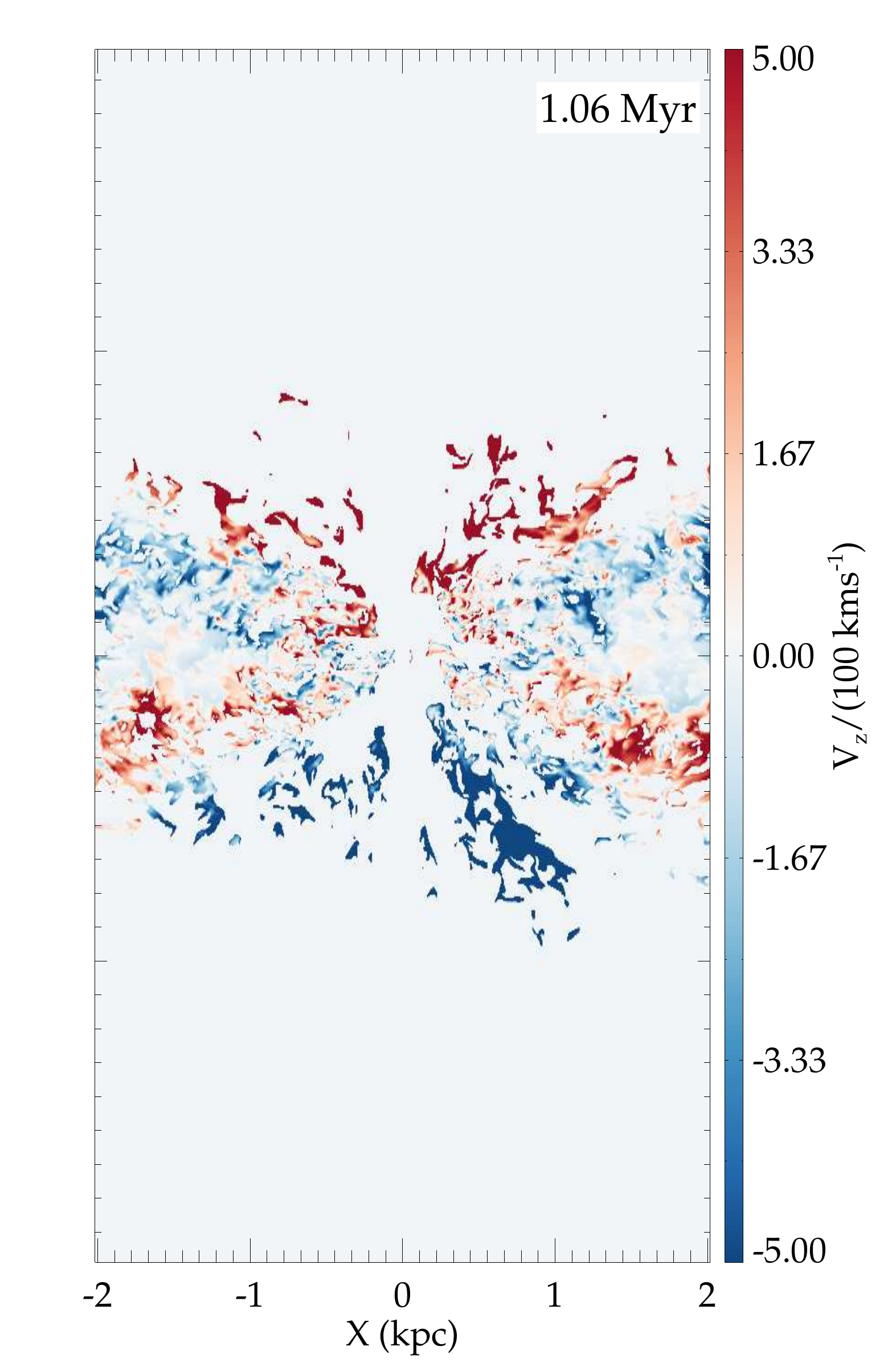}\vspace{-0.cm}
	\caption{\small \textbf{Top}: Density ($\log(n [\cc])$) at different times in the $X-Z$ plane for simulation B with $P_{\rm j}=10^{45} \ergs$, $n_{w0}=200 \cc$ (see Table~\ref{tab.params}), launched along the $Z$ axis. The white and magenta lines denotes contours of $\beta=0.5c$ and $\beta=0.9c$ respectively. \textbf{Bottom}: The vertical component of the velocity $V_z$ normalised to $100 \kms$ for the dense gas (defined here as $n>0.1 \cc$). The jet launches a central outflow, but the pressure from the energy bubble drives an inflow at the outer edges.}
	\label{fig.p45dir00}
\end{figure*}
In the case of the simulations with jets perpendicular to the disk\footnote{Simulations A, B, F, G \& H, as in Table~\ref{tab.sims}.}, although briefly frustrated by the turbulent ISM, the jets breaks out quickly, within a few 100 kyr. This is much shorter than the longer ($\sim 1 $ Myr) confinement time scales observed in simulations with spherically distributed gas and similar gas density and jet power \citep{mukherjee16a,mukherjee17a}. 

Figure~\ref{fig.p45dir00} shows the density on a logarithmic scale ($\log(n [\cc])$) and vertical velocity $V_z$ normalised to $100 \kms$ in the $X-Z$ plane for simulation B. The white and yellow dotted line in the density image represent contours of $\Gamma \sim 2$ and $\sim 1.05$ respectively, corresponding to speeds of $\beta \sim 0.9c$ and $\sim0.5c$ respectively. The contours denote the region of the flow with bulk relativistic velocities, and trace the evolution of the jet.

The figures show the evolution of the jets after break out from the disk. The jets create high pressure energy bubbles driving a strong shock through the ambient medium. The jets and the energy bubble expand rapidly into the ambient medium, sweeping up the gas in the forward shock and evacuating cavities above and below the disk. Initially, the swept up gas lies in a thin shell at the forward shock, which at later stages shows signature of Rayleigh-Taylor instabilities at the contact discontinuity. The low densities ($\sim 0.1-0.2 \cc$) and high temperatures ($T\sim10^7$K) of the swept up gas, imply cooling times of $\sim 1.4\times10^8$ yr. Thus the shell of halo gas is not expected to feed back into the galaxy for times greater than 100 Myr.

\begin{figure*}
	\centering
	\includegraphics[width = 6.5cm, keepaspectratio] 
	{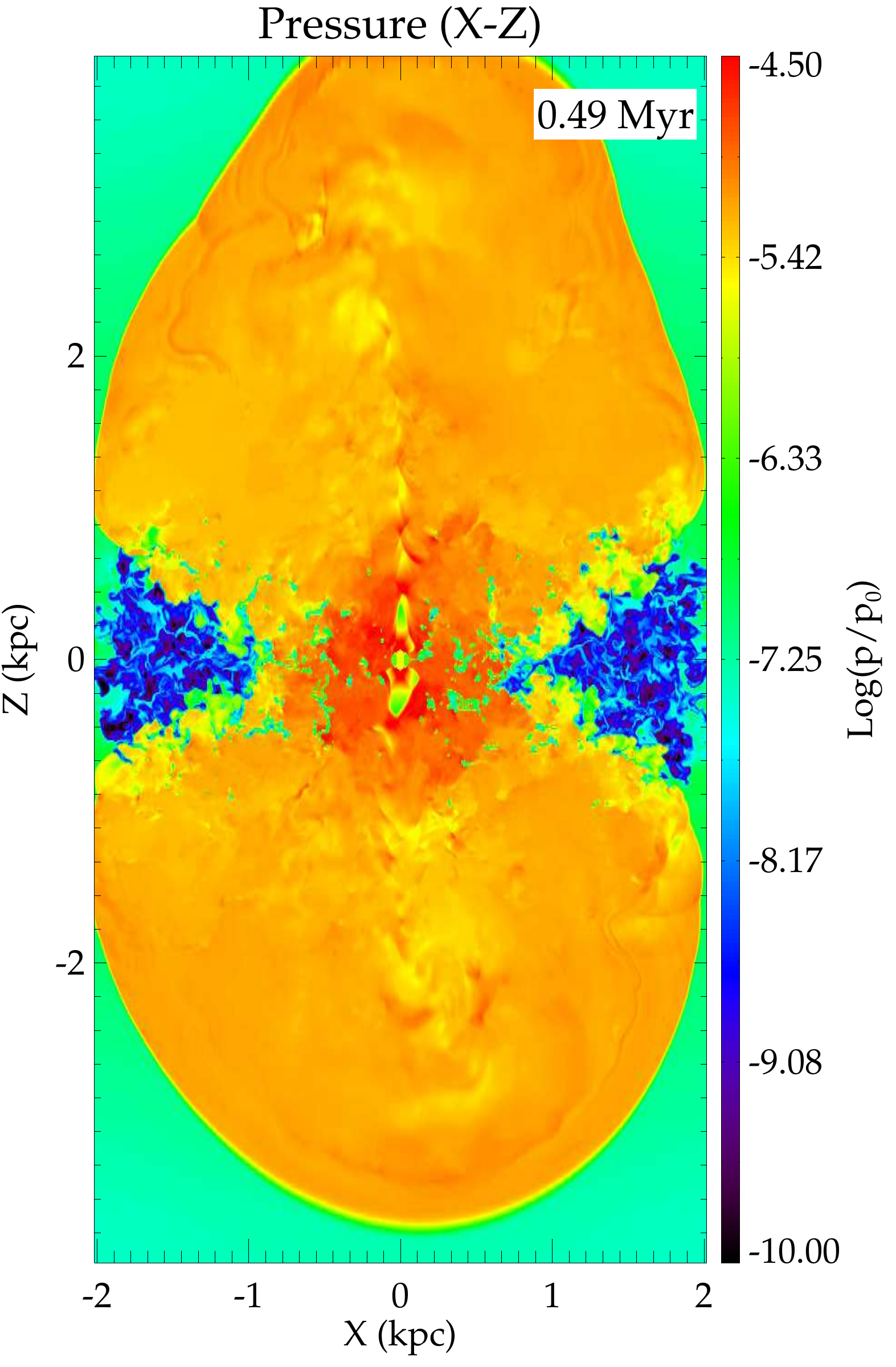}\vspace{-0cm}
	\includegraphics[width = 6.5cm, keepaspectratio] 
	{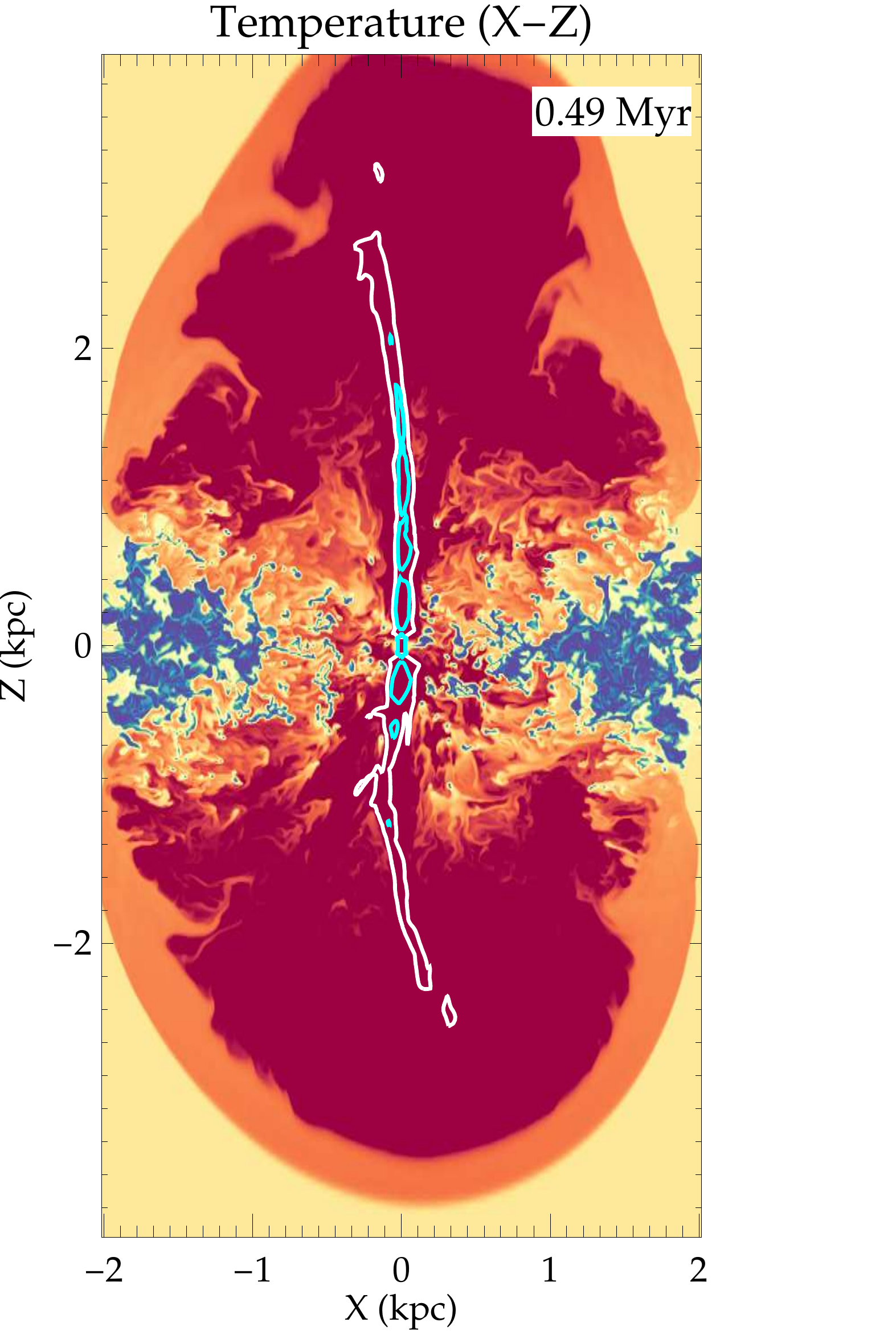}\vspace{-0cm}\hspace{-2cm}
	\includegraphics[width = 6.5cm, keepaspectratio] 
	{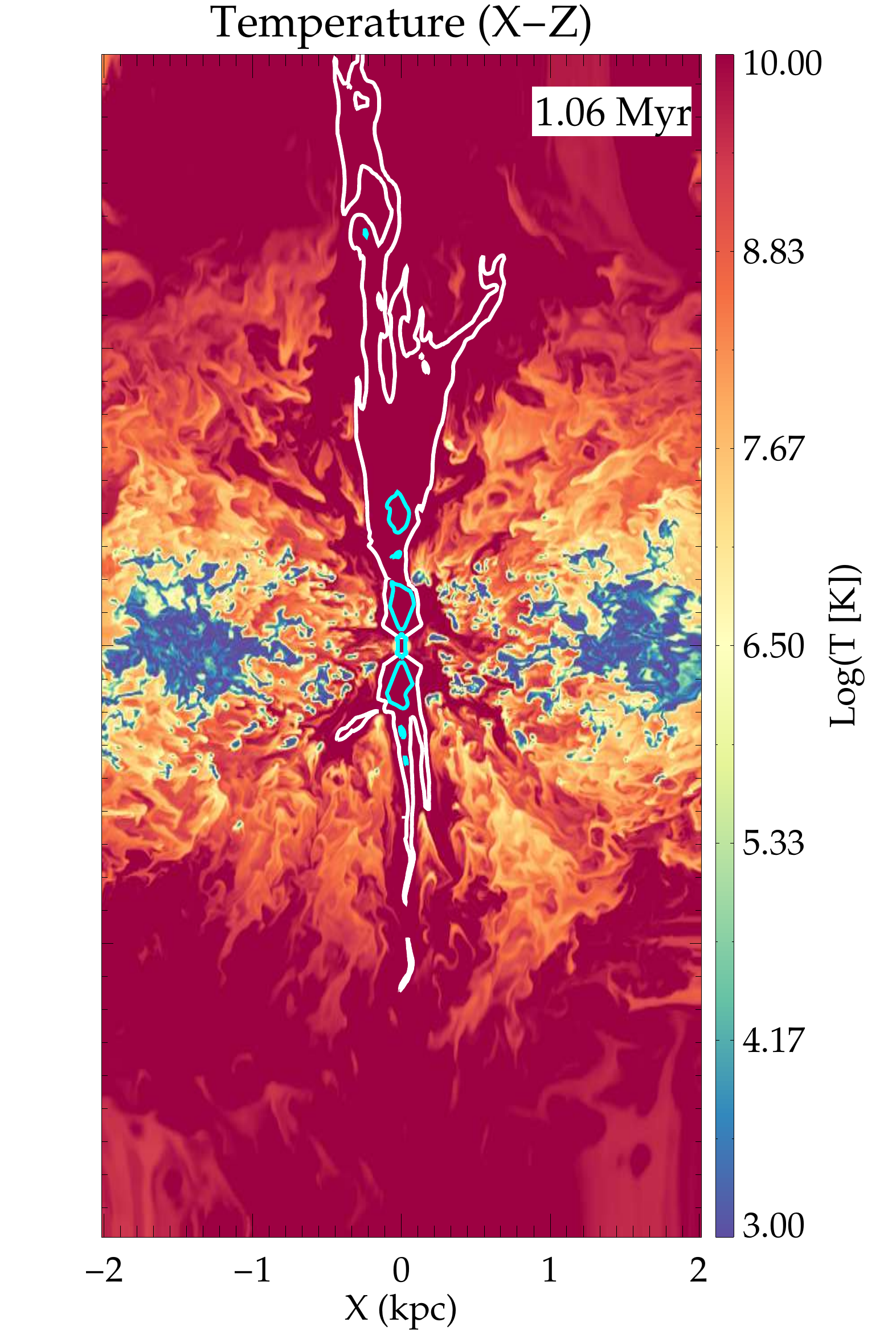}\vspace{-0.cm}\hspace{-2cm}
	\caption{\small \textbf{Left:} Pressure in the $X-Z$ plane normalised to $p_0=\rho_0 c^2=9.2 \times 10^{-4} \mbox{dynes cm}^{-2}$, where $\rho_0=0.6165 m_a$, $m_a$ being the atomic mass unit. The pressure inside the bubble is nearly homogeneous. Along the jet axis multiple recollimation shocks can be seen up to a height of $\sim 1$ kpc. \textbf{Middle and Right:} Temperature in the $X-Z$ plane at times corresponding to the middle and right panel of Fig.~\ref{fig.p45dir00}. The white and cyan lines represent contours of $\beta=0.5c$ and $\beta=0.9c$ respectively.  }
	\label{fig.p45dir00temp}
\end{figure*}
\begin{figure}
	\centering
	\includegraphics[width = 6.5cm, keepaspectratio] 
	{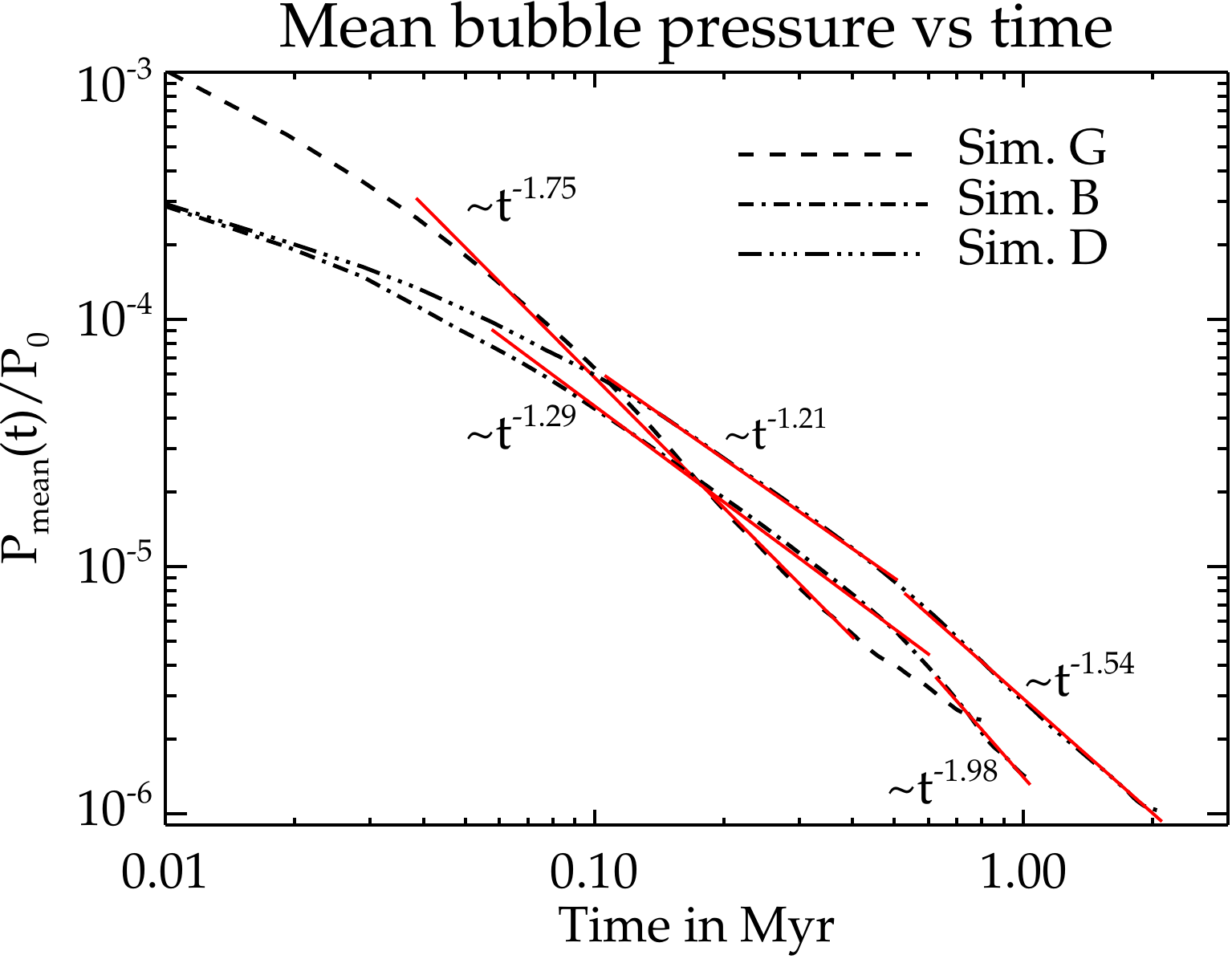}
	\caption{\small The black lines denote the evolution of the mean pressure (volume averaged) inside the energy bubble, as a function of time. Simulations B ($P_{\rm j}=10^{45} \ergs, \theta=0^\circ$), D ($P_{\rm j}=10^{45} \ergs, \theta=45^\circ$) and G ($P_{\rm j}=10^{46} \ergs, \theta=0^\circ$) are shown. The solid red lines are fits to the portions of the plots which can be well described by a power-law. The best fit exponent of the time for the different sections are presented above the respective plots. }
	\label{fig.meanpres}
\end{figure}
The gas inside the bubble is composed of a mixture of non-thermal plasma, (whose mass fraction is given by the jet tracer), and outflowing thermal gas dredged up from the disk around the central nucleus. The pressure inside the bubble is nearly uniform, as shown in a snap shot in the left panel of Fig.~\ref{fig.p45dir00temp}. The mean pressure inside the bubble decreases with time (see Fig.~\ref{fig.meanpres}) as a result of the adiabatic expansion of the bubble as well as cooling losses. The time evolution of the pressure is a power-law  with a steeper slope at later stages. As discussed later in Appendix~\ref{append.bub}, the pressure evolution is close to that described by the self-similar solutions of an adiabatically expanding energy bubble. The departure from the ideal relations of the bubble pressure vs time is the result of: 1) energy losses from cooling and 2)  non-spherical growth of the bubbles. 

The interaction of the jet with the gas clumps and the large scale energy bubble have a significant effect on the gas kinematics as we outline below. These simulations have the following features:
\begin{enumerate}
\item\textbf{Vertical outflow from the nuclear region:}
	\begin{table}
\centering
\caption{Mass ejected beyond 1.5 kpc}\label{tab.masseject}
\begin{tabular}{| c | c |}
\hline
Simulation  & Mass ejected$^a$ \\ 
\hline \\
A           & $3.71\times10^3 M_\odot$ \\
B           & $2.17\times10^5 M_\odot$ \\
C           & $6.92\times10^4 M_\odot$ \\
D           & $4.11\times10^6 M_\odot$ \\
E           & $3.55\times10^6 M_\odot$ \\
F           & $2.23\times10^6 M_\odot$ \\
G           & $4.33\times10^6 M_\odot$ \\
H           & $1.98\times10^6 M_\odot$ \\
\hline
\end{tabular} 
\flushleft
$^a$ The amount of mass in $z>1.5$ kpc, with cloud tracer $>0.98$. The values presented are at times corresponding to the end of the simulation, with the initial value at $t=0$ subtracted.
\end{table}
		In the central region ($r \lesssim 500 \pc$) the jets launch an outflow of shocked gas.  Clouds accelerated by the interaction with the jet are ejected with speeds $\gtrsim 200 \kms$. The outflowing clumps form cometary tails because of ablation by the surrounding flow. The diffuse ablated material is swept away at higher speeds exceeding $\gtrsim 500 \kms$ \citep[as also noted in ][]{mukherjee16a}.  At late stages the outflowing material is seen to extend beyond $1 \kpc$ from the centre, forming diffuse filamentary structures. The total amount of gas ejected from the disk to heights greater than $\gtrsim 1.5$ kpc, is between $\sim 2\times10^5 M_\odot - 4 \times 10^6 M_\odot$. The mass of the ejected gas increases with greater jet power (as shown in Table~\ref{tab.masseject}), as well as inclination which we discuss later. The amount of mass ejected is a small fraction of the total mass in the disk. This is because the cold dense rapidly cooling clumps are more resilient to feedback, as also discussed in earlier works \citep{nayakshin12a,mukherjee16a}. It must be noted that inadequate spatial resolution results in numerical diffusion of the ablated material \citep{cooper08a}. Hence the velocities of the ablated material should be considered as an upper limit. \footnote{High resolution simulations of wind-filament interaction \citep{barragan16a,barragan18a} show that numerical convergence requires 120 resolution elements along the filament.}

\item\textbf{Outward radial flows in the disk plane:}
\begin{figure*}
	\centering
	\includegraphics[width = 6.5cm, keepaspectratio] 
	{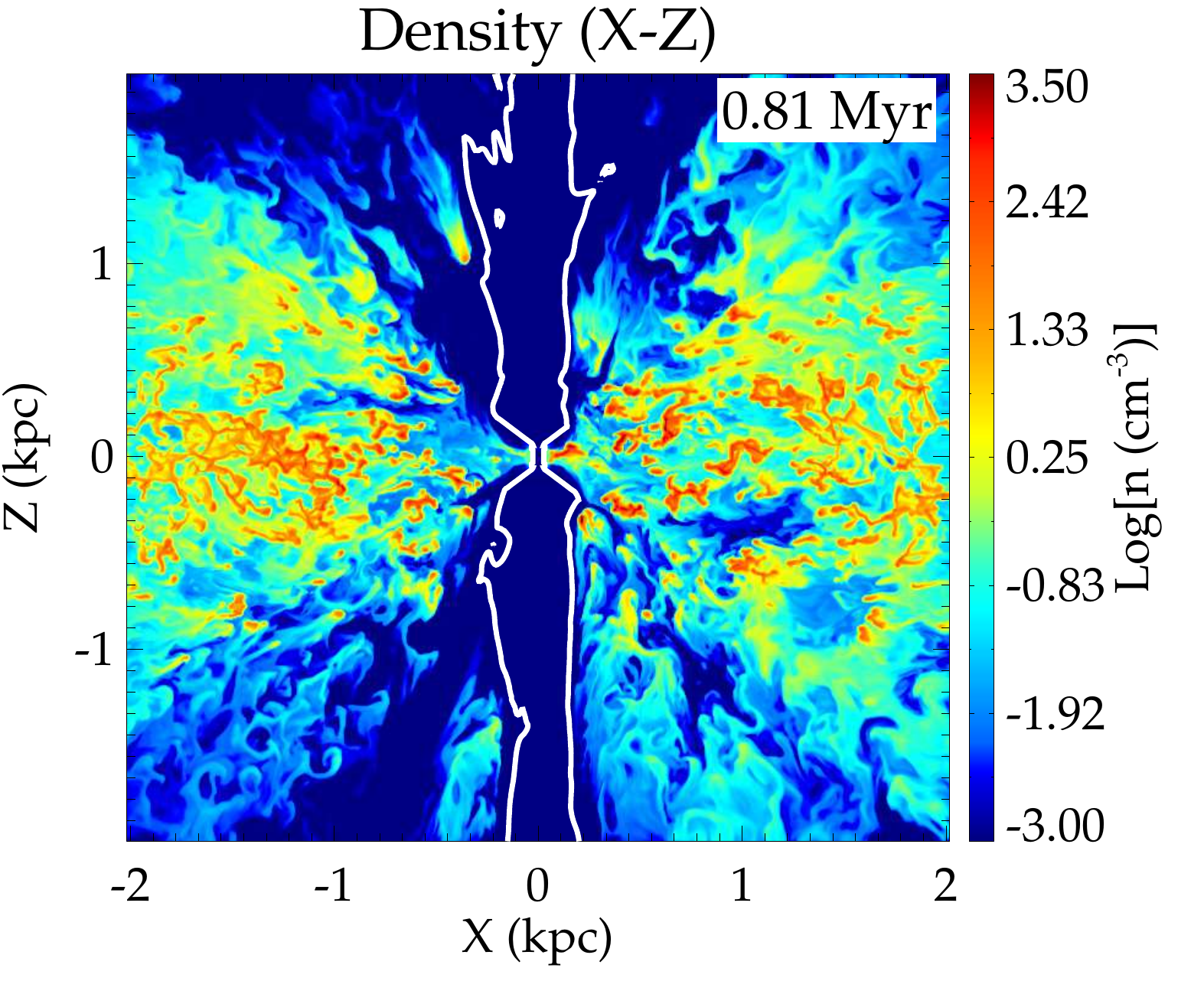}\vspace{-0cm}\hspace{-0.1cm}
	\includegraphics[width = 6.5cm, keepaspectratio] 
	{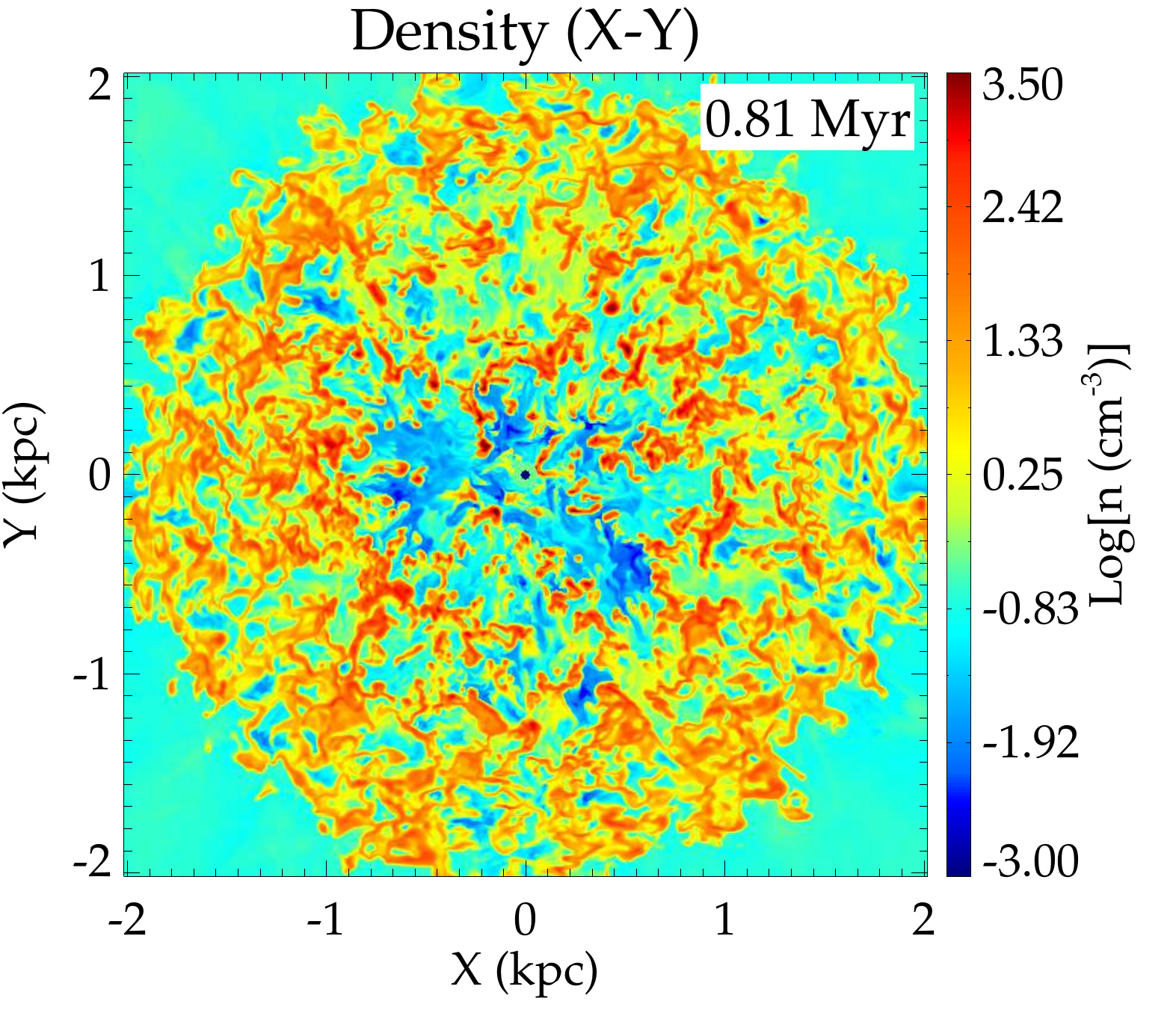}\vspace{-0cm}\hspace{-0.1cm}
	\includegraphics[width = 6.5cm, keepaspectratio] 
	{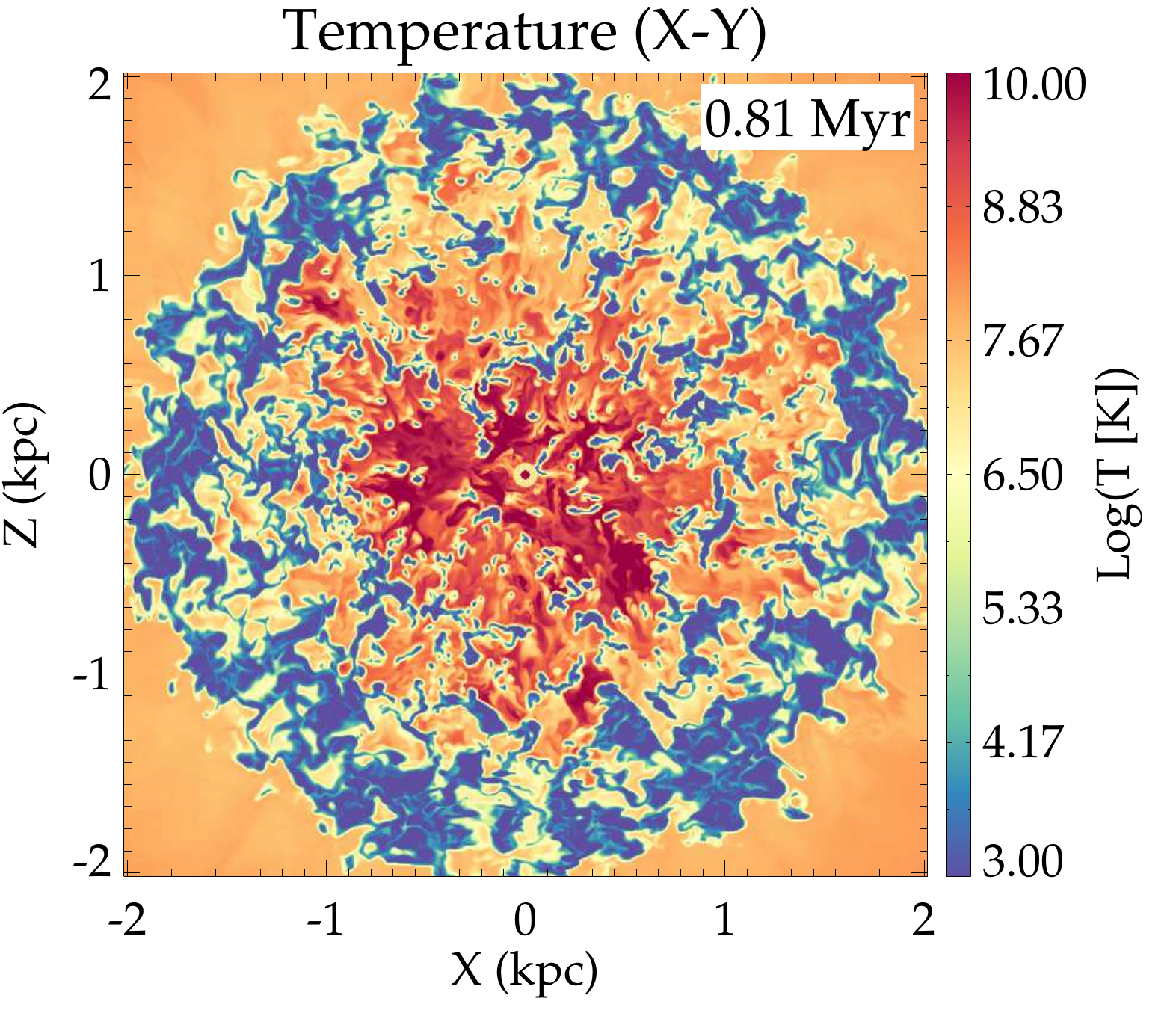}\vspace{-0.cm}\hspace{-0.1cm}
	\includegraphics[width = 6.5cm, keepaspectratio] 
	{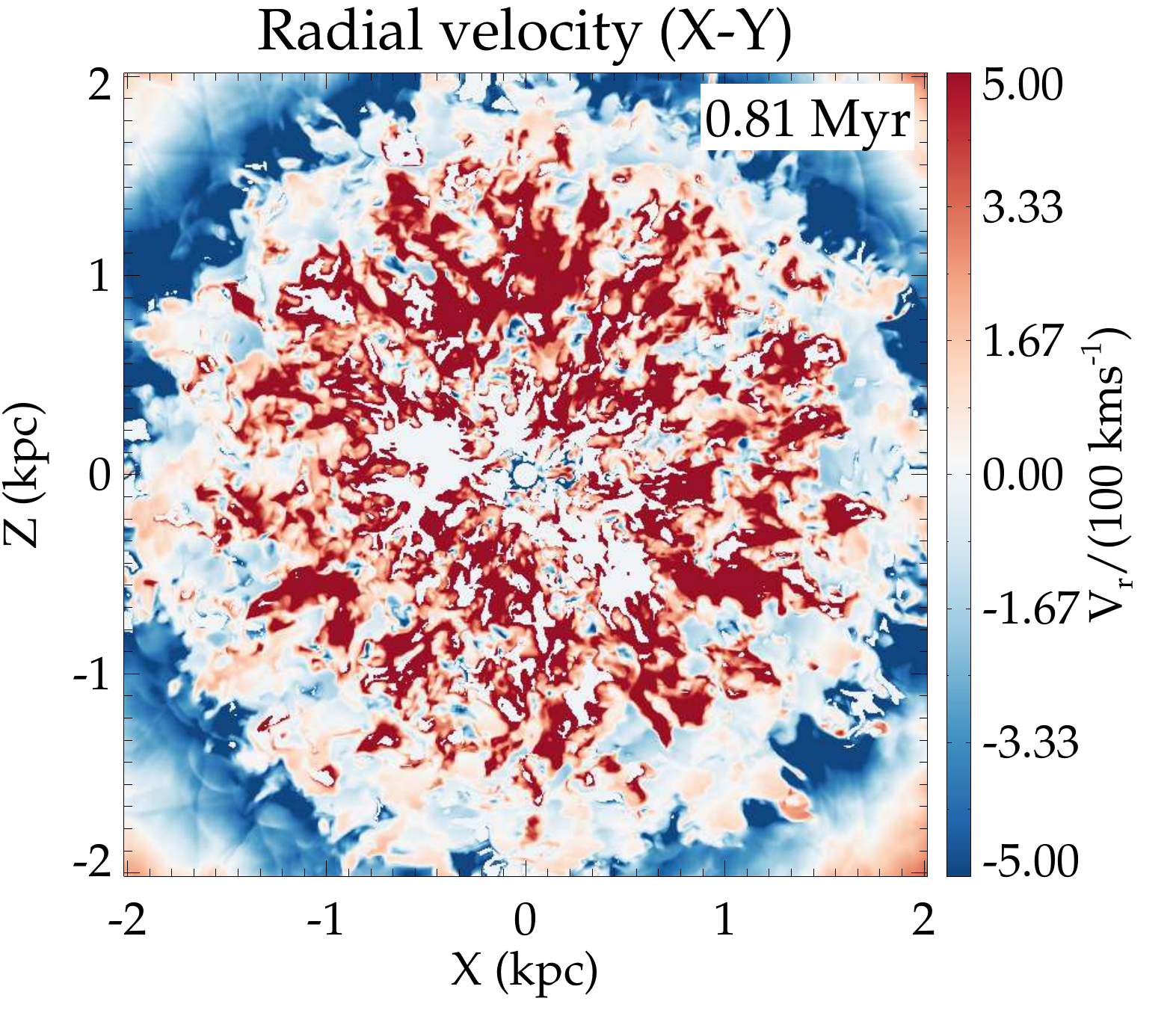}\vspace{-0cm}
	\caption{\small \textbf{Top Left:} The density in $X-Z$ plane for simulation G with $P_{\rm j}=10^{46} \ergs$, $n_{w0}=200 \cc$. The image is zoomed onto the central region, focusing on the dense disk. The white line denotes contours of constant $\beta=0.9c$, representing the relativistic component of the jet. \textbf{Top Right:} The density in $X-Y$ plane for simulation G showing the evacuation of the cavity. \textbf{Bottom Left:}  Temperature in $X-Y$ plane for simulation G, showing the radial spread of the shocks driven by the energy bubble. \textbf{Bottom Right:} Cylindrical radial velocity in the $X-Y$ plane, normalised to $100 \kms$. }
	\label{fig.p46dir00}
\end{figure*}
As the jet-driven energy bubble expands, its internal pressure drives strong outward radial flows into the plane of the disk. Fig~\ref{fig.p46dir00} shows the density, temperature and radial velocity in the $X-Y$ plane for simulation G, with a jet kinetic power of $\sim 10^{46} \ergs$. The snapshots correspond to a time ($\sim 0.81$ Myr) when the jet has evolved significantly and the energy bubble has spread over the disk. The shocks from the jet driven outflow engulf nearly all of the disk. The central region ($r \sim 1.5$ kpc) shows strong radial outflow with velocities exceeding $\sim 500\kms$.  

\begin{figure}
	\centering
	\includegraphics[width = 6.5cm, keepaspectratio] 
	{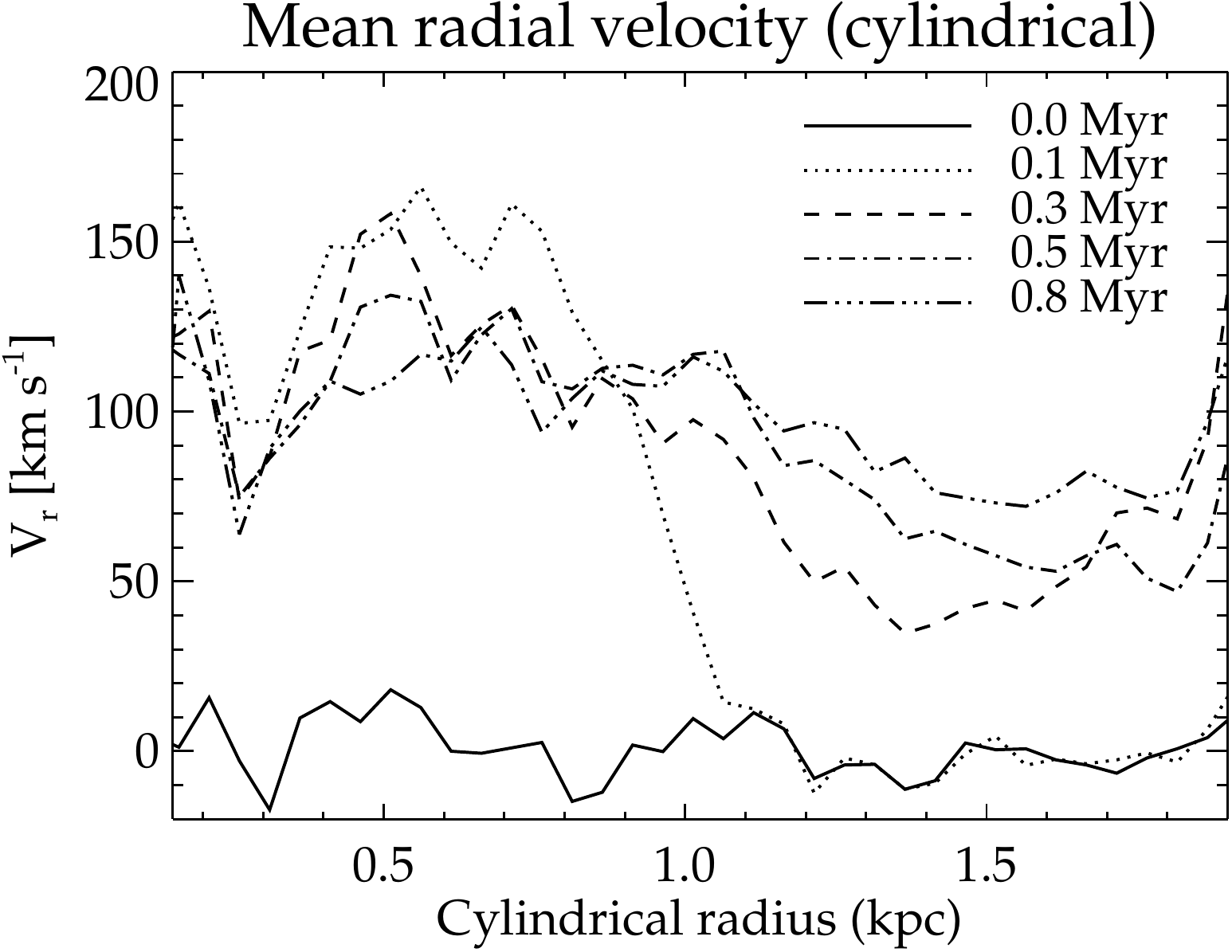}
	\caption{\small Evolution of the mean cylindrical radial velocity (mass weighted) as a function of radius, at different times for simulation G.}
	\label{fig.meanvr}
\end{figure}
\begin{figure}
	\centering
	\includegraphics[width = 7cm, keepaspectratio] 
	{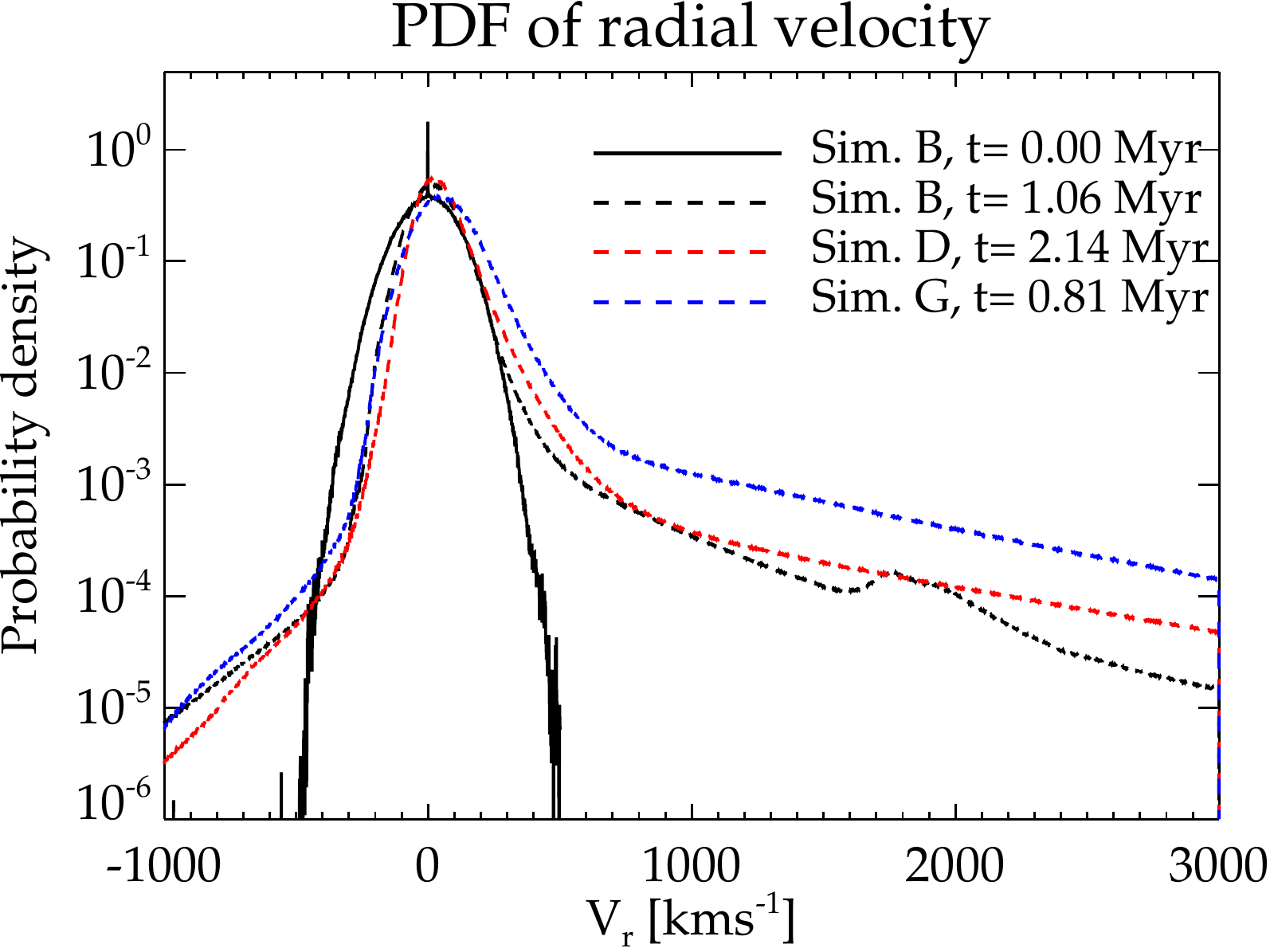}
	\caption{\small The mass-weighted probability distribution function (PDF) of the radial velocity for simulations B, D and G. The black solid line shows the PDF at $t=0$. All the simulations start from an identical ISM. The PDF shows an extended tail at high velocities due to the radial outflows driven by the jet.}
	\label{fig.velpdf}
\end{figure}
		Fig.~\ref{fig.meanvr} shows the evolution of the mass weighted mean radial velocity as a function of radius at different times for simulation G, and Fig.~\ref{fig.velpdf} shows the mass-weighted probability distribution function of the radial velocity for simulations B, D and G. At the beginning $\bar{V}_r=0 \kms$ as there is no net radial flow in the disk. This also results in a velocity PDF (shown in black in Fig.~\ref{fig.velpdf}) symmetric about 0. At $t\sim 100$ kyr, the central region has a radial velocity of $\sim 150 \kms$ up to $\lesssim 1 \kpc$, which is the approximate extent of the energy bubble inside the disk at that time. At later stages, the radial outflows extend to the entire disk and the velocity PDF develops an extended tail of high radial velocity ($\gtrsim 1000 \kms$).

The radially driven outflows evacuate the central region (radius $\sim 1 \kpc$) of the initial dense volume filling gas. The evacuated region is composed of non-thermal plasma originating from the jet together with remnant filaments of gas clouds, and material ablated from them. A similar result has been discussed in \citet{gaibler12a}, who pointed out that the initial blast wave of the jet-driven bubble, while confined to the disk, causes the lateral evacuation of disk material. In our simulations, the vertical extent of the density at the jet launch sites is small ($\lesssim 200 \pc$). Hence the jet confinement times are smaller than those  in \citet{gaibler12a}. However, even after jet break out, which happens fairly quickly for our simulations ($t \lesssim 100$ kyr), the cavity continues to expand due to the lateral expansion of the jet driven bubble.

\item \textbf{Non-circular gas motions due to jet driven outflows:}
\begin{figure}
	\centering
	\includegraphics[width = 7cm, keepaspectratio] 
	{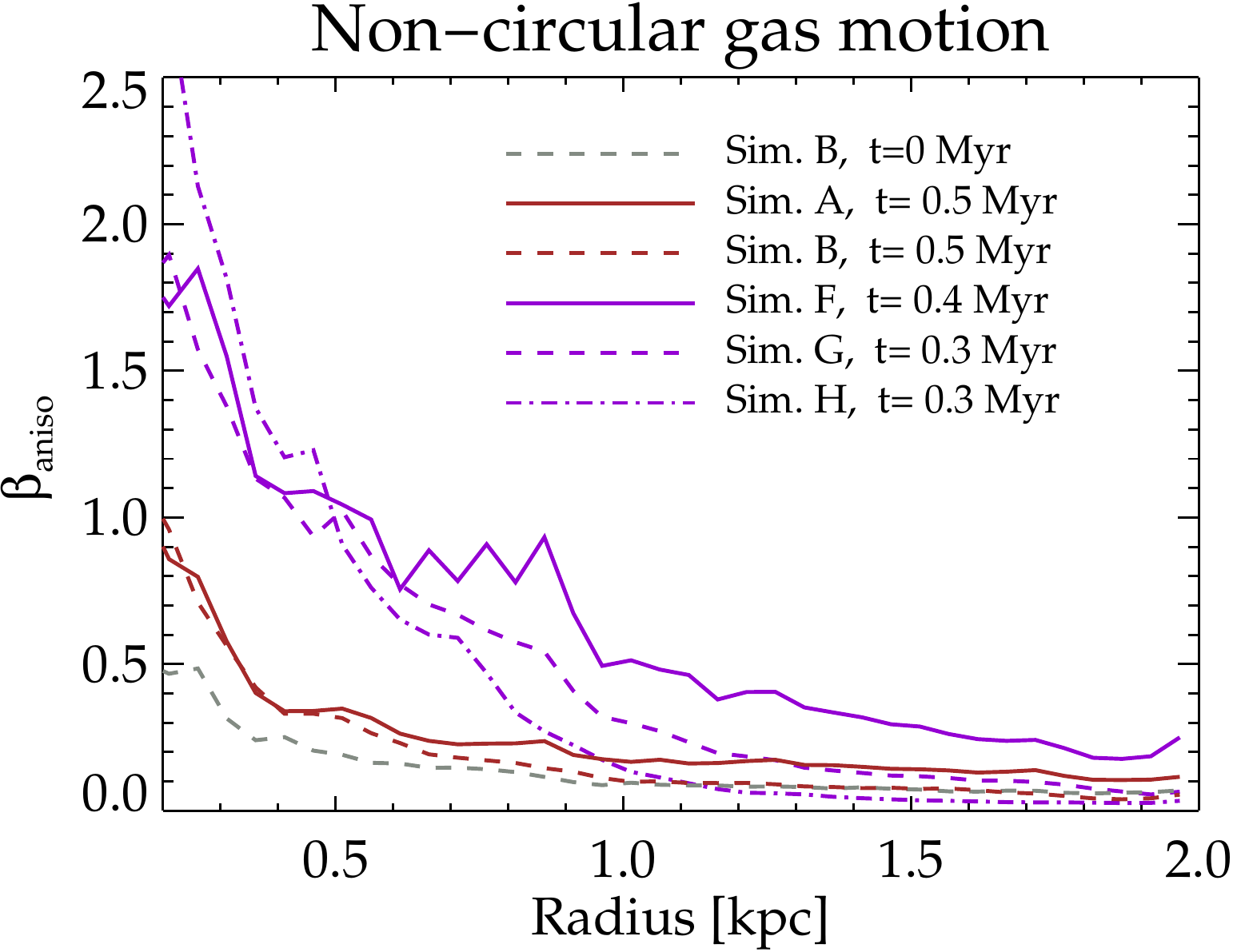}
	\caption{\small The ratio $\beta_{\rm aniso}$ as defined in eq.~\ref{eq.beta_aniso}. The central regions ($\lesssim 500 \pc$) show significant non-circular motions ($\beta _{\rm aniso} \gtrsim 1$). The plots have been generated at times similar to that in Fig.~\ref{fig.dispersion}. }
	\label{fig.beta}
\end{figure}
The jet-driven outflows result in departure of the gas kinematics from the rotation initially established. In Fig.~\ref{fig.beta} we plot the radial variation of the anisotropy parameter which we define here as
\begin{equation}
\beta_{\rm aniso}= \frac{\langle V_r^2 \rangle + \langle V_z^2 \rangle}{2 \langle V_\phi^2 \rangle}, \label{eq.beta_aniso} 
\end{equation}
where $\langle V_i^2 \rangle$ implies a mass weighted average of the square of the $i$-th velocity component. A rotating disk has $\beta _{\rm aniso} \sim 0$, whereas $\beta _{\rm aniso} \gtrsim 1$ implies significant non-circular motions\footnote{The definition of the anisotropy parameter employed here is better suited to describe the non-radial motion in a gas disk. This is different from the standard definition of the anisotropy parameter, often used to quantify stellar kinematics in elliptical galaxies \citep[e.g.][]{binney95a}}. The grey dashed line denotes $\beta _{\rm aniso}$ at $t=0$ for simulation B. Although for a purely circular velocity field $\beta _{\rm aniso}$ is expected to be 0,  the initial small non-zero value ($\beta_{\rm anisp} \lesssim 0.5$) results from the turbulent velocity field imposed on the rotating disk. The values of $\beta _{\rm aniso}$ for other simulations are similar at $t=0$. At later times $\beta _{\rm aniso}$ significantly increases in the inner kpc, as a result of the outward radial and vertical outflows induced by the jet, as discussed in the earlier sections. Thus, the central region has significant non-circular motions of the gas clouds, which are being directly impacted by the jet.

\item\textbf{Inflows into the disk driven by the energy bubble:}
The evolving high pressure bubbles spread over the disk, driving shocks and inflows into the disk towards the mid-plane; this is apparent in the velocity plots of Fig.~\ref{fig.p45dir00}. This behaviour contrasts with the outflows launched by the jet from the central nucleus. From the temperature plots in the second and third panels of Fig.~\ref{fig.p45dir00temp} and the corresponding velocity maps in Fig.~\ref{fig.p45dir00}, we see that the shocks driven by the high pressure bubble penetrate the disk with inflow speeds of $\sim 500 \kms$. Shocks also spread radially outwards from the centre due to the lateral expansion of the bubble. Consequently, the cold component of the disk is gradually engulfed with shocks driven by the bubble, which raises the temperature of the gas to $\gtrsim 10^5$K. Such a mechanism is usually expected to negatively affect star formation inside the disk. However, there may also be an enhancement in star formation as a result of the increased pressure compressing the disk \citep[e.g.][]{bieri16a}, and also from the density enhancement resulting from radiative shocks \citep{gaibler12a}. We discuss this further in Sec.~\ref{sec.sfr}.

\item\textbf{Turbulence in the disk}\label{sec.turb}
\begin{figure*}
	\centering
	\includegraphics[width = 6.5cm, keepaspectratio] 
	{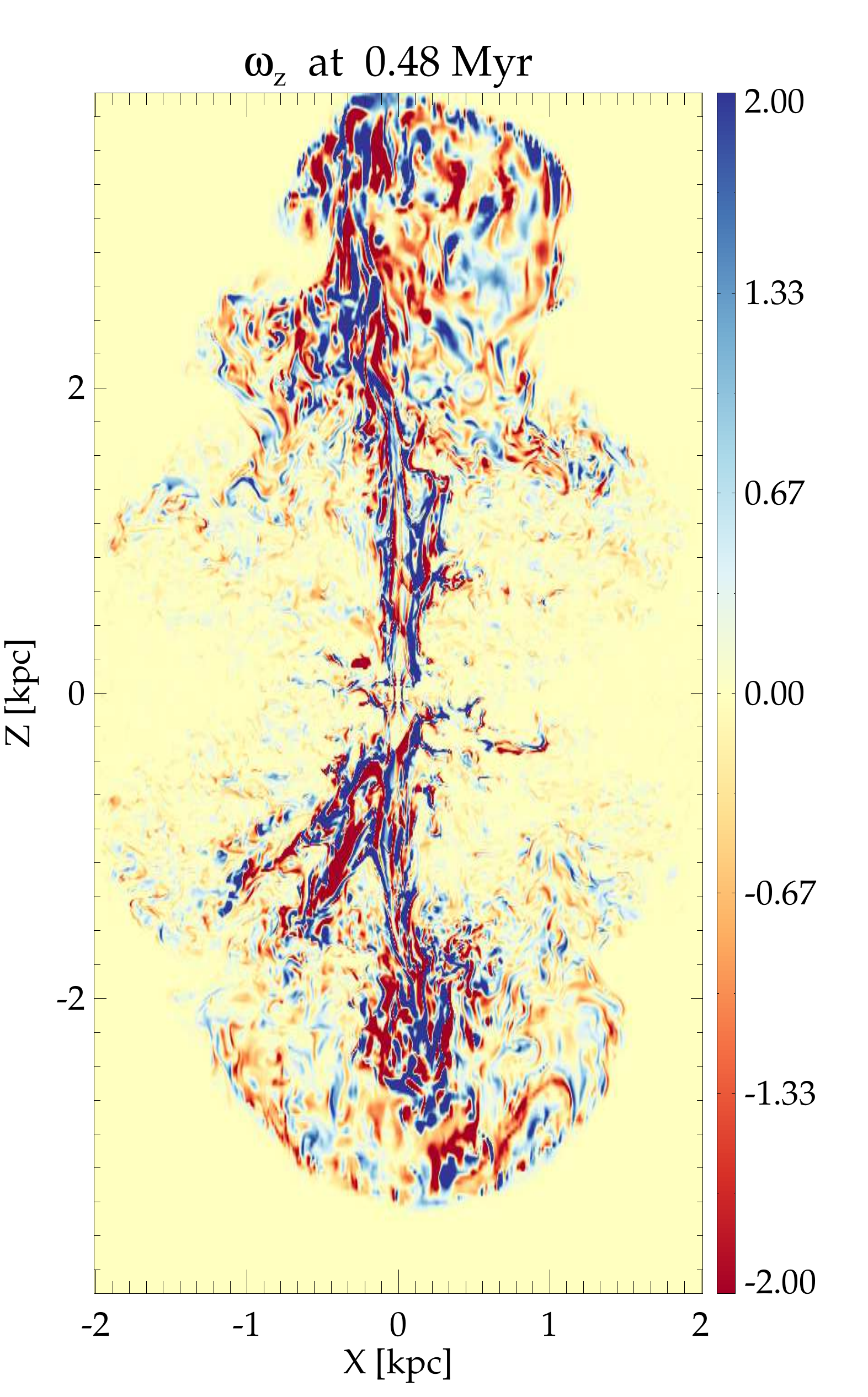}\vspace{-0cm}\hspace{-0.1cm}
	\includegraphics[width = 6.5cm, keepaspectratio] 
	{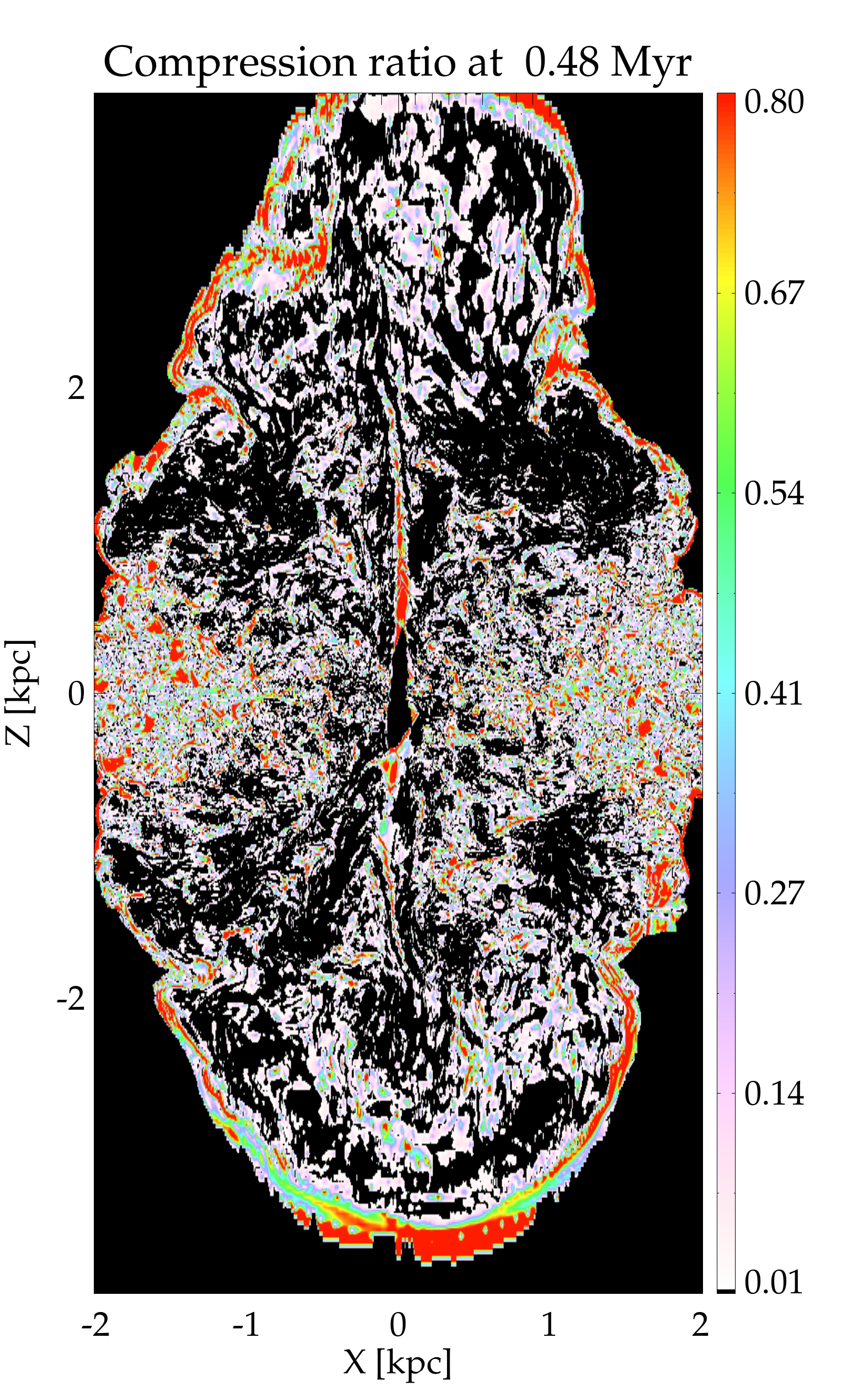}\vspace{-0.cm}\hspace{-0.1cm}
	\caption{\small \textbf{Left:} Vertical component of the vorticity ($\omega = \nabla \times \mathbf{v}$) at time $t=0.48$ Myr in the X--Z plane. \textbf{Right:} Compression ratio $r_c = |\nabla \cdot \mathbf{v}|^2 /( |\nabla \cdot \mathbf{v}|^2 + |\nabla \times \mathbf{v}|^2)$.
	}
	\label{fig.compress}
\end{figure*}
The jet-induced flows, driven into the disk, shock the initially cold cloud clumps and induce turbulence. To probe the relative strength of compressible ($\nabla \times \mathbf{v} = 0$) and incompressible motions ($\nabla \cdot \mathbf{v} = 0$), we present in Fig.~\ref{fig.compress} the compression ratio defined as \citep[e.g. see][]{iapichino11a}: $r_c = |\nabla \cdot \mathbf{v}|^2 /( |\nabla \cdot \mathbf{v}|^2 + |\nabla \times \mathbf{v}|^2)$. Similar to \citet{bourne17a}, we find that the lobe is dominated by incompressible turbulence with low compression ratio ($\sim 0.1-0.2$). The outer bow shock has has a higher compression ratio $\sim 0.8$. Within the disk, there are certain regions with higher $r_c$ where the shocks from the energy bubble impinge on the disk, or due to cloud-cloud collisions. However, over-all, the disk is dominated by incompressible turbulence as well. The mean value of the compression ratio is $r_c \sim 0.17$ in the jet lobes and $r_c \sim 0.2$ in the disk.

Compressible motions result from shocks generated by the expanding bubble and sound waves and weak shocks due to the turbulent motions in the cocoon. Incompressible flows result from turbulence and shear flows generated by the jet-ISM interaction. While compressible modes can decay, being carried by sound waves, incompressible modes can persist for a longer time \citep{reynolds15a} contributing to the turbulent energy budget in the gas.

\begin{figure}
	\centering
	\includegraphics[width = 7.5cm, keepaspectratio] 
	{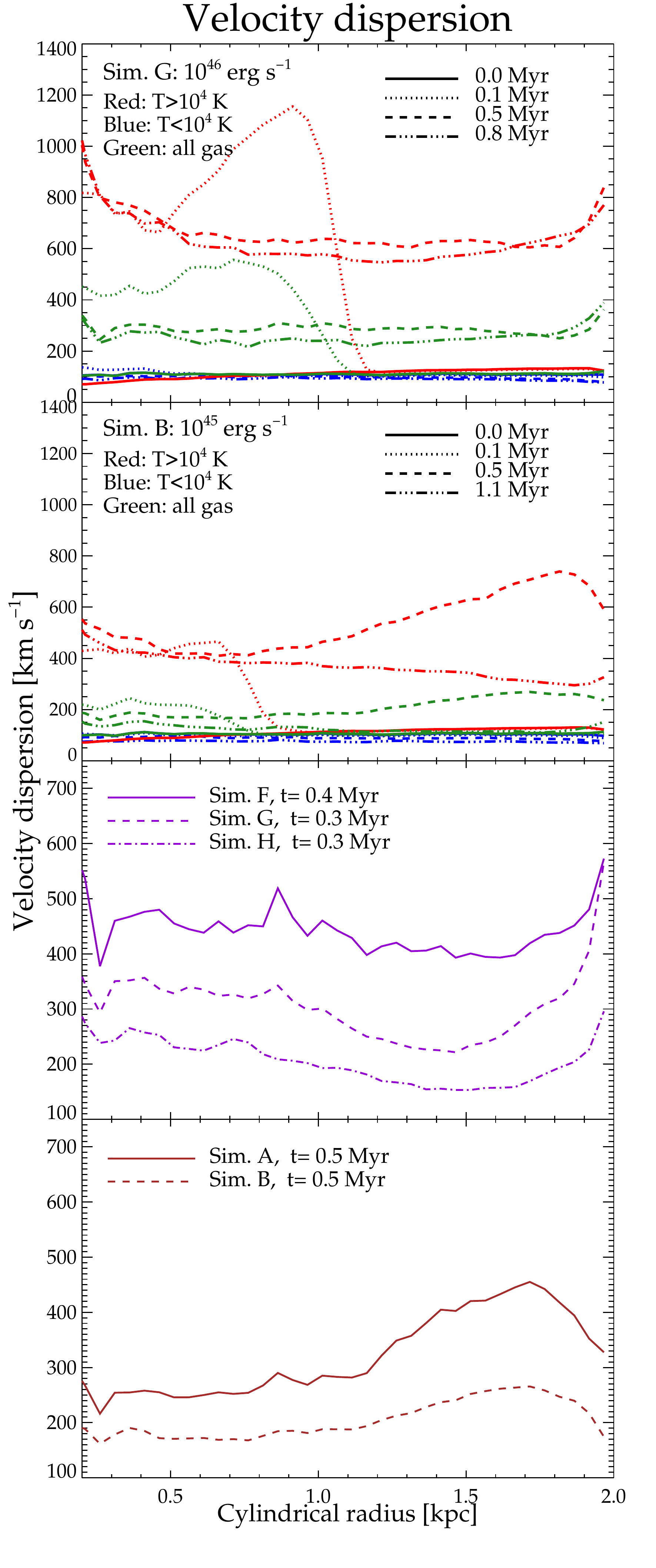}
		\caption{\small \textbf{Top:} Mass-weighted velocity dispersion in the disk as a function of radius at different times for simulation G ($P_{\rm j}=10^{46} \ergs$). The red lines show the velocity dispersion for gas with $T>10^4$K representing the shocked gas. The blue lines are for gas with $T<10^4$ K, representing the cold undisturbed component. The green lines show the dispersion for all mass, without phase separation. \textbf{Second:} Same as above, for simulation B ($P_{\rm j}=10^{45} \ergs$). \textbf{Third and Fourth:} Total velocity dispersion for each simulation, following the spread of the energy bubble over the disk. The third panel shows simulations (F,G,H) with $P_{\rm j}=10^{46} \ergs$ but different mean densities ($n_{w0}=100,200,400 \cc$). The fourth panel is for $P_{\rm j}=10^{45} \ergs$ (A,B) with mean densities $n_{w0}=100,200\cc$ respectively.
		}
	\label{fig.dispersion}
\end{figure}
Fig.~\ref{fig.dispersion} shows the increase in dispersion in the disk due to jet induced turbulence. We compute the mass weighted mean velocity dispersion $\tilde{\sigma}^2=\frac{1}{3}\left(\tilde{\sigma}_r^2+\tilde{\sigma}_\phi^2+\tilde{\sigma}_z^2\right)$, where the dispersion for each component is computed as
\begin{align}
	\tilde{V}_i&=\frac{1}{M}\Sigma \rho V_i  \nonumber \\
	\tilde{\sigma}_i^2 &=\frac{1}{M} \Sigma \rho \left(V_i-\tilde{V}_i\right)^2
\end{align}
where the index $i$ corresponds to the $(x,y,z)$ components.
The dispersion has been computed in annular rings as a function of cylindrical radius. This is strictly appropriate only for azimuthally symmetric disks.  For jets pointed in the $Z$ direction one expects that the mean jet-induced motions will on average be azimuthally independent. Nevertheless, for inclined jet as well, an azimuthal average conveys a good sense of the radial dependence of the jet-induced velocity dispersion.

We compute the dispersions separately for gas with temperatures above and below $T=10^4$ K, to distinguish the shocked gas from the undisturbed gas, which is initially at a temperature corresponding to the cooling floor. This is done to investigate the effect on the kinematics of the different gas phases, as the jet driven shocks create a multi-phase environment (as discussed in Paper I). The top two panels of Fig.~\ref{fig.dispersion} show the radial distribution of the mean dispersion at different times for simulations G ($P_{\rm j}=10^{46} \ergs$) and B ($P_{\rm j}=10^{45} \ergs$) respectively. The dispersion of the shocked gas is represented in red, the cold undisturbed component in blue and the total dispersion including all gas phases is in green. We see that the dispersion in the shocked gas is increased by several times its initial value, indicating the strong turbulence induced by the shocks driven into the disk. For simulation B the dispersion is raised to $\sim 400 \kms$ throughout the disk. For simulation G with a higher jet power, the value is higher at $\sim 600 \kms$. 
		
The dispersion in the disk varies with time, depending on the evolution of the energy bubble.  At $t=0.1$ Myr, only the central region ($\lesssim 1$ kpc) has an enhanced dispersion, as it demarcates the location of the energy bubble. At later times the dispersion is the enhanced throughout the disk. The dispersion in the central region decreases with time. This occurs because a) the turbulence naturally decays with time, b) the mass from the central region is evacuated by the jet and the computed dispersion being mass-weighted diminishes. Overall, at late stages, after the jet driven shocks have penetrated the disk, the dispersion increases on average. The cold undisturbed component remains steady at the initial value of $\sim 80 \kms$, with a slight decrease due to the inherent decay of turbulence, as discussed earlier in Sec.~\ref{sec.twophase}.

The last two panels of Fig.~\ref{fig.dispersion} presents the radial distribution of the velocity dispersion (mass-weighted) for various simulations. The dispersion is computed at a time when the jets have significantly evolved, and the energy bubble has spread over the disk within the simulation domain. The simulations with jet power $P_{\rm j}=10^{46} \ergs$ (F,G,H) are presented in purple in the third panel, and those with $P_{\rm j}=10^{45} \ergs$ (A,B) are in brown in the last panel. The dispersions have been computed for all gas masses, irrespective of temperature or phase.

We firstly notice that with a factor of 10 increase in jet power, the mean dispersion increases by a factor of $2-3$ for the runs with similar mean density. Overall, the dispersion in the disk increases to several hundred kilometres per second. Secondly, the runs with lower mean density ($n_{w0} \sim 100 \cc$) have a higher dispersion for the same jet power. Gas clumps with lower density are more susceptible to fragmentation into smaller clumps as a result of hydrodynamic instabilities, which then further interact with the jet-driven flows, raising the overall dispersion of the disk. The dispersion in simulation H with $n_{w0} \sim 400 \cc$ is not very different from that in simulation G with $n_{w0} \sim 200 \cc$.

\begin{figure}
	\centering
	\includegraphics[width = 8.cm, keepaspectratio] 
	{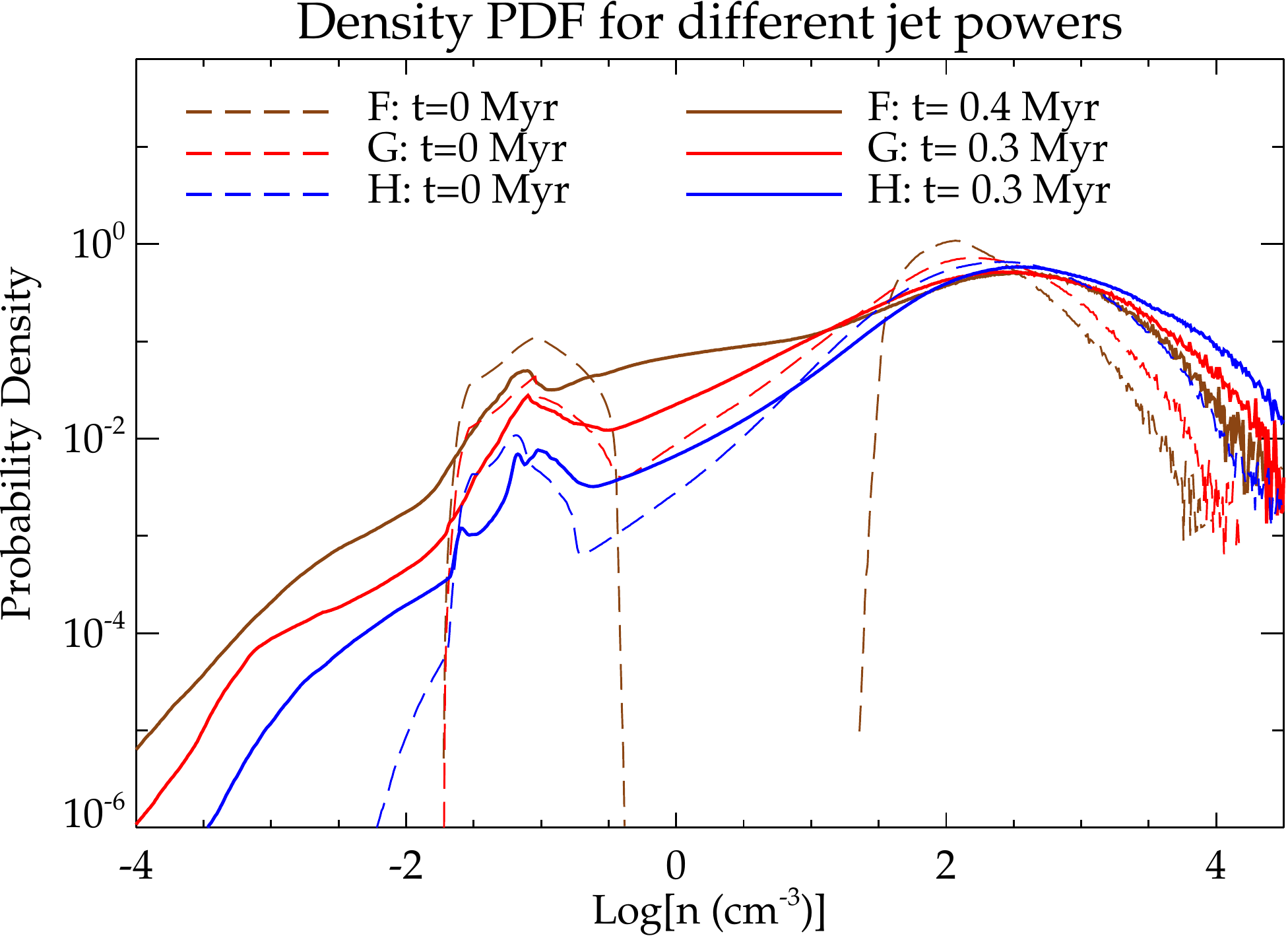}
	\caption{\small  Mass-weighted density probability density function for simulations F, G an H. These simulations have the same kinetic power ($P_{\rm j}=10^{46}\ergs$) but differ in density ($n_{w0}:100,200,400\cc$ for F,G,H respectively). The PDF at $t=0$ is presented as a dashed line. All simulations show an increase in the PDF at the higher densities $\gtrsim 100 \cc$ from its initial value. The lower density ISM in simulation F has a higher PDF at lower densities, due to greater cloud ablation. }
	\label{fig.comparepdf}
\end{figure}
The above is well demonstrated in Fig.~\ref{fig.comparepdf}, which shows the mass weighted probability distribution function (PDF) of the density of simulations F, G and H. A mass weighted density PDF, $P_m(\rho)$, highlights the regions with higher mass i.e. the gas disk in this case. Thus, it is a  better tool to describe the nature of the density distribution in the gas disk than a volume weighted PDF\footnote{See \citet{li03a} for a  discussion on mass vs volume weighted PDFs.}. The dashed lines represent the PDF at the start of the individual simulations. The solid lines show the density PDF at the times corresponding to those in Fig.~\ref{fig.dispersion}. 

Beyond a critical density of $\sim 100-300 \cc$, all the simulations show an increase in the PDF from its starting value. This results from the compression and enhancement of densities in the post shock regions. The shocks driven into the clouds quickly become radiative resulting in a shell of high density surrounding the clouds, making them less prone to ablation. Similar results have also been discussed in \citet{mukherjee17a} and \citet{sutherland07a}. However, simulation F with $n_{w0}=100 \cc$, is more extended at lower densities ($n\lesssim 50 \cc$), compared to that of G ($n_{w0}=200 \cc$) and H ($n_{w0}=400 \cc$). This implies that the lower density clouds in simulation H are more easily dispersed, compared to the denser ISM where clouds remain more intact. The enhanced cloud ablation also raises the velocity dispersion in the turbulent gas. The present simulations do not include self-gravity. The gas more prone to ablation below the critical density of $\sim 100-300 \cc$ is not expected to be affected by its self-gravity, as the free-fall time scales are much larger than the dynamical scales at such densities. However, self-gravity may enhance the PDF at the high density end. We plan to explore this in a future work.

\end{enumerate}

\subsection{Inclined jets}\label{sec.inclined}
\begin{figure*}
	\centering
	\includegraphics[width = 5.cm, keepaspectratio] 
	{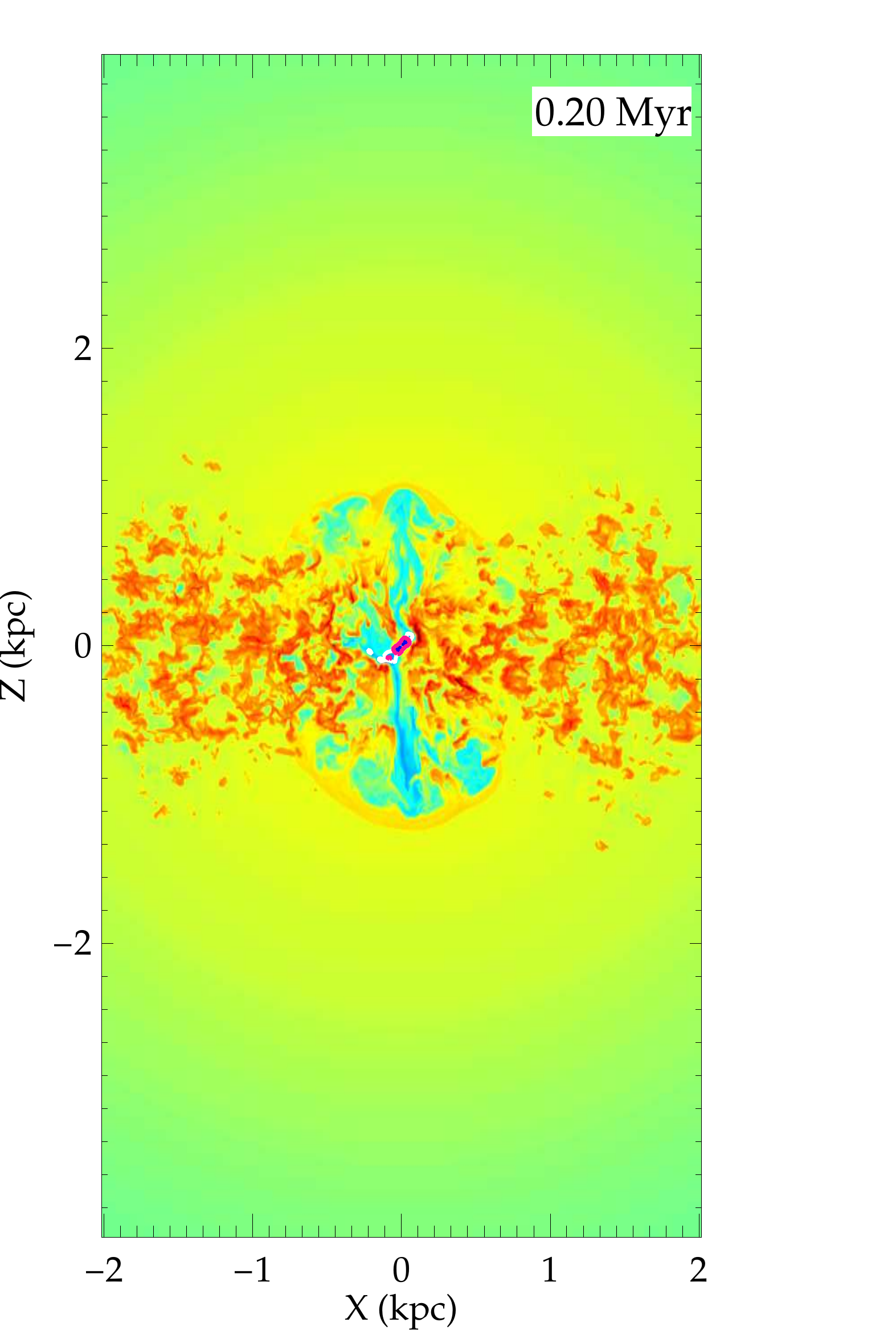}\vspace{-0cm}\hspace{-1.6cm}
	\includegraphics[width = 5.cm, keepaspectratio] 
	{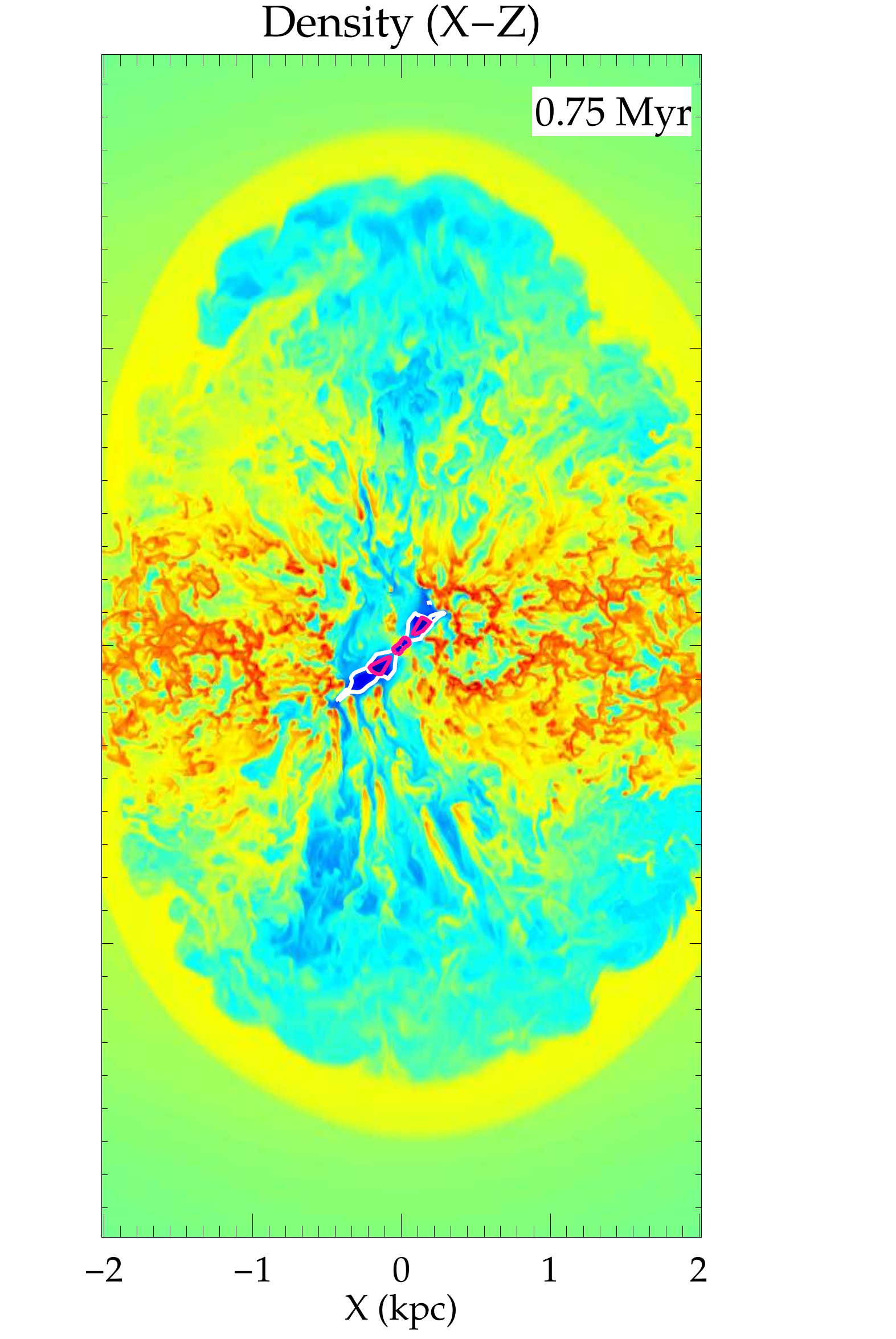}\vspace{-0.cm}\hspace{-1.6cm}
	\includegraphics[width = 5.cm, keepaspectratio] 
	{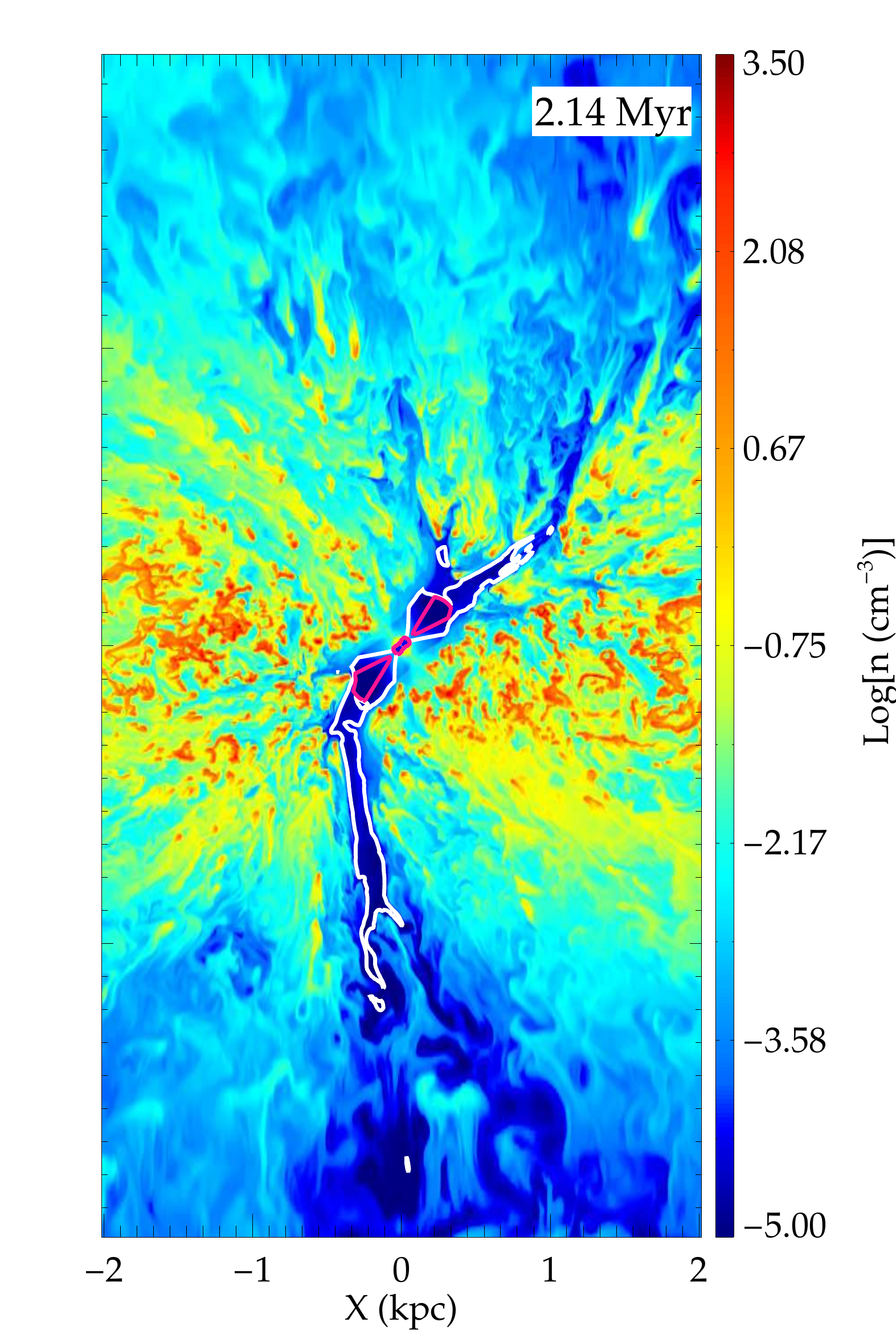}\vspace{-0cm}\hspace{-0.cm}
	\includegraphics[width = 5.cm, keepaspectratio] 
	{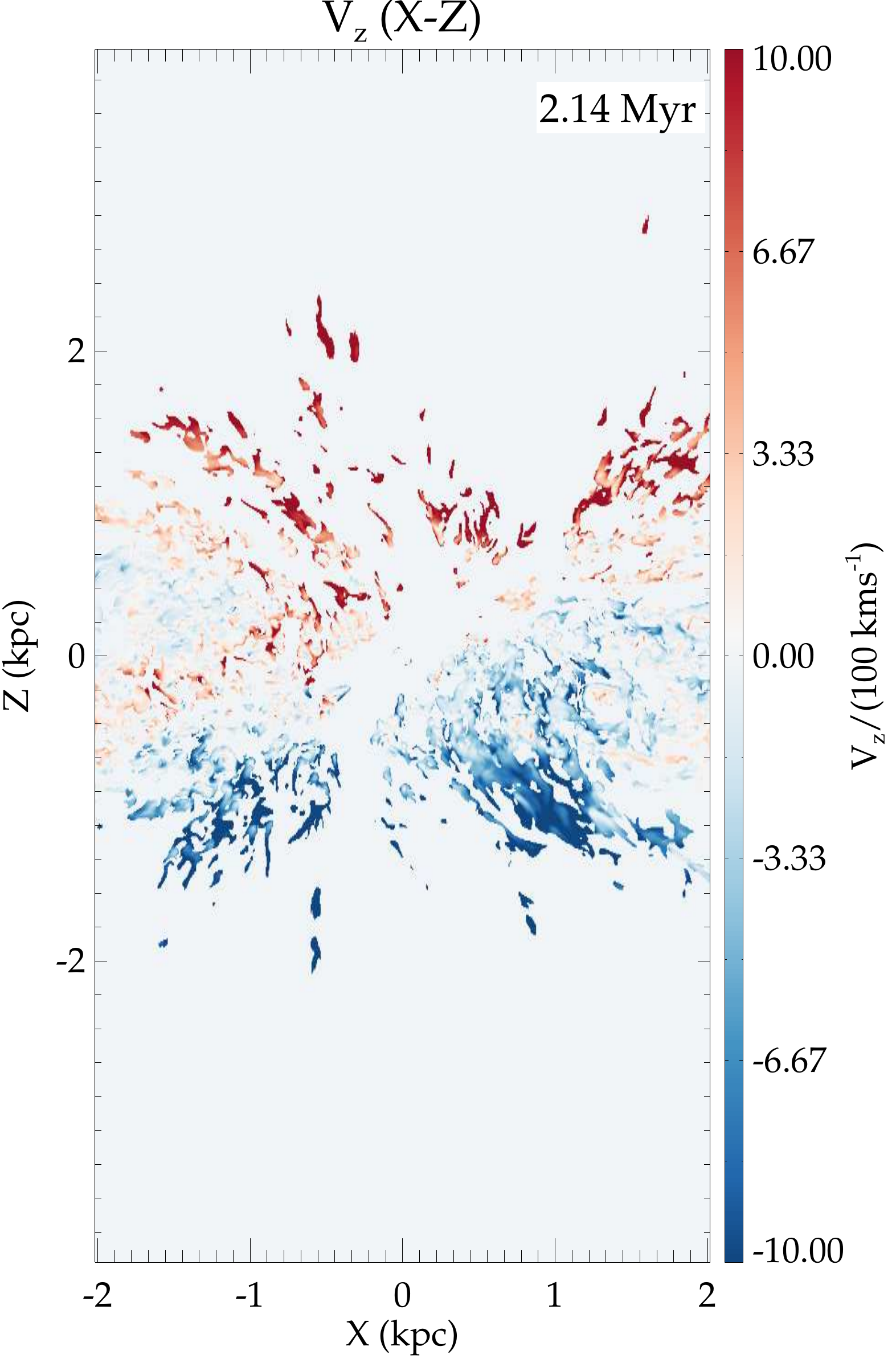}\vspace{-0.cm}
	\caption{\small \textbf{Left 3 panels:} Density ($\log(n [\cc])$) in the $X-Z$ plane at different times for simulation D with $P_{\rm j}=10^{45} \ergs$, $\theta _{\rm inc}=45^\circ$ (Table~\ref{tab.params}). The contours denote regions with relativistic gas motions with white giving contours of $\beta = 0.4c$ and $\beta=0.9c$ in magenta. The inclined jets decelerate, launching a spherically evolving bubble along the minor axis. \textbf{Right:} Vertical component of the velocity $V_z$ in units of  $100 \kms$ for gas with density $n > 0.1 \cc$ at $t\sim 2.14$ Myr, the last density panel in the previous set. The inclined jet couples more with the ISM, resulting in greater outflow from the disk.  }
	\label{fig.p45dir45}
\end{figure*}
Recent papers \citep{gallimore06a,combes16a} have proposed that there is a lack of preferred orientation of jets with respect to the plane of the disk (or the galaxy's major axis). There are a number of galaxies in which the jets are inclined to the disk normal (e.g. NGC 3079, \citet{cecil01a}, NGC 1052 \citet{dopita15a}), including galaxies in which the jet is in the plane of the disk \citep[e.g. IC~5063, ][]{morganti15a}. In order to study such interactions we have performed simulations (simulations~C,D and E), for which the jet axis is inclined towards the disk. Fig.~\ref{fig.p45dir45} shows the evolution of the density in the $X-Z$ plane for simulation D where the jet is inclined at $45^\circ$ to the plane of the disk.
 \citet{zubovas13a} and \citet{zubovas17a},
The evolution of jets is markedly different in this case, compared to the cases of the perpendicular jets discussed earlier. Since the jets encounter a larger column depth of clouds along their path, they interact more strongly with the dense ISM. The jets do not immediately clear a channel along the inclined axis of launch. Rather, the dense clouds deflect the jets, which then decelerate and launch a sub-relativistic outflow along the minor axis following the path of least resistance (see Fig.~\ref{fig.p45dir45}). 

\begin{figure*}
	\centering
	\includegraphics[width = 5.cm, keepaspectratio] 
	{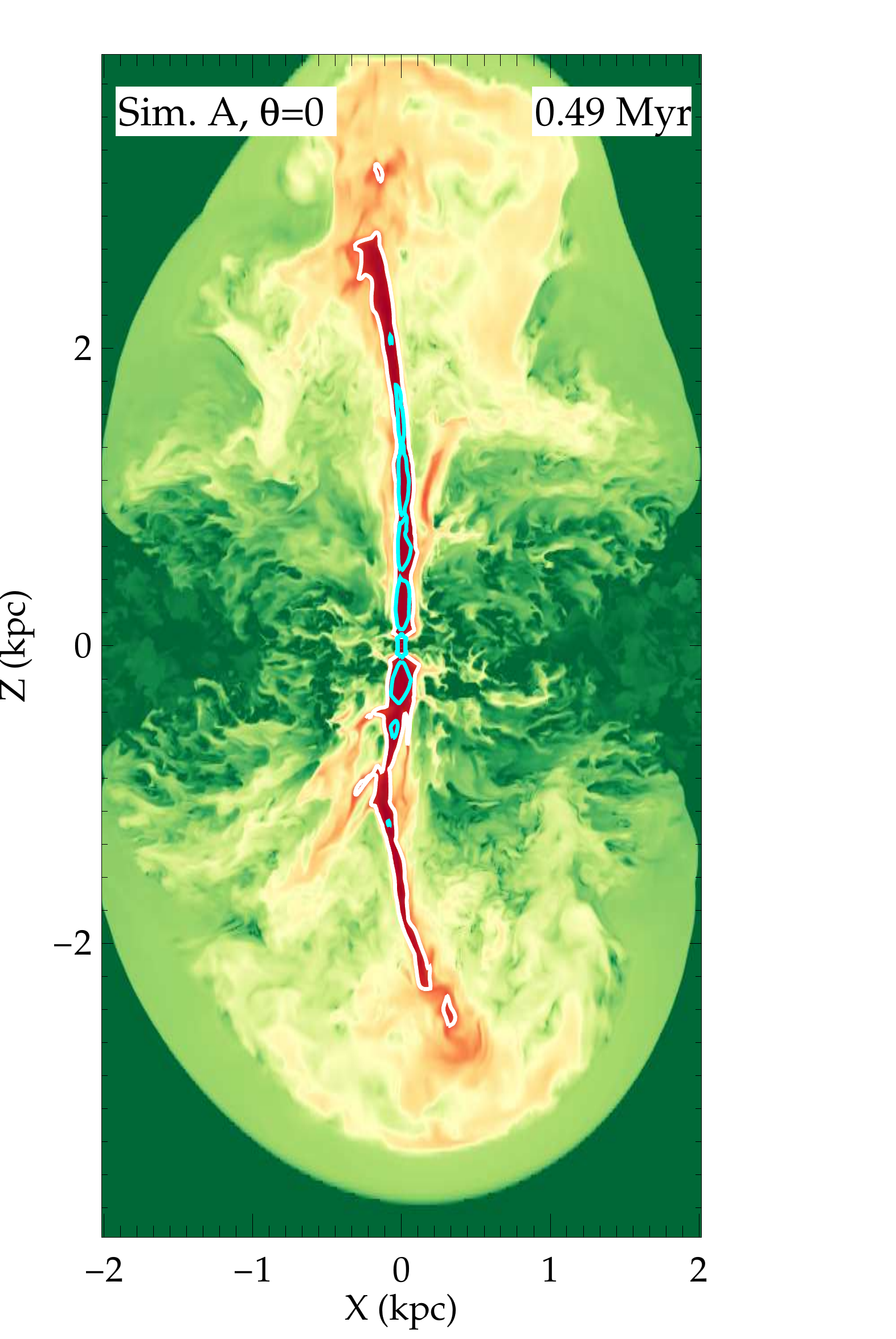}\vspace{-0cm}\hspace{-1.6cm}
	\includegraphics[width = 5.cm, keepaspectratio] 
	{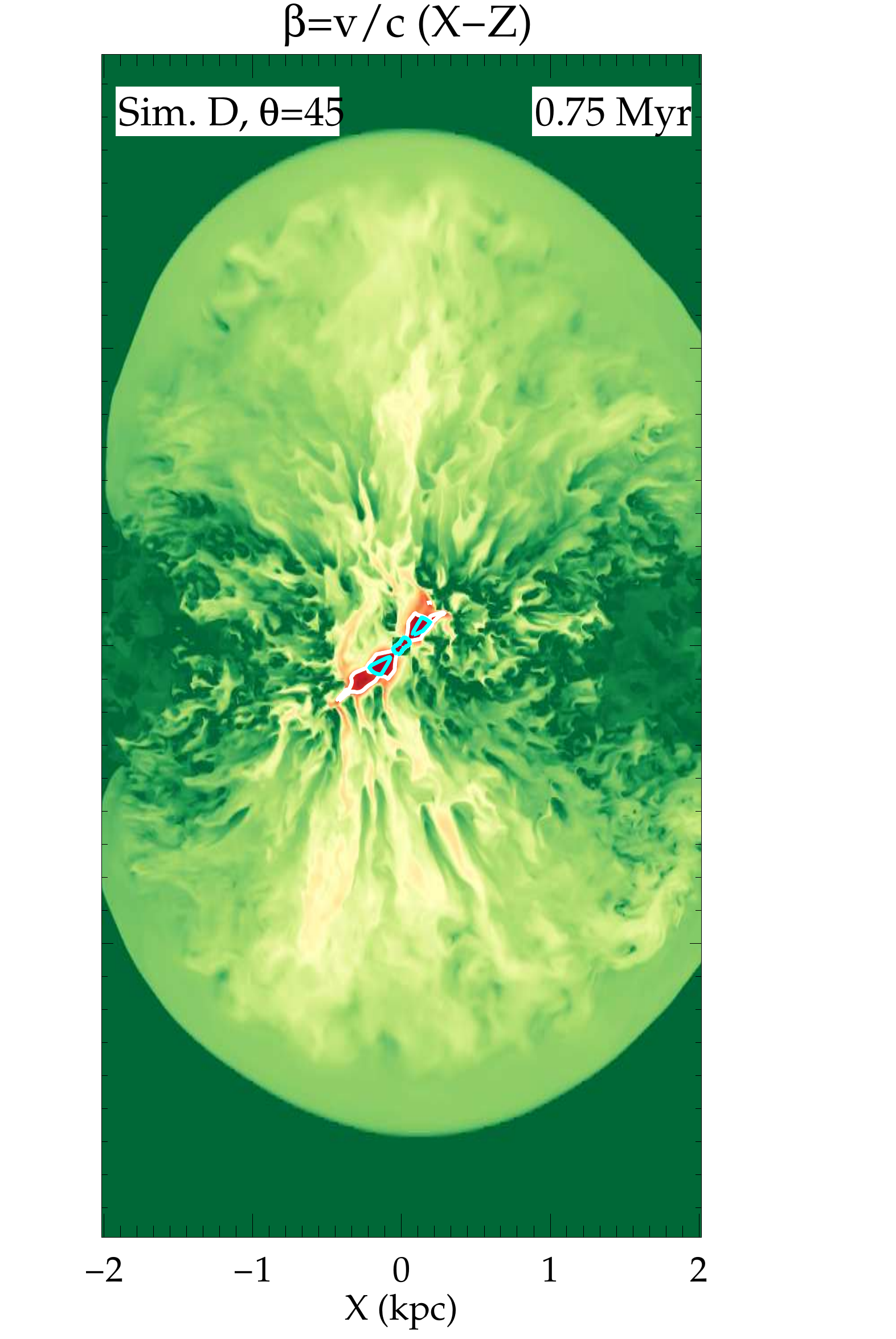}\vspace{-0.cm}\hspace{-1.6cm}
	\includegraphics[width = 5.cm, keepaspectratio] 
	{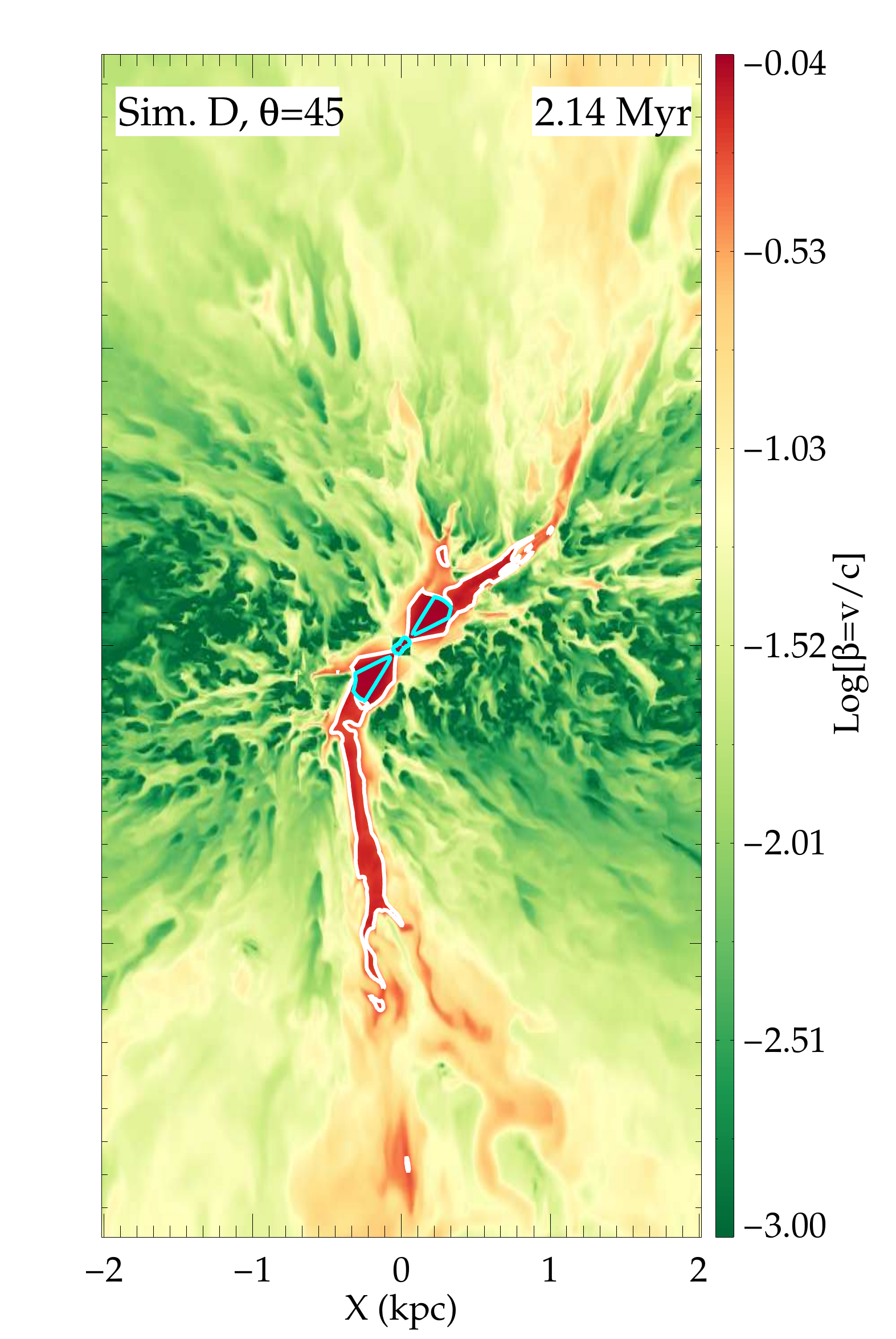}\vspace{-0cm}
	\caption{\small \textbf{Left:} Evolution of the velocity magnitude, $\beta = v/c$, for simulation B with $\theta _{\rm inc}=0^\circ$. \textbf{Middle and Right:} Evolution of $\beta$ at two different times for simulation D with $\theta _{\rm inc}=45^\circ$. The times correspond to the middle and right panels of Fig.~\ref{fig.p45dir45}. The contours (white for $\beta = 0.4c$ and cyan for $\beta=0.9c$) denote regions with relativistic gas motions. The figures show sub-relativistic wider-angled outflows in the inclined simulations, as compared to the vertical relativistic outflows for jets launched perpendicular to the disk.}
	\label{fig.p45dir45beta}
\end{figure*}
Fig.~\ref{fig.p45dir45beta} shows the time evolution of the gas speed ($\beta=v/c$). The first panel shows the evolution of the jet for simulation~B with the launch axis perpendicular to the disk. It can be clearly seen that the the main jet stream remains relativistic with $\beta \sim 0.9 c$ up to the hot spots, where it becomes turbulent and decelerates. The second and third panels show the evolution of the velocity for simulation D with the jet inclined at $45^\circ$. The jets remain relativistic in the central region ($\lesssim 200 \pc$), where they evacuate a cavity. Beyond the central cavity, as the jets interact with the clouds, they quickly decelerate and launch a sub-relativistic outflow ($v \sim 0.01-0.1$c) of non-thermal plasma. 

\begin{figure*}
	\centering
	\includegraphics[width = 6.cm, keepaspectratio] 
	{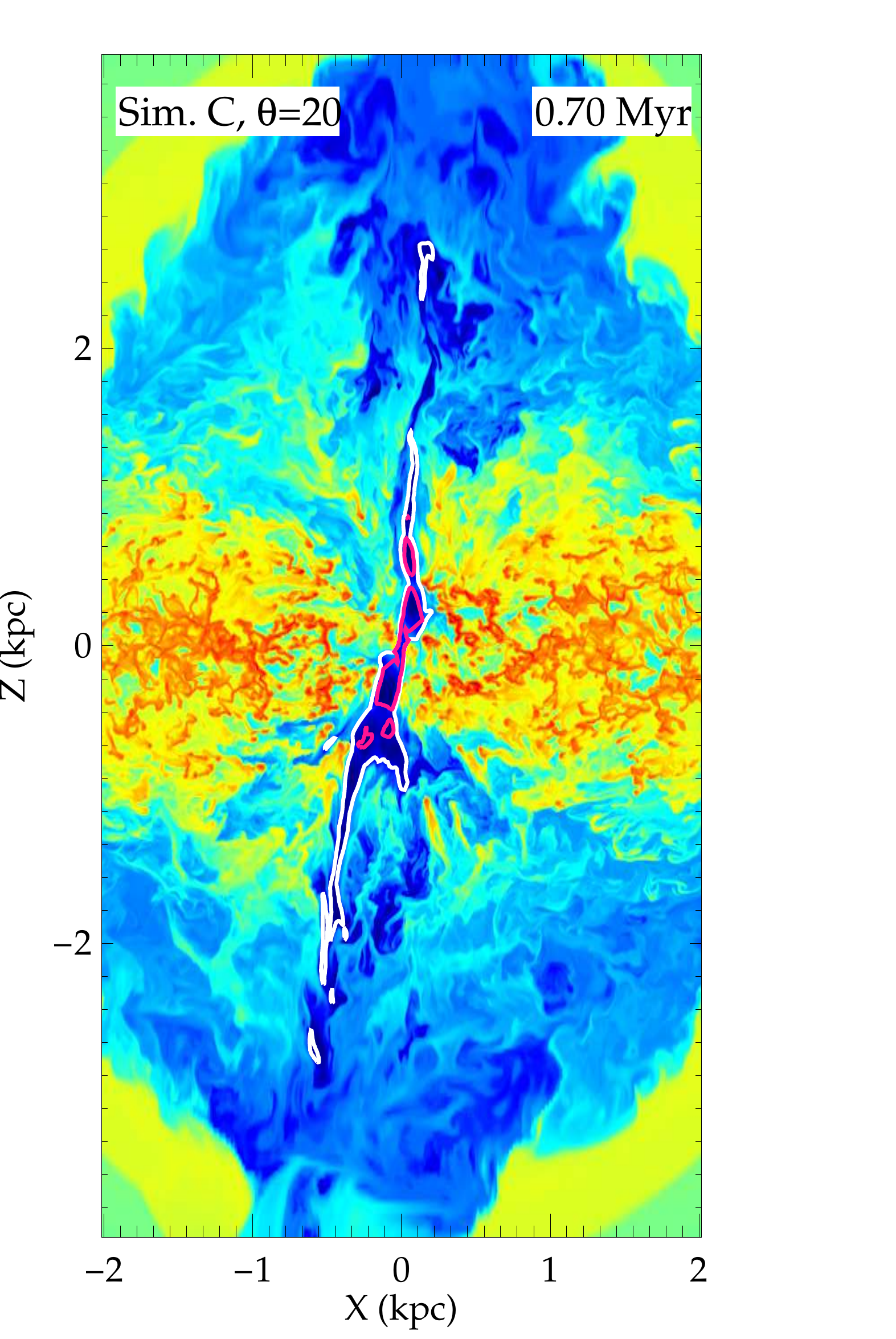}\hspace{-1.8cm}
	\includegraphics[width = 6.cm, keepaspectratio] 
	{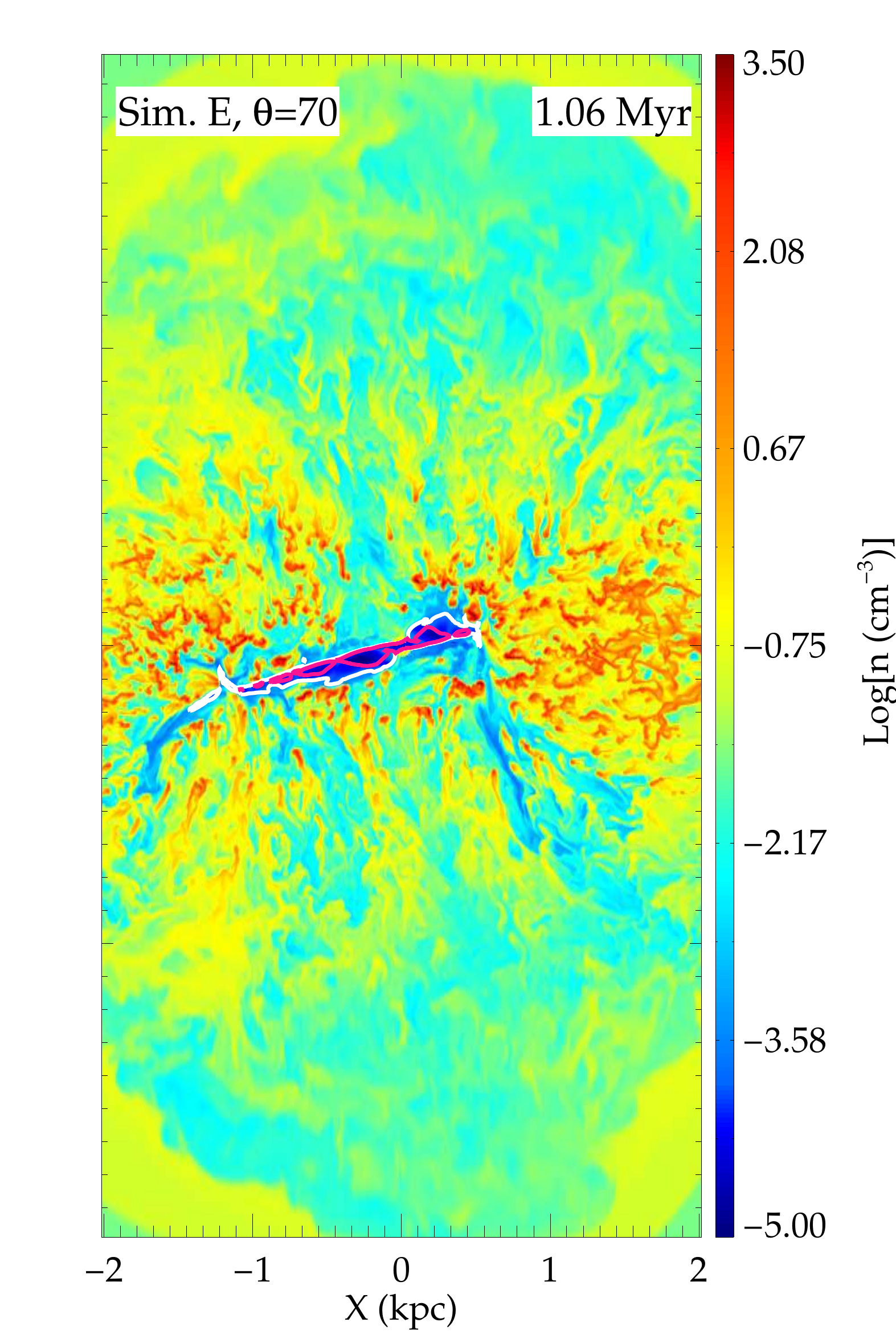}
	\includegraphics[width = 6.cm, keepaspectratio] 
	{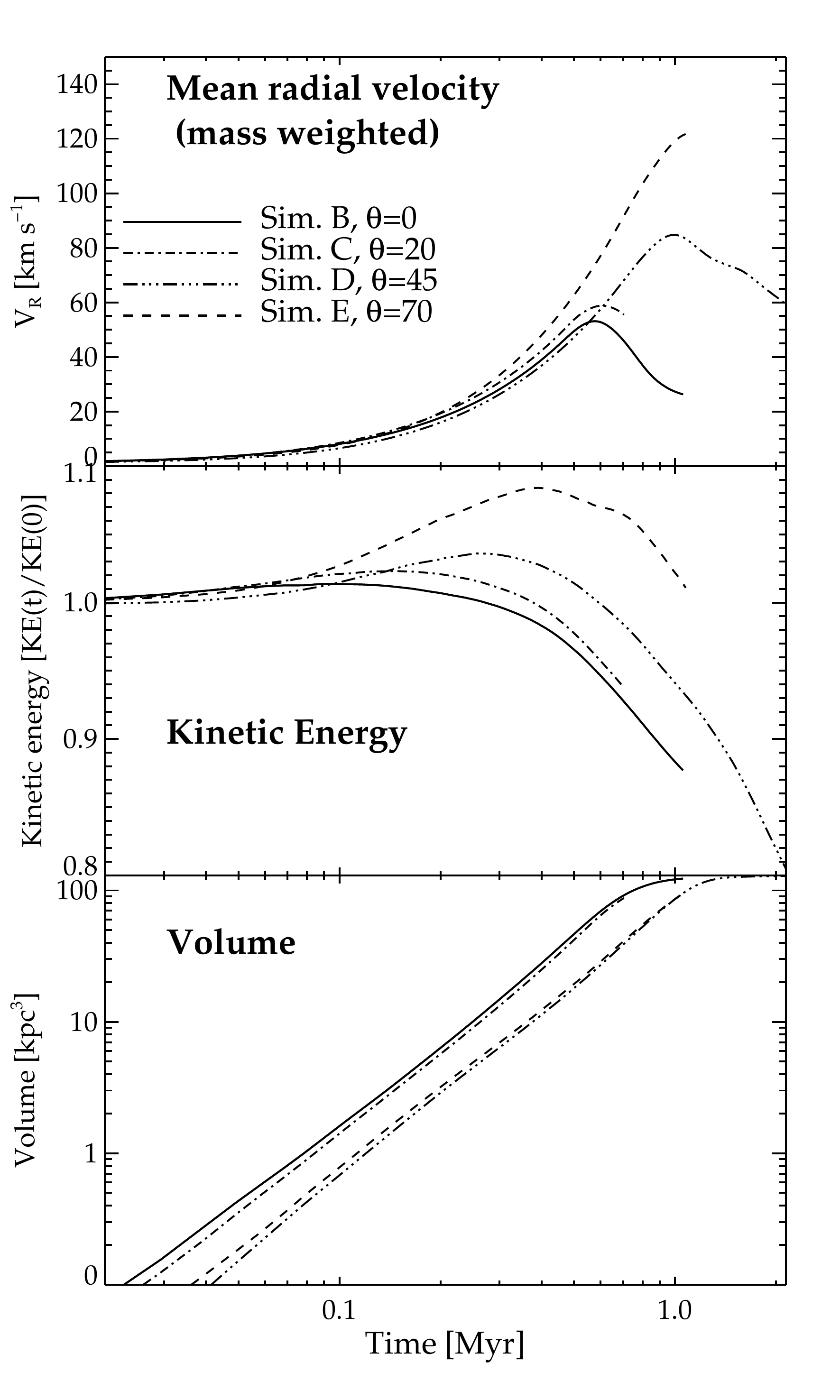}
	\caption{\small \textbf{Left and Middle:} Density in the $X-Z$ plane for simulations C and E with $\theta_{\rm inc}=20^\circ$ and $70^\circ$ respectively. \textbf{Right:} A three panel plot showing the time evolution of the mass-weighted mean radial (spherical) velocity, the kinetic energy of the dense gas ($n > 1 \cc$) and volume of the energy bubble in units of kpc$^3$, for simulations with different angles of inclination. Computational cells with a jet tracer value higher than $1.e-8$ are considered to form the bubble.  }
	\label{fig.p45dirothers}
\end{figure*}

The wind creates a spherical bubble which clears the ambient gas as it rises, similar to the vertical jets. However, the evolution of the bubble is more spherical in morphology, driven by the wide angle wind, when compared to the ellipsoidal bubbles created by the vertical jets. The evolution of the bubble is also slower in the case of the inclined jets (as shown later in Fig.~\ref{fig.p45dirothers}). At later stages, the northern jet approximately proceeds along the axis of launch to some extent. The southern jet however, is deflected due to strong interactions with a cloud and is bent towards the minor axis, along which it continues into the previously evacuated cavity.

The strong interaction of the jet with the dense gas in the disk results in dense gas being vertically lifted off the disk, as well as in the radially driven outflows described earlier. The rightmost panel in Fig.~\ref{fig.p45dir45} shows the vertical velocity in units of $100 \kms$ at $t\sim 2.14$ Myr. Clouds outflowing at $\sim 1000 \kms$ can be seen being driven to heights of $\sim 1.5-2$ kpc, forming cometary tails of stripped gas. With the jets inclined, the interacting clouds experience an outward vertical component of the imparted momentum resulting in an outflow away from the disk. This is in contrast to the results of simulation B, where, despite the entrainment of clouds near the core of the jet, the dominant effect is the compression of the disk from the energy bubbles, resulting in inflows at the outer edges ($\gtrsim 500 \pc$) of the disk.

The strength of the interaction increases with the angle of inclination as the jet drills through a larger gas column. Fig.~\ref{fig.p45dirothers} shows the density slices for simulations D and E, with jet inclinations  $\theta_{\rm inc}=20^\circ$ and $70^\circ$ respectively. Simulation C proceeds similarly to that of B, the inclination angle being $\sim 20^\circ$. In simulation E, the jet is frustrated inside the disk, similar to that of simulation~C (with $\theta _{\rm inc}=45^\circ$). The vertical outflow of clouds is more widespread as the jet ploughs through the disk. The third plot shows the evolution of the mean spherical radial velocity (mass weighted), the kinetic energy in the dense gas (for $n\gtrsim 0.1 \cc$) normalised to its initial value and the volume of the energy bubble for simulations with different angles of inclinations (simulations~B,C,D and E). 

From the evolution of the mean radial velocity (mass weighted) and kinetic energy, it is clear that jets aligned more closely to the disk interact more strongly. For simulations B and C with smaller inclination angles (measured from the disk-normal), the mean radial velocity increases until about $t\sim 0.5$ Myr, decreasing thereafter as the jet-ISM interaction decreases in intensity. For simulations D and E, the mean radial velocity increases for a longer time ($\sim 1$ Myr), as the jets are frustrated inside the disk. The kinetic energy similarly shows an initial increase followed by a decrease. For simulations E and D, with jets more inclined to the disk, both the mean radial velocity and kinetic energy are significantly higher than in simulations B and C. Similarly, the evolution of the volume of the energy bubble presented in the third panel shows that simulations D and E closely follow each other, and evolve at a much slower rate than that of B and C. This is because of the lower velocity of the wide-angled sub-relativistic winds, which power the bubbles formed from the inclined jets. 

\begin{figure}
	\centering
	\includegraphics[width = 7.cm, keepaspectratio] 
	{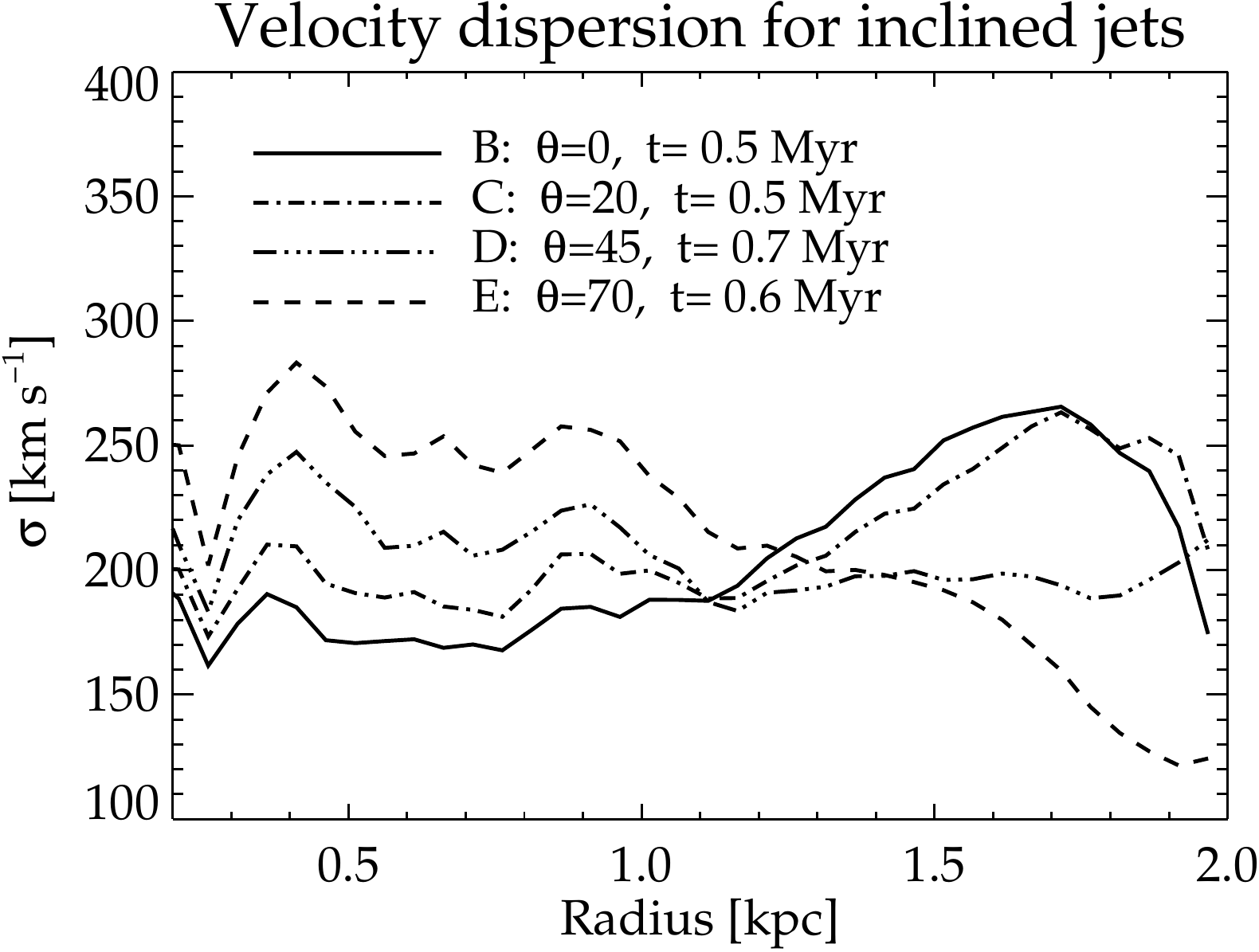}
	\caption{\small The mass-weighted velocity dispersion as a function of radius for simulations B,C,D and E. The times have been chosen such that the jet driven energy bubble has spread over the disk. }
	\label{fig.dispersion_inclined}
\end{figure}
The jet-driven outflows do not unbind a significant fraction of the ISM (by mass), as also discussed in \citet{mukherjee16a}. The mean radial velocity, though higher for simulations D and E, is much less than the velocity dispersion of the galaxy\footnote{It must be noted here that the mass weighted mean radial velocity is averaged over the entire ISM. This includes the undisturbed dense cores and, thus, does not truly reflect the velocity of individual out-flowing filaments, which can be much higher, as can be seen in Fig.~\ref{fig.p45dir00}, Fig.~\ref{fig.p45dir45} and Fig.~\ref{fig.p45dir45beta}}. 

Instead, the jet-ISM interactions induces local turbulence within the disk, as evidenced by the higher kinetic energy of simulation E, as compared to others. This also raises the velocity dispersion of the gas disk. In Fig.~\ref{fig.dispersion_inclined} we present the radial variation of the velocity dispersion for simulations B, C, D and E. In the central regions, the resultant velocity dispersion of simulation E was found to be higher than that of simulation~B by a factor of $\sim 3$. To summarize, jets more closely inclined with the disk, couple more strongly with the ISM, over a longer time, inducing more local turbulence.

\subsection{Qualitative similarities with some observed galaxies}
\subsubsection{Shocked disk in  NGC 1052:} 
\begin{figure}
	\centering
	\includegraphics[width = 8cm, keepaspectratio] 
	{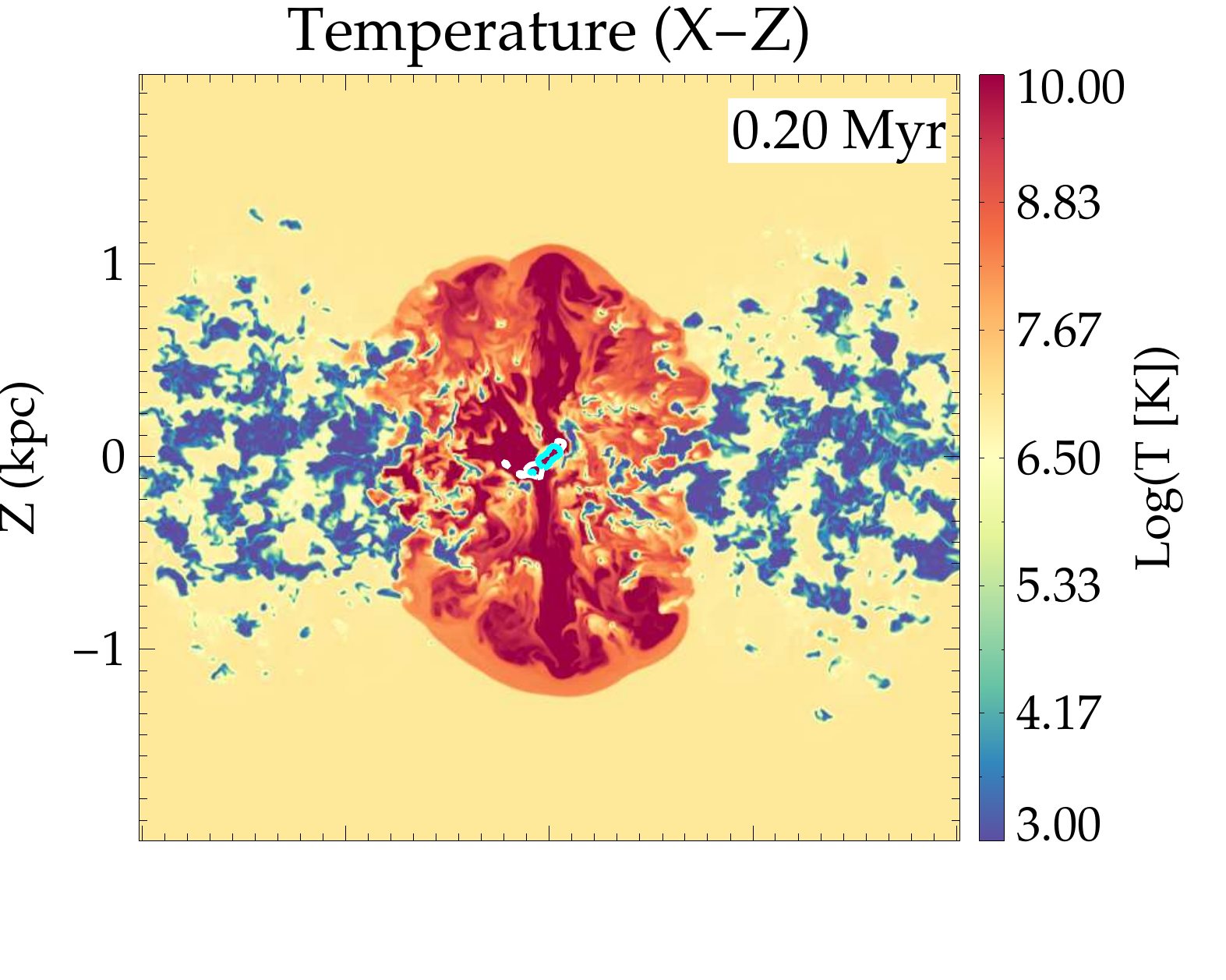}\vspace{-1.4cm}
	\includegraphics[width = 8cm, keepaspectratio] 
	{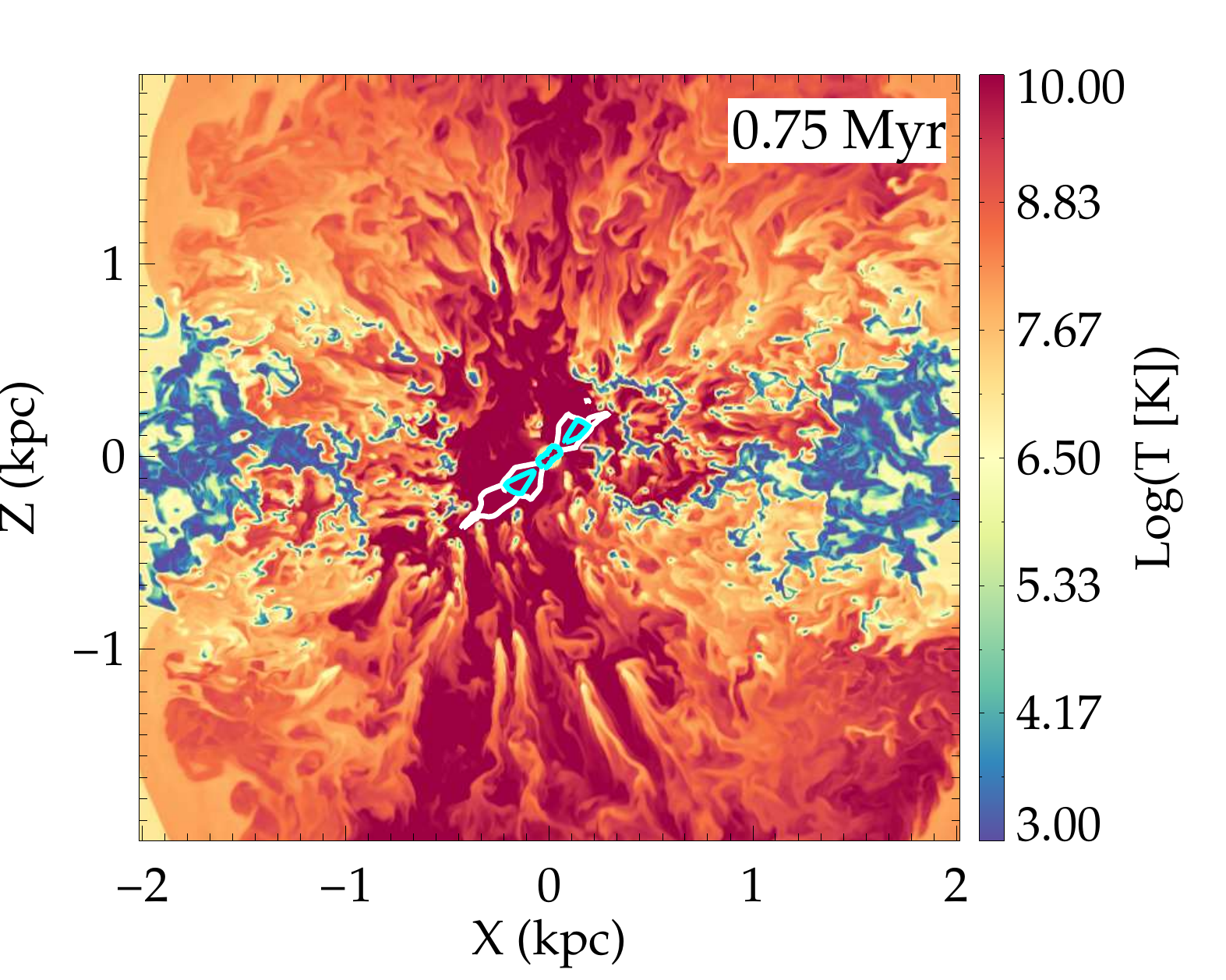}
	\caption{\small Temperature in $X-Z$ plane for simulation D ($P_{\rm j}=10^{45} \ergs$, $n_{w0}=200 \cc$, $\theta _{\rm inc}=45^\circ$). The jet head is surrounded by an arc of dense clouds with shocked surface layers. Shocks from the jet driven energy bubble spread through the disk. The  bubble created on top of the disk is  composed of high temperature non-thermal plasma, as well as shocked thermal gas from ablated clouds. The contours in white ($\beta=0.4c$) and magenta ($\beta=0.9c$) denote regions with relativistic gas motions.}
	\label{fig.tempp45}
\end{figure}

NGC 1052 is a nearby ($z=0.0045$) LINER\footnote{Low Ionisation Nuclear Emission line Region, \citet{heckman80a}} galaxy with a prominent radio jet \citep{claussen98a,wrobel84a}. Observational studies \citep{kadler04a,kadler04b,sugai05a,dopita15a} have shown signatures of an inclined radio jet strongly interacting with its environment. We list below some of the key  inferences drawn from these observational signatures.
	\begin{enumerate} 
		\item The inner sub-arcsecond region shows strong OIII emission \citep{sugai05a}, characterized by two  ridges extending east-northeast and west-south-southwest. The spectrum shows signatures of two components emitting in OIII, a narrow low-velocity component that is spatially extended  and a less extended component with higher velocity, with values reaching $\sim 1000 \kms$ \citep[see sec. 3.2 of][]{sugai05a}.  
		\item Mapping the OIII in different emission channels reveals spatially distinct components with a central core where the outflow originates and two others which appear as working surfaces of a jet driven shock \citep[Fig. 4 of][]{sugai05a}.

		\item The outflow axis (which follows the jet) is inclined to the rotation axis of the disk by $\sim 50^\circ$ \citep{sugai05a,dopita15a}. 
		\item Above and below the disk there exists outflowing $H\alpha$ bubbles with diameters of $\sim 1.5$ kpc, extending along the minor axis of the galaxy \citep{dopita15a}. The bubbles have an internal dispersion of $\sim 150 \kms$. 
		\item The gas in the disk in strongly disturbed with a high velocity dispersion $\sim 300 \kms$. 
		\item \citet{kadler04a} have found spatially extended Xray emission which is thermal in origin (with $T\sim 0.4-0.5$ keV). The extent of the Xray emitting region is co-incident with radio emission at 1.5 GHz \citep{kadler04a,kadler04b} and the turbulent gas disk, pointing to the presence of strongly shocked gas \citep{dopita15a}. 
	\end{enumerate}

\citet{dopita15a} have put forward a model of a jet inclined to a gas disk to explain the above \citep[see schematic in Fig.~7 of ][]{dopita15a}.  Our simulations of inclined jets, specifically simulation D with the jet inclined at $45^\circ$, qualitatively address several of these results and confirms the suggestion put forth in \citet{dopita15a}. As the jet impinges on the disk, it shocks and ablates the clouds. The resulting morphology is that of an arc near the head of the jets (see Fig.~\ref{fig.tempp45}). Such an arc of shocked clouds, can indeed appear as a ridge of OIII emission, similar to the bow-shock shaped emission profile observed in the high velocity channel in Fig.~4 of \citet{sugai05a}. 

As the jet decelerates, it launches a slower moving bubble of hot gas that rises along the minor axis (see discussion Sec.~\ref{sec.inclined} and Fig.~\ref{fig.p45dir45beta}). Such a low density hot bubble is similar to the observed $H\alpha$ bubbles reported in \citet{dopita15a}. The two velocity components in the OIII emission may be explained by a) shocks permeating the surfaces of slower-moving dense clouds before they are accelerated, and b) emission from the  fragments ablated from the clouds, which are swept outwards at speeds of $\sim 1000 \kms$, as shown in Fig.~\ref{fig.p45dir45}. 

Although the jet head is confined within the inner few 100 pc (see Fig.~\ref{fig.tempp45}), the energy bubble created by the jet proceeds farther. Thus the region of shocked gas extends a few kpc beyond the collimated part of the jet. This implies that the observed high velocity dispersion in the gas disk in NGC~1052 is driven by turbulence induced by the jet, as proposed by \citet{dopita15a}, and also demonstrated in Fig.~\ref{fig.dispersion_inclined}. 

Thus, although simulation D is not a precise model of NGC~1052, it shows remarkable qualitative similarities with the observed kinematics and morphology of the shocked gas in the galaxy. This strengthens the argument that wide-scale shocks driven by the jet and the expanding energy bubble may be dominant contributors to the observed emission from NGC~1052 \citep[as suggested by spectral studies in ][]{dopita15a} and to the  kinematics over scales of a few kpc away from the nucleus. 

\subsubsection{The superbubble in NGC 3079:} 
NGC 3079 is another galaxy which shows kpc scale bubbles arising from a nuclear outflow \citep{duric88a,Irwin88a,veilleux94a,cecil01a}.  One striking feature of this galaxy is the presence of ionized filaments extending to about a kpc from the disk, tracing the contours of the bubble \citep{cecil01a}. \citet{cecil01a} model the bubbles as arising from the wide-angle mechanically driven nuclear outflow. The source of the outflow has been debated in the literature. The central nucleus does indeed harbour star forming molecular gas \citep{lawrence85a}, pointing to the possibility of a starburst driven wind \citep{veilleux94a,sofue01a,shafi15a}. However, this is disputed in \citet{hawarden95a} who favour an AGN\footnote{NGC 3079 is classified as a seyfert 2 \citep{ford86a} with a LINER nucleus.} driven wind.

\begin{figure}
	\centering
	\includegraphics[width = 8.5cm, keepaspectratio] 
	{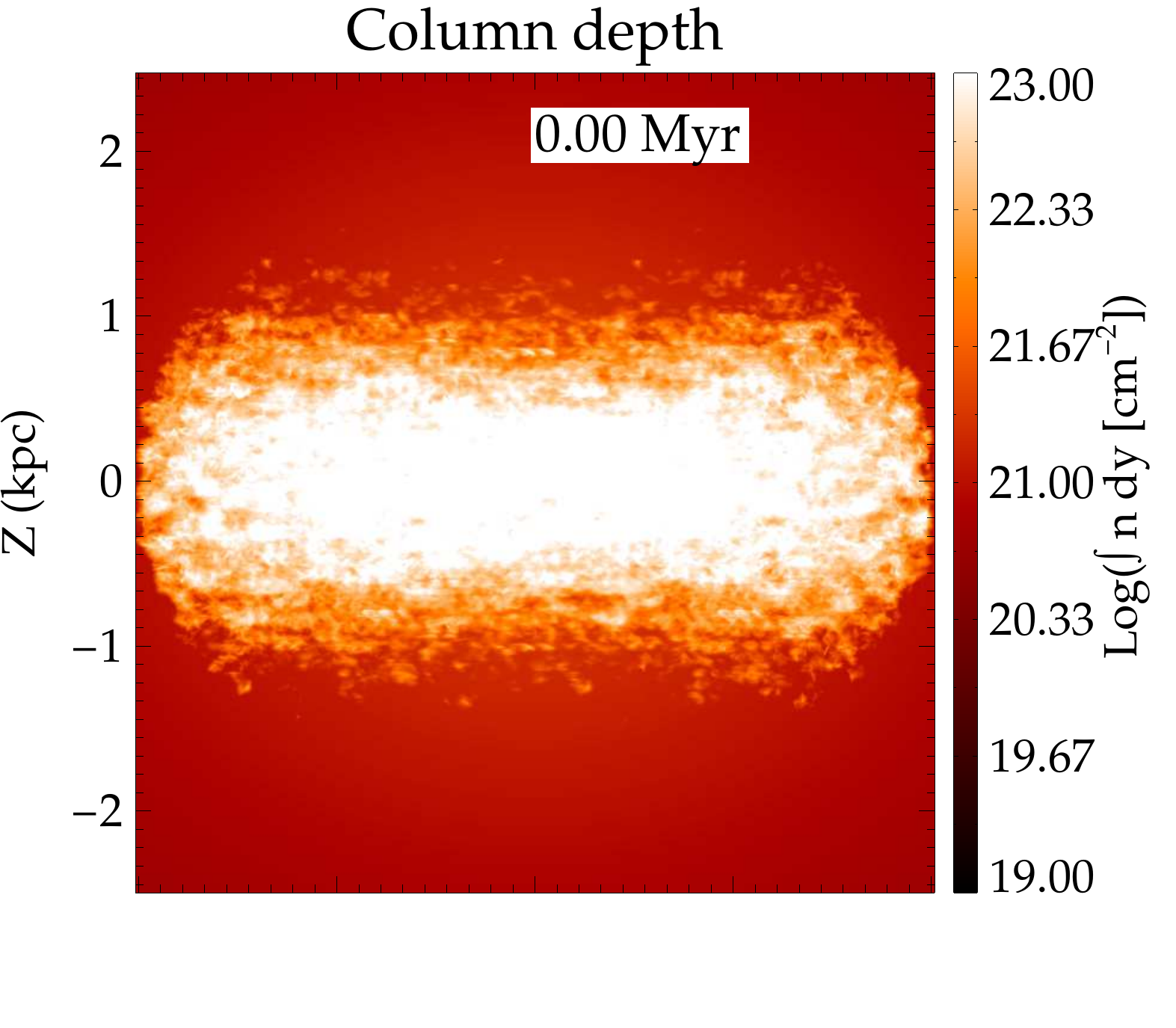}\vspace{-1.5cm}
	\includegraphics[width = 8.5cm, keepaspectratio] 
	{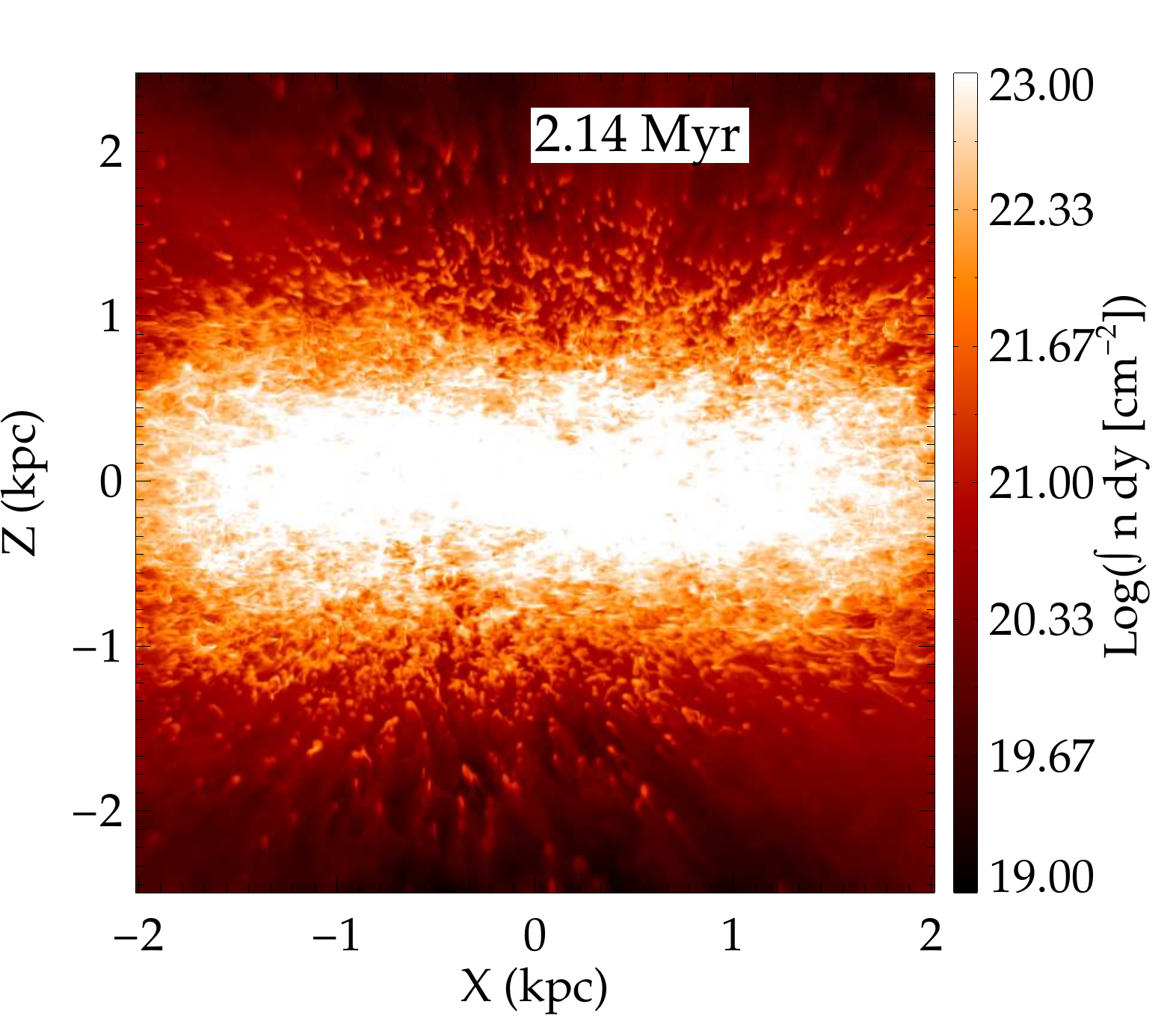}
	\caption{\small Column density along the $Y$ axis for simulation D ($P_{\rm j}=10^{45} \ergs$, $n_{w0}=200 \cc$, $\theta _{\rm inc}=45^\circ$). The top panel shows the disk at beginning before the jet is injected. The bottom panel is at $t\sim 2.14$ Myr when the energy bubble has developed (see Fig.~\ref{fig.p45dir45}).  }
	\label{fig.column}
\end{figure}
Nevertheless, the galaxy harbours a jet whose orientation is misaligned with that of the large scale outflow (bubbles) by $\sim 65^\circ$ \citep{Irwin88a,trotter98a}. \citet{Irwin88a} have proposed that such a jet itself can inflate a bubble to form the observed structures. This is similar to our results for inclined jets, where the jets remain relativistic in the central cavity, beyond which they decelerate and launch a slower outflow along the minor axis. As shown in Fig.~\ref{fig.p45dir45} and Fig.~\ref{fig.p45dirothers} the inclined jets launch a vertical outflow that creates a super bubble with a morphology usually expected from a nuclear star burst driven bubble \citep{veilleux05a,cooper08a,vijayan18a}. 

The column density plots in Fig.~\ref{fig.column} show filamentary structures protruding from the disk, extending to heights of $\sim 1$ kpc from the dense central disk. This is similar to the simulations of filamentary outflows from a purely starburst driven wind \citep{cooper08a}. We note that our results do not model the outflows in NGC~3079 precisely. However, our results do qualitatively demonstrate that, as an inclined jet interacts with the disk, it can also drive a wide angle outflow with gas clouds/filaments being lifted from the disk. Thus jets inclined to the disk can significantly affect the ISM over a radius of a few kpc. 

Similar results have also been explored recently in other simulations such as those \citep{dugan17a,cielo17a}. Jets inclined to the disk may in fact be more common, as the the orientation of the jet with respect to the large scale disk may not be related, and in fact suggested to be random in \citet{gallimore06a,combes16a}. Such interactions have also been observed in other systems such as NGC~4258 \citep{ogle14a} and 3C~293 \citep{mahony16a}. Another system with a kpc scale bubble, possibly inflated by a jet, is NGC 6764 \cite{hota06a,croston08b}. Wide angled radio bubbles extend along the minor axis of galaxy, similar to the morphology of bubbles in simulation D.

\subsubsection{Enhanced turbulent dispersion in gas disks}
An increase in disk turbulence as a result of the wide-spread effect of jet-feedback has been observed in some sources where direct interactions of jets and a turbulent ISM has been detected. Such results are similar to what we find from our simulations, as we outline below.
\begin{itemize}
\item\emph{3C 326:}
In 3C 326 the emission line kinematics of the multi-phase ISM  show very high local dispersion, with a FWHM of $\sim 600 \kms$ in the ionized gas \citep{nesvadba10a} and warm H2 \citep{nesvadba11b}, and $\sim 350 \kms$ in the CO (1-0). The H2 (1-0) emission lines show complex line features requiring multiple Gaussian components to fit the profile. The authors conclude in these papers that the emission is primarily shock excited. Such broad complex line profiles indicate that the jets significantly couple with the entire disk, increasing its turbulence. This is analogous to the increased velocity dispersion in the shocked gas ($\sim 300-600 \kms$) presented in Fig.~\ref{fig.dispersion}. The shocked gas in our simulations is hot and ionized (with $T\sim 10^5$ K). Thus our results are in agreement with the dispersion observed in the ionized components in 3C 326. Since our simulations lack molecular chemistry, we are unable to follow the cooling of the shocked gas to form molecules. However it is likely that the shocked components with high velocity dispersion will show enhanced gas kinematics in the molecular phase when allowed to cool, as we also argue in \citet{mukherjee18a}.

\item\emph{3C 293:}
Similarly, 3C 293 hosts a powerful radio galaxy where fast outflows have been detected in ionised \citep{emonts05a,mahony16a} and neutral \citep{morganti03a,mahony13a} gas, along with emission of thermal X-rays from shocked gas close to the jet \citep{lanz15b}. \citet{mahony16} find that the ionized gas exhibits a large width in the emission lines (FWHM $\sim 200-600 \kms$) along slits positioned both along the jet axis and also along slits displaced from the jet axis. 
Such enhanced kinematics away from the jet axis, is analogous to our results that a jet-driven energy bubble can have a wide-spread effect on the disk, at a few kpc away from the central axis.
\end{itemize}

\section{Impact on star formation rate}\label{sec.sfr}
\subsection{Positive feedback}
Relativistic jets can both promote star formation via density enhancements from shocks (positive feedback) or quench it by mass expulsion or by creating turbulence. Shocks from jets increase the density in the radiative post-shock gas, facilitating collapse. Our current numerical set up lacks an explicit mechanism to follow star formation within the gas disks.  However, one can approximately estimate the star formation rate within a computational volume as 
\begin{equation}
	\mbox{SFR} =\epsilon_{\rm SFR} (\rho d^3x)/t_{\rm ff}\label{eq.sfr}
\end{equation}
Here $t_{\rm ff}= \left(3 \pi/(32 G \rho )\right)^{1/2}$ is the free fall time. $\epsilon_{\rm SFR}$ is an efficiency factor that defines the fraction of gas within a volume that is converted to stars.   A similar definition of star formation rate has been widely used to estimate the star formation rate in computational simulations of galaxy evolution \citep{springel03a,dubois08a} and also in recent studies of positive feedback \citep[such as ][]{gaibler12a,bieri15a,dugan17a,bieri16a}.  

For the results presented here we compute the star formation rate for computational cells with density above a threshold $n>100 \cc$ \citep[similar to][]{gabor14a}, to select the dense regions  with free-fall time scales of a few million years, which will be prone to star formation. Such densities are typical of the mean density in giant molecular clouds \citep{mckee99a}. The value of $\epsilon_{\rm SFR}$ depends on the local physical conditions of the gas cloud such as internal velocity dispersion, magnetic fields and gas dispersal from stellar outflows. Such processes are sub-grid for our simulations since we do not resolve the relevant scales. Typically a value of $\epsilon_{\rm SFR} \sim 0.01-0.1$, used in numerical simulations of star formation in molecular clouds \citep{federrath12a,federrath13a}, matches well with observations. Similar values have also been estimated from observations of star formation rates in high density star forming regions in the Milky Way Galaxy \citep{krumholz07c}. 

\begin{figure}
	\centering
	\includegraphics[width = 7cm, keepaspectratio] 
	{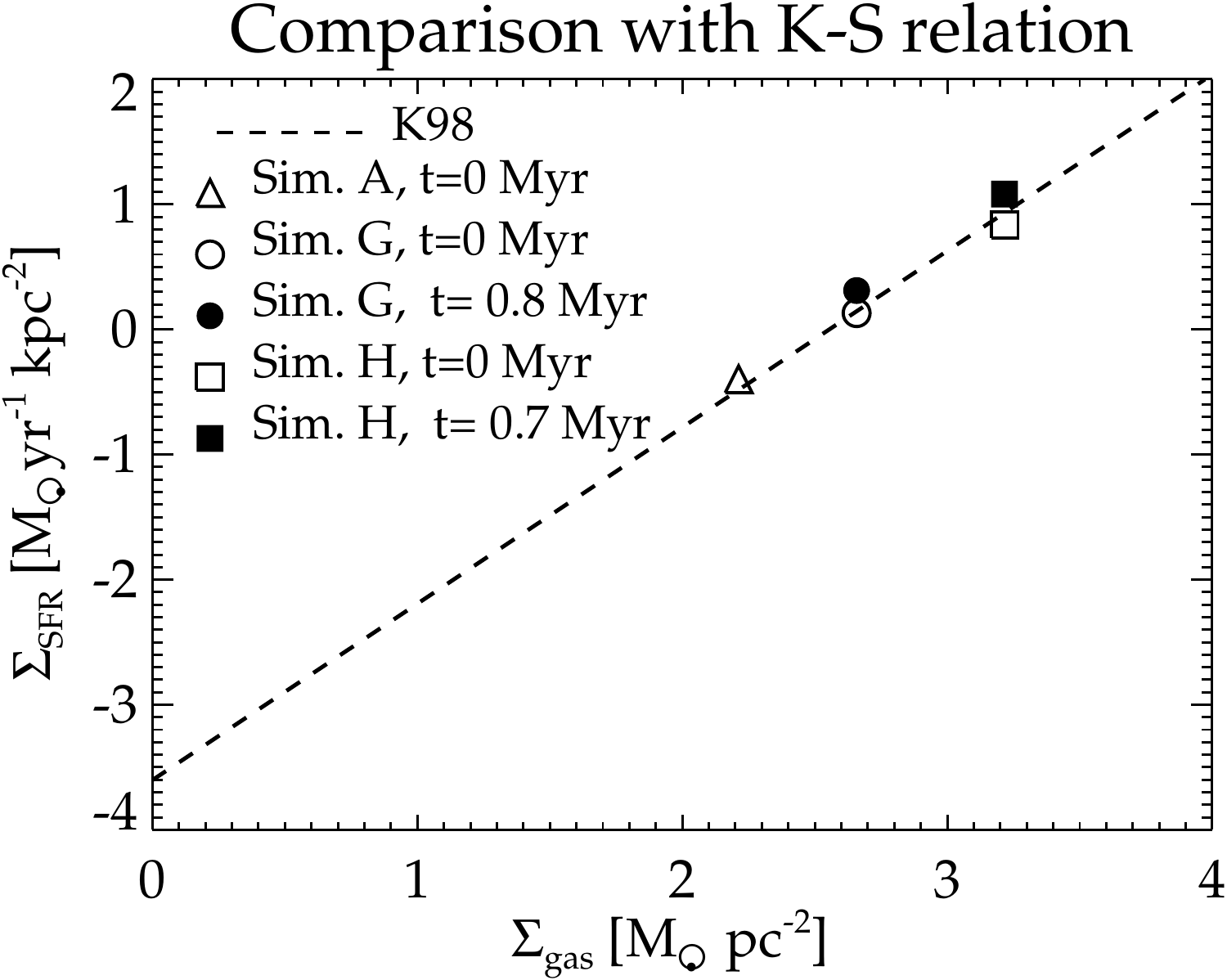}
	\caption{\small The mean star formation rate surface density (total SFR/area of disk) compared with mean gas surface density (total mass/area of disk). The dotted line shows the Kennicutt-Schmidt relation \citep{kennicutt98a}. The initial star formation rates ($t=0$) for simulations A, G and H are shown in partially filled symbols. The filled symbols for simulations G and H show the final star formation rates at the end of the simulation. Other simulations have initial SFR similar to those of A, G or H, being started from a similar initial set up, as discussed in the text.}
	\label{fig.ksplot}
\end{figure}
\begin{table}
\centering
\caption{Star formation rates at $t=0$}\label{tab.sfr}
\begin{tabular}{| c | c |}
\hline
	Simulation	   & SFR ($t=0$) 	\\
Label	   & ($M_\odot \mbox{yr}^{-1}$)$^a$	 \\
\hline
A  			&  5.02  \\
B 			&  17.44 \\
D  			&  17.44 \\
E  			&  17.44 \\
F  			&  5.02  \\
G  			&  16.99 \\
H  			&  86.02 \\
\hline
\end{tabular} 
\flushleft
$^a$ Assuming $\epsilon_{\rm SFR}=0.015$ in eq.~\ref{eq.sfr}. The values scale linearly with $\epsilon_{\rm SFR}$.
\end{table}
For our work, we find that assuming a value of $\epsilon_{\rm SFR}=0.015$ yields beginning star formation rates which match well with the Kennicutt-Schmidt relation \citep{kennicutt98a}. This is shown in Fig.~\ref{fig.ksplot} where we present the star formation rate densities of simulations A, G and H. Simulations B, D, G have similar initial configurations (with $n_{w0}=200 \cc$), and hence similar SFR. Similarly simulations A and F with $n_{w0}=100 \cc$ share identical starting conditions in the gas disk. The initial star formation rates at the start of different simulations, discussed later in this section, is presented in Table~\ref{tab.sfr}.

The above definition of star formation rate used in this work \citep[and others such as ][]{gaibler12a,bieri15a,dugan17a,bieri16a} does not depend on the temperature of the gas. An accurate estimate of star formation, must however, account for local Jeans length, ensure a converging flow and that the gravitational potential energy is higher than the local turbulent kinetic energy \citep[][]{federrath10a}. However, although the absolute magnitudes of star formation rates may change on employing a more robust prescription for star formation, the simple star formation law assumed here, allows for ready comparison of the relative changes in the SFR between the different simulations, with varying mean densities, jet power and orientation, as presented in this work. For the assumed definition, the resultant star formation rate matches well with the Kennicutt-Schmidt relation \citep{kennicutt98a}. This ensures that the initial SFR, based on which the relative strength of positive feedback estimates are made, are realistic.

\begin{figure}
	\centering
	\includegraphics[width = 9cm, keepaspectratio] 
	{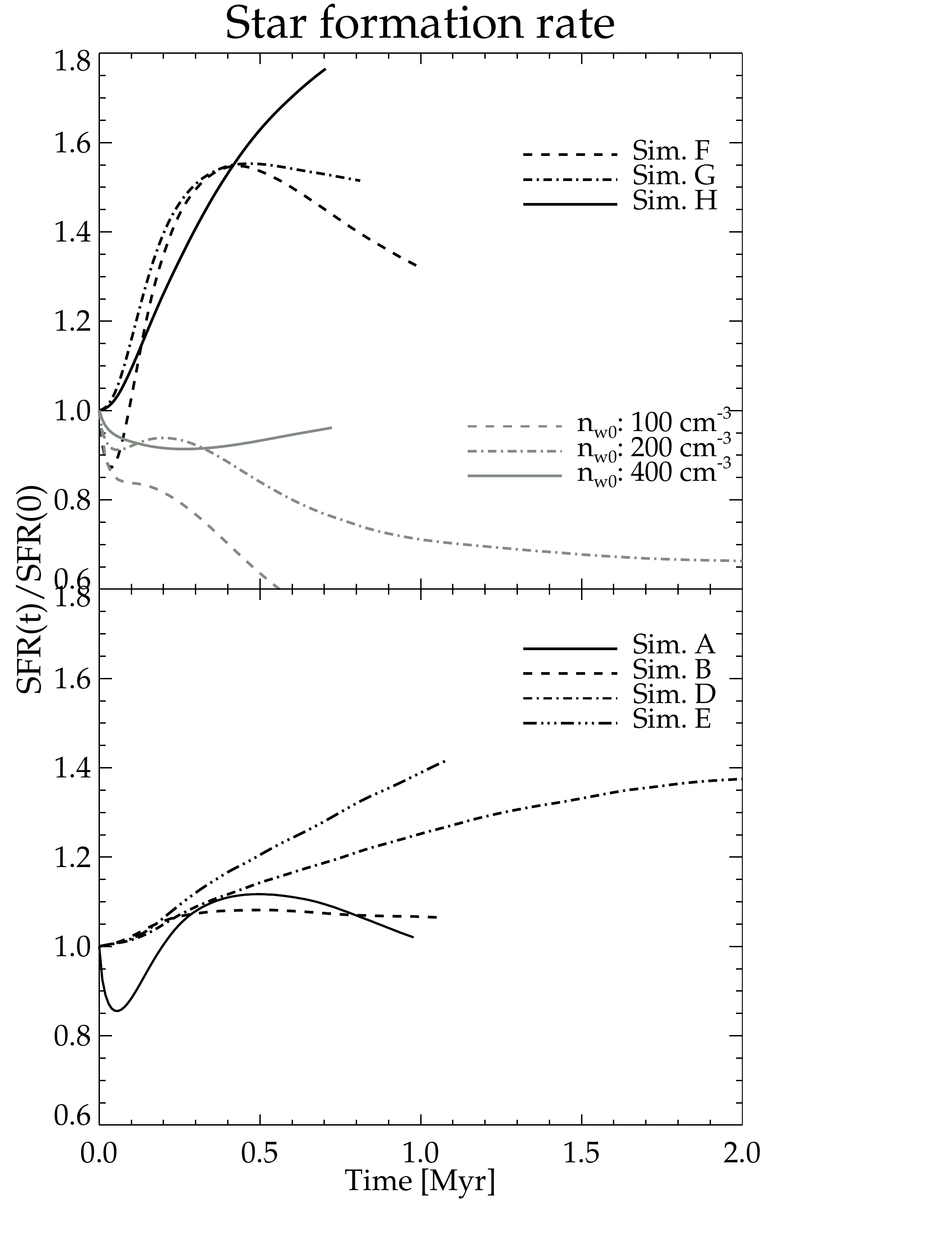}
	\caption{\small Evolution of the star formation rate (defined in eq.~\ref{eq.sfr}) for different simulations. The values of the SFR are normalised to their initial values (presented in Table~\ref{tab.sfr}). \textbf{Top:} Simulations F, G and H with $P_{\rm j}=10^{46} \ergs$ in black. In grey we present the evolution of the SFR without the jets for disks, with densities $n_{w0}=100, 200, 400 \cc$ respectively. \textbf{Bottom:} SFR for simulations A, B, D and E.	Simulations G and H  show a substantial increase in SFR. Simulations A and F with $n_{w0}=100 \cc$ decline after an initial increase. Simulations D and E with jets inclined to the disk ($\theta_{\rm inc} =45^\circ, 70^\circ$ respectively) show an increasing trend in the star formation. }
	\label{fig.sfrtotal}
\end{figure}
Following the above relation, we compute the evolution of the SFR in the  gas disks in our simulations. In Fig.~\ref{fig.sfrtotal} we present the star formation rate of different simulations normalised to their initial values (see Table~\ref{tab.sfr}).  We present the relative changes in the star formation rates to better highlight the effect of the jet on the star formation rate in each simulation. We see that the evolution of the star formation strongly depends on the jet power, ISM density and jet-orientation. 

Simulations B with jet power $\sim 10^{45} \ergs$ show only a modest increase in the over-all star formation rate from its initial value. Simulation A with $n_{w0} \sim 100 \cc$ shows an initial decline before catching up to its original value and declining again. Simulations D and E with the jet inclined to the disk shows a gradual increase by $\sim 40\%$ at late times. Simulation E with a higher inclination angle ($\theta_{\rm inc}=75^\circ$) shows a higher increase in SFR. Jets inclined to the disk couple directly with a larger column of gas, resulting in higher SFR estimates. 

Jets with power $10^{46} \ergs$ show a much sharper increase in the star formation rate. Simulation H with a higher density ($n_{w0} \sim 400 \cc$) shows an increase by a factor of 1.8 during the run time of the simulation, with an increasing trend. Simulation G (with $n_{w0}=200\cc$) rises initially and flattens at $\sim 1.6$. However, simulation F with $n_{w0}=100\cc$ shows a decline in SFR after an initial increase. This is likely because of shearing of clouds as the jets engulf the disk.

Thus it is evident that higher jet powers can, in fact, potentially enhance star formation as they drive stronger shocks into the ISM, resulting in higher density compression. In the top panel, we show the evolution of the SFR for the disks with the jet turned off (presented in grey). Without the jet, the SFR decreases with time. Thus any enhancement in SFR, is purely due to the impact of the jets.  These findings are consistent with the recent results on positive feedback such as \citet{gaibler12a}, \citet{dugan17a} and \citet{bieri16a}. However, disks with lower ISM densities show a decline in star formation after an initial increase due to destruction of clouds by the jet driven flows.

\begin{figure*}
	\centering
	\hspace{-0.2cm}
	\includegraphics[width = 9cm, keepaspectratio] 
	{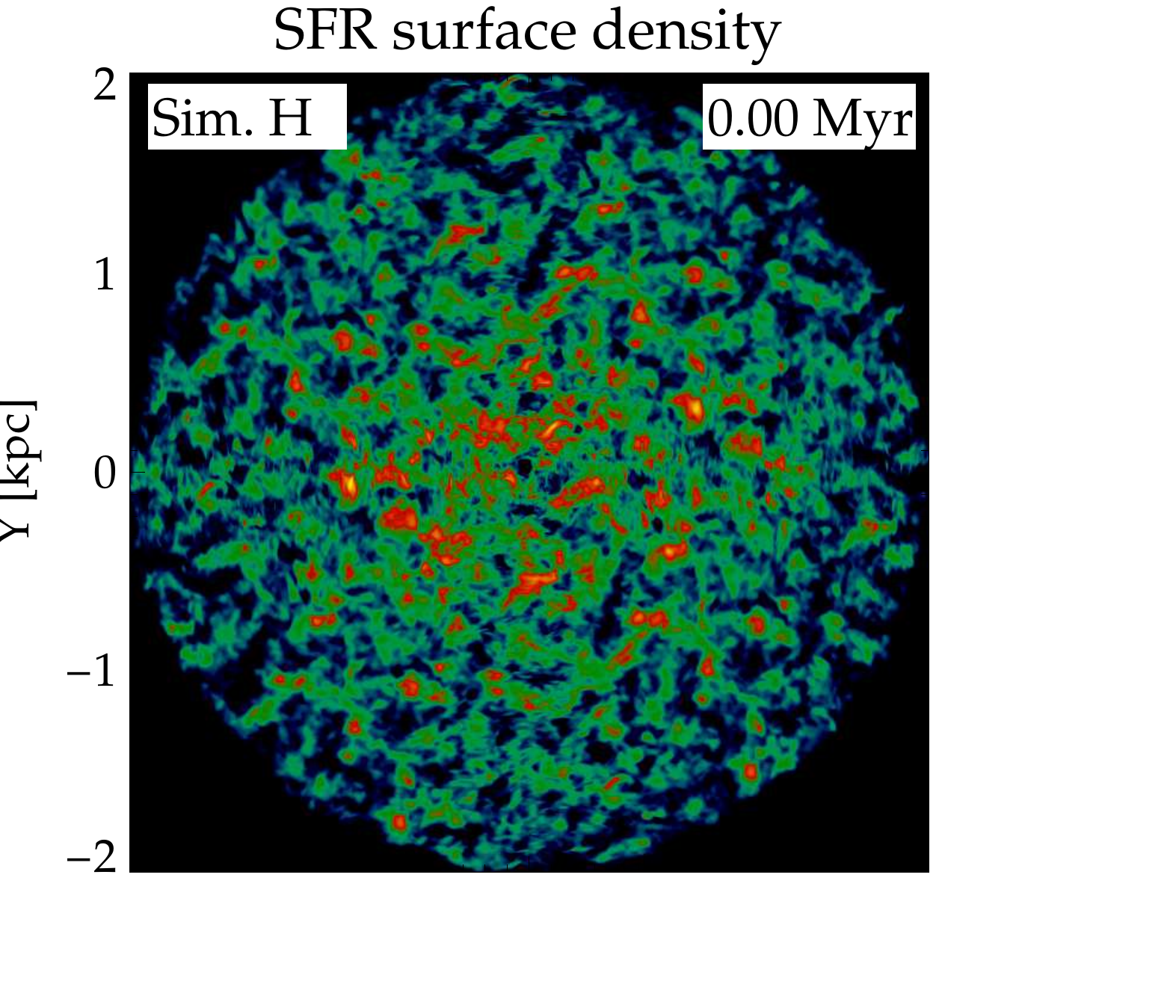}\vspace{-0.8cm}\hspace{-3cm}
	\includegraphics[width = 9cm, keepaspectratio] 
	{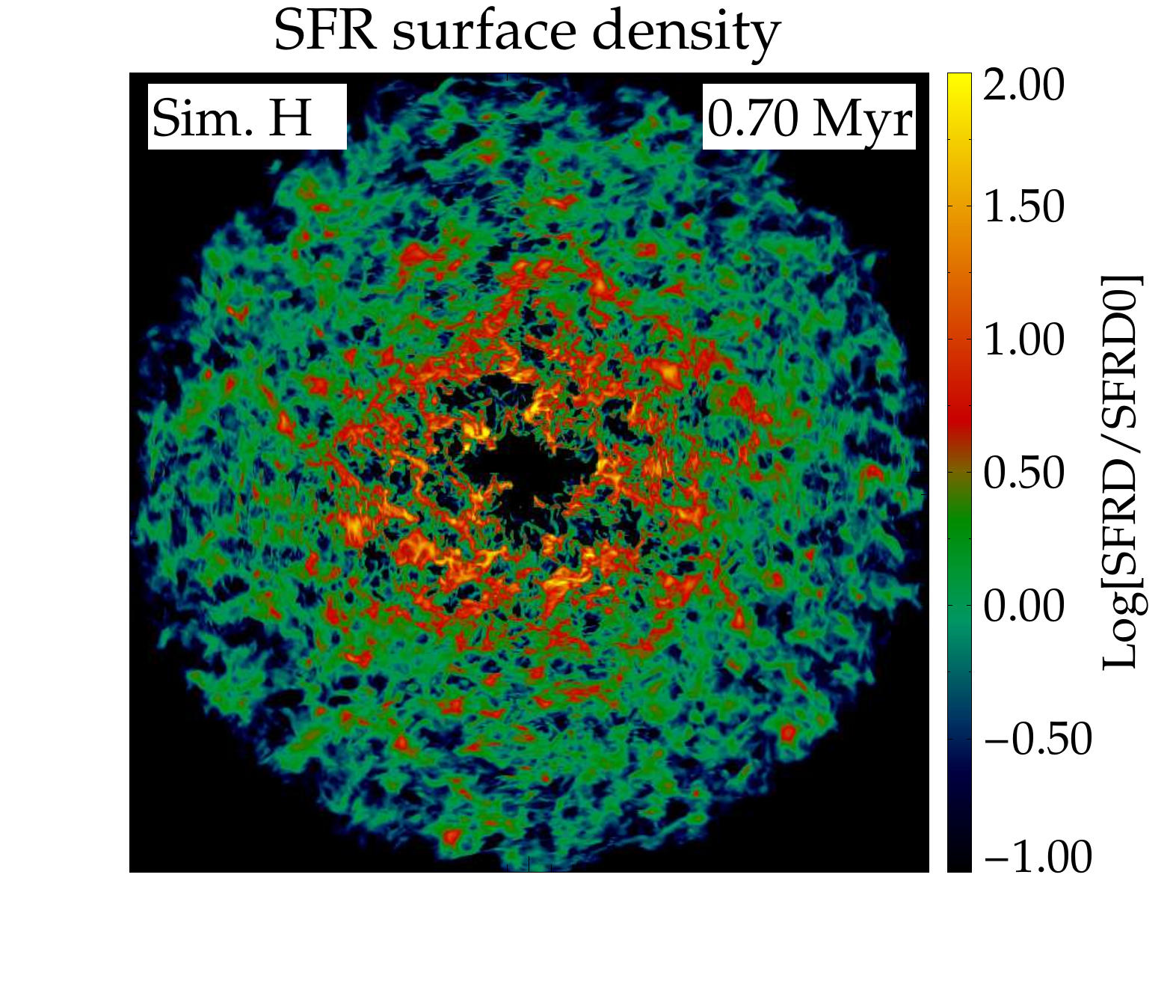}\vspace{-0.8cm}
	\includegraphics[width = 9cm, keepaspectratio] 
	{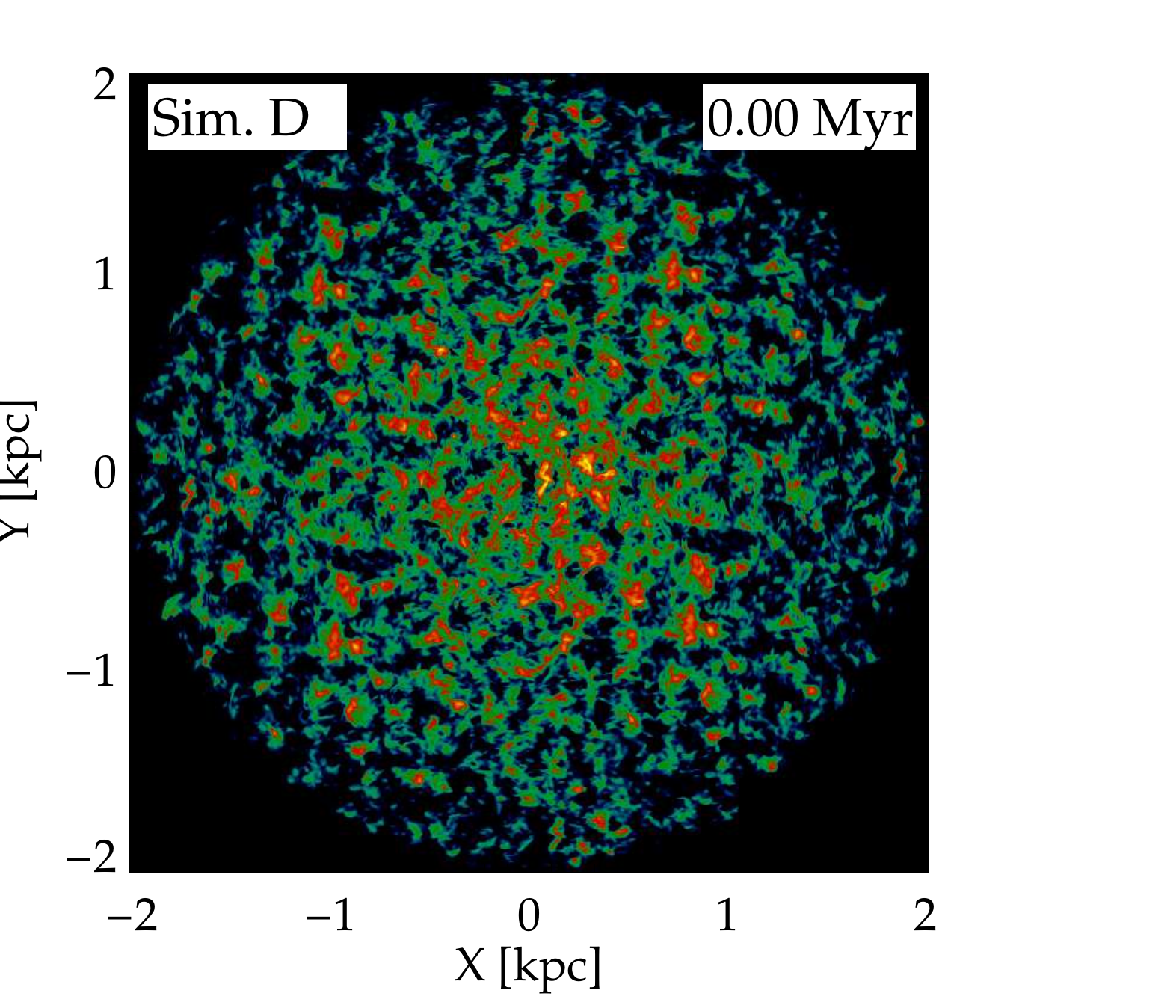}\vspace{-0.cm}\hspace{-3cm}
	\includegraphics[width = 9cm, keepaspectratio] 
	{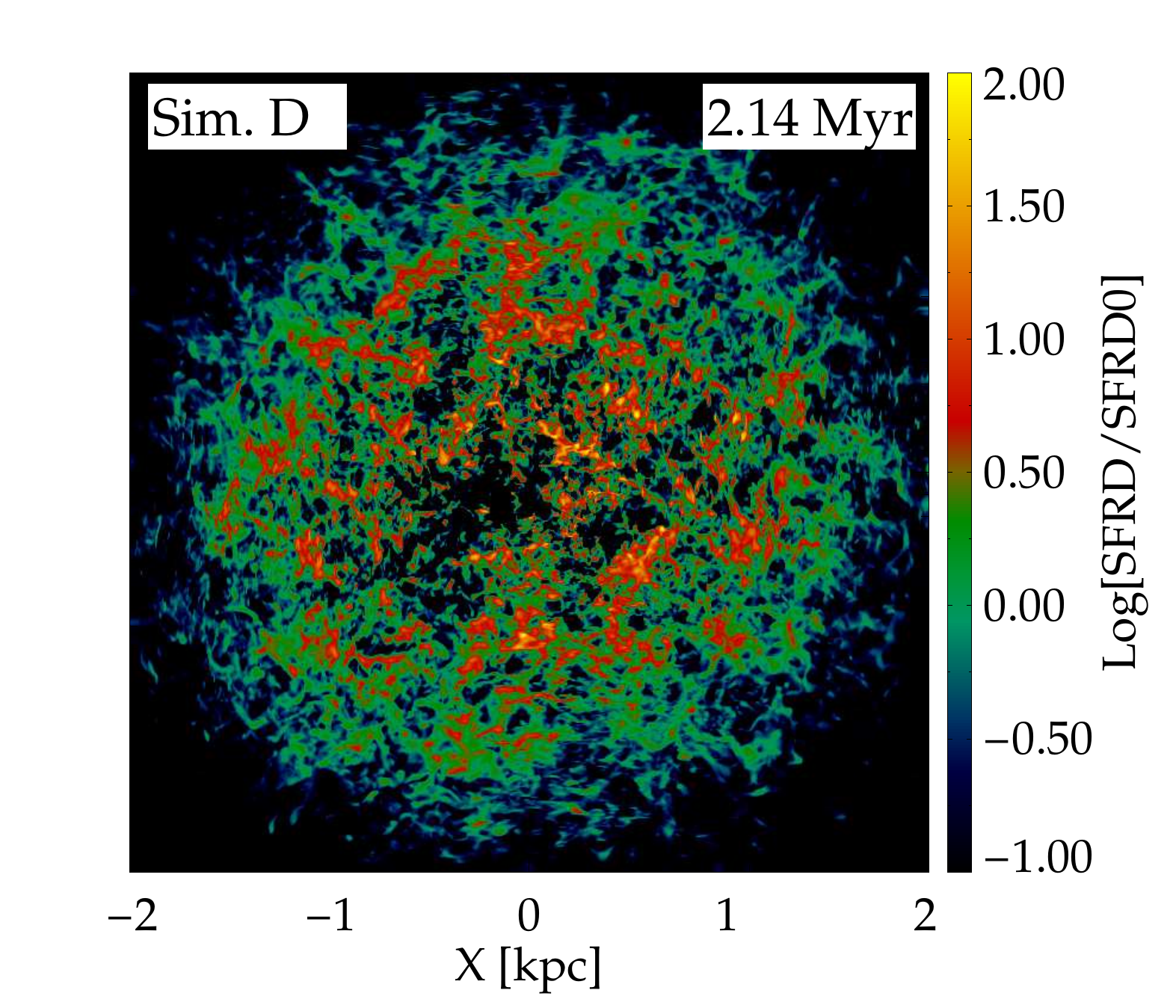}\vspace{-0cm}
	\caption{\small \textbf{Top:} Star formation rate density (SFRD) for simulation H ($P_{\rm j}=10^{46} \ergs$,$\theta_{\rm inc}=0^\circ$), defined as star formation rate (eq.~\ref{eq.sfr}) per unit area. The SFRD is normalised to the mean star formation rate density (SFRD0), defined as the SFR(t=0)/$(\pi (2 [\mbox{kpc}])^2)$, where the radius of the disk is $\sim 2$ kpc. \textbf{Bottom:} Same as above but for simulation D ($P_{\rm j}=10^{45} \ergs$, $\theta_{\rm inc}=45^\circ$) inclined at $45^\circ$ to the disk. The figures show the relative change in star formation rate density under the influence of the jet. Simulation H shows an enhancement in SFRD in a circular ring around the central cavity. Simulation D shows patchy increase in SFRD with a elongated quenched region (south-west) tracking the channel carved by the jet. }
	\label{fig.sfrmap}
\end{figure*}
Although the jets can affect the disk as a whole, jet induced star formation is likely to occur closest to the jet axis, where the pressure from the jet is highest, leading to larger compression of shocked gas. To demonstrate this we present in Fig~\ref{fig.sfrmap} the star formation rate surface density (SFRD) in the $X-Y$ plane, i.e. star formation rate per unit area, defined as: $\int (\rho/t_{\rm ff}) dz$. The SFRD presented in Fig.~\ref{fig.sfrmap} is normalised  to a mean star formation rate density, defined as the total star formation rate at $t=0$ divided by the area of the disk ($=4 \pi$ kpc$^2$, for a disk of radius 2 kpc). Thus Fig.~\ref{fig.sfrmap} shows the relative change in star formation at different locations in the disk to highlight the effect of jet activity on the SFR.

For simulation H, we see a strong enhancement in the star formation rate in the regions surrounding the central cavity ($\lesssim 1 \kpc$). This would imply that a star-forming ring may be created around the jet axis, with a young stellar population. Similar results have also been presented in \citet{gaibler12a}. Although the jet has progressed considerably by this time, and the jet driven bubble has spread over the disk, we do not see  appreciable enhancement of star formation rate (or density) at the outer edges. 

For simulation D, the central cavity is not symmetric, as the jet is inclined to the disk and drills through the sides before eventually breaking out. Such a disruption of the disk is clearly visible as an extended region of quenched star formation (cavity). Patches of enhanced star formation are found at the outer edges. Thus our results show that a relativistic jet can both quench star formation in the central regions due to destruction of clouds, as well as enhance star formation rates in the immediate surroundings due to compression from the jet driven shocks.

\begin{figure}
	\centering
	\includegraphics[width = 7cm, keepaspectratio] 
	{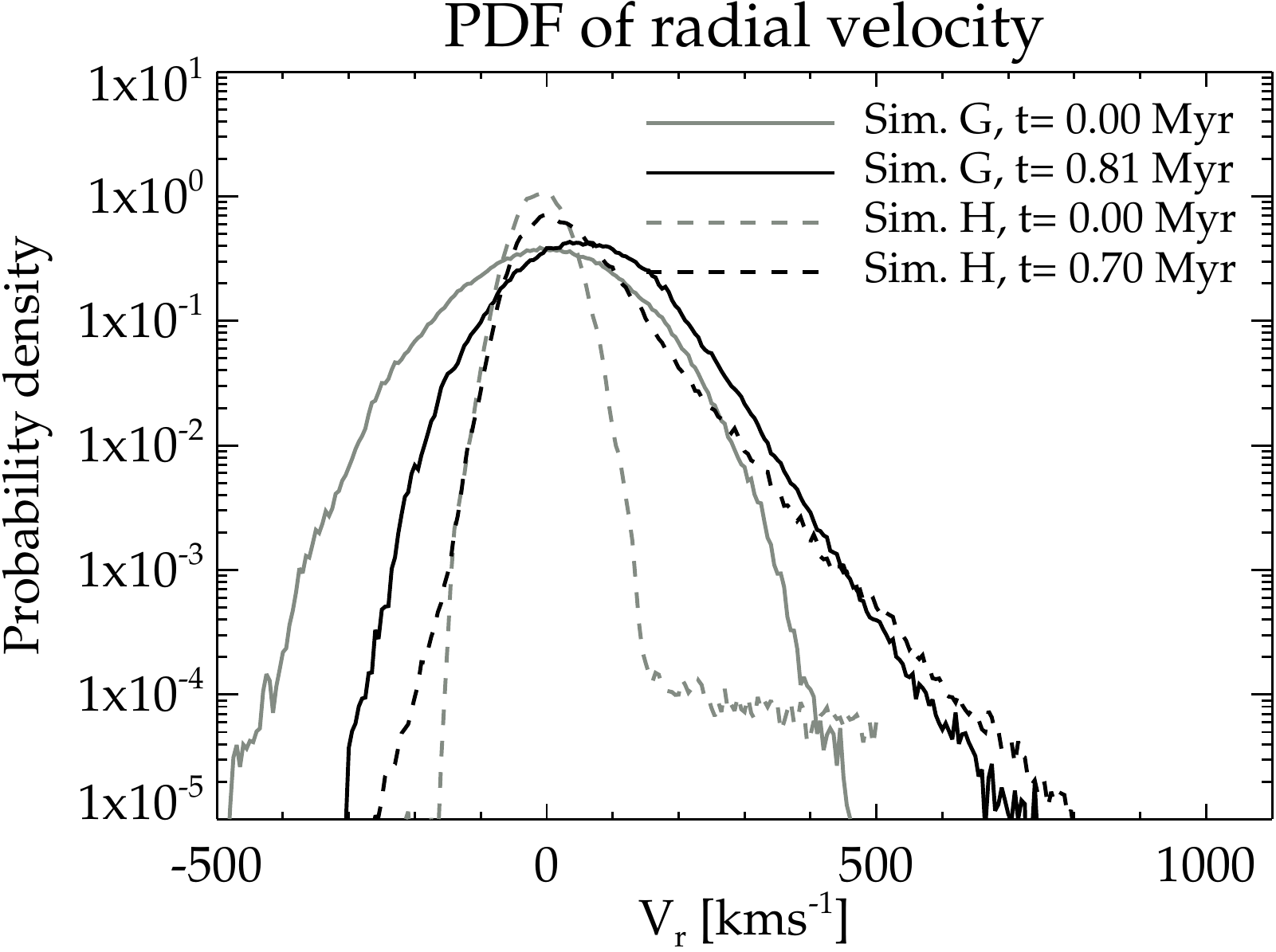}
	\caption{\small Probability density function (PDF) of the cylindrical radial velocity weighted by the star formation rate (given by eq.~\ref{eq.sfr}, for $n>100 \cc$), for simulations G and H. The grey line denotes the PDF at $t=0$, which is symmetric about $V_r=0 \kms$. In black we show the PDF when the jet has evolved. The PDF is asymmetric, with a tail extending beyond $V_r \gtrsim 500 \kms$. The PDF is an approximate indicator of the kinematics of the stars expected to form as a result of positive feedback.}.
	\label{fig.starpdf}
\end{figure}
Stars formed as a result of positive feedback will have motions which are significantly perturbed from the regular stellar kinematics (circular rotation in the present case). Similar results have been presented earlier by \citet{dugan14a}, \citet{zubovas13a} and \citet{zubovas17a}, where stars formed as a result of positive feedback were found to have non-circular orbits with high radial velocities. In Fig.~\ref{fig.starpdf} we present the  PDF of the cylindrical radial velocity of the gas in simulations G and H (which show a strong increase in SFR), weighted by the star formation rate given by eq.~\ref{eq.sfr}, for gas with density $n>100 \cc$. The grey lines denoting the PDF at $t=0$ are symmetric about $V-r=0 \kms$. As the jet evolves (black lines), the velocity PDF develops an extended tail, up to $Vr \lesssim 500 \kms$. This is similar to Fig.~\ref{fig.velpdf}, with smaller upper limit of the radial velocity due to the choice of the density threshold of $n > 100 \cc$. Although the simulations presented here lack explicit numerical schemes to track star formation in-situ \citep[as done in ][]{gaibler12a,dugan14a}, the SFR weighted velocity PDF demonstrates that positive feedback is expected to give rise to hyper-velocity stars.

\subsection{Negative feedback}
The above considerations of star formation rate are based purely on arguments of collapse induced by density enhancement. However, the star formation rate  also significantly depends on the local turbulent velocity dispersion \citep[e.g. see][ for a review]{federrath12a}. The increased turbulence can provide additional support against collapse. True collapse will set in at spatial scales where turbulent velocity fluctuations are less than of the order of the local sound speed, such that global turbulent support does not affect the local dynamics \citep{vazquez03a,federrath10,federrath12a}. Such scales are comparable to (or smaller than) the Jeans length \footnote{The Jeans length is 
\[\lambda _J = (\pi c_s)/\sqrt{G \rho} \sim 1.38 \mbox{ pc } \left(\frac{c_s}{0.2 \kms}\right)\left(\frac{n}{10^3 \cc} \right)^{-1/2},\] where $c_s=\left(\frac{\gamma k_B T}{\mu m_a}\right)^{1/2}$ is the speed of sound. A value of $\mu \sim 2$ has been assumed.} which our simulations do not resolve. However we note that the enhanced turbulence induced in the disk (as discussed earlier in Sec.~\ref{sec.vertical}, point \ref{sec.turb}), will also add to turbulent support preventing collapse. 

If we assume that the turbulence in the disk behaves according to the standard theory of a scale invariant cascade \citep[e.g.][]{kolmogorov41a}, the velocity dispersion ($\sigma_v$) at smaller length scales $l$, depends on dispersion ($\sigma_V$) at larger scales $L$ as: $\sigma _v \sim \sigma_V \left(l/L\right)^{\zeta}$, \citep[e.g. eq 12 of ][]{federrath12a}. Numerical simulations \citep{kritsuk07a,schmidt09a} and measurements of velocity structure function in giant molecular clouds \citep{brunt03a,heyer04a} have shown that the exponent to vary is $\zeta \sim 0.4-0.5$. For length scales $\lesssim 1$ pc, at which collapse may occur, the velocity dispersion would thus be $\sigma _v \sim 12 \kms \left(\sigma _V/400 \kms \right) \left(1 \mbox{ pc}/1 \kpc\right)^{0.5}$. This is still much higher than the thermal sound speed expected in the cold molecular gas ($c_s \sim 0.2 \left(T/10 \mbox{ K}\right)^{1/2} \left(\mu/2\right)^{-1/2}\kms$). This implies that turbulent support may in fact prevent collapse at such scales resulting in negative feedback.

\begin{figure}
	\centering
	\includegraphics[width = 7cm, keepaspectratio] 
	{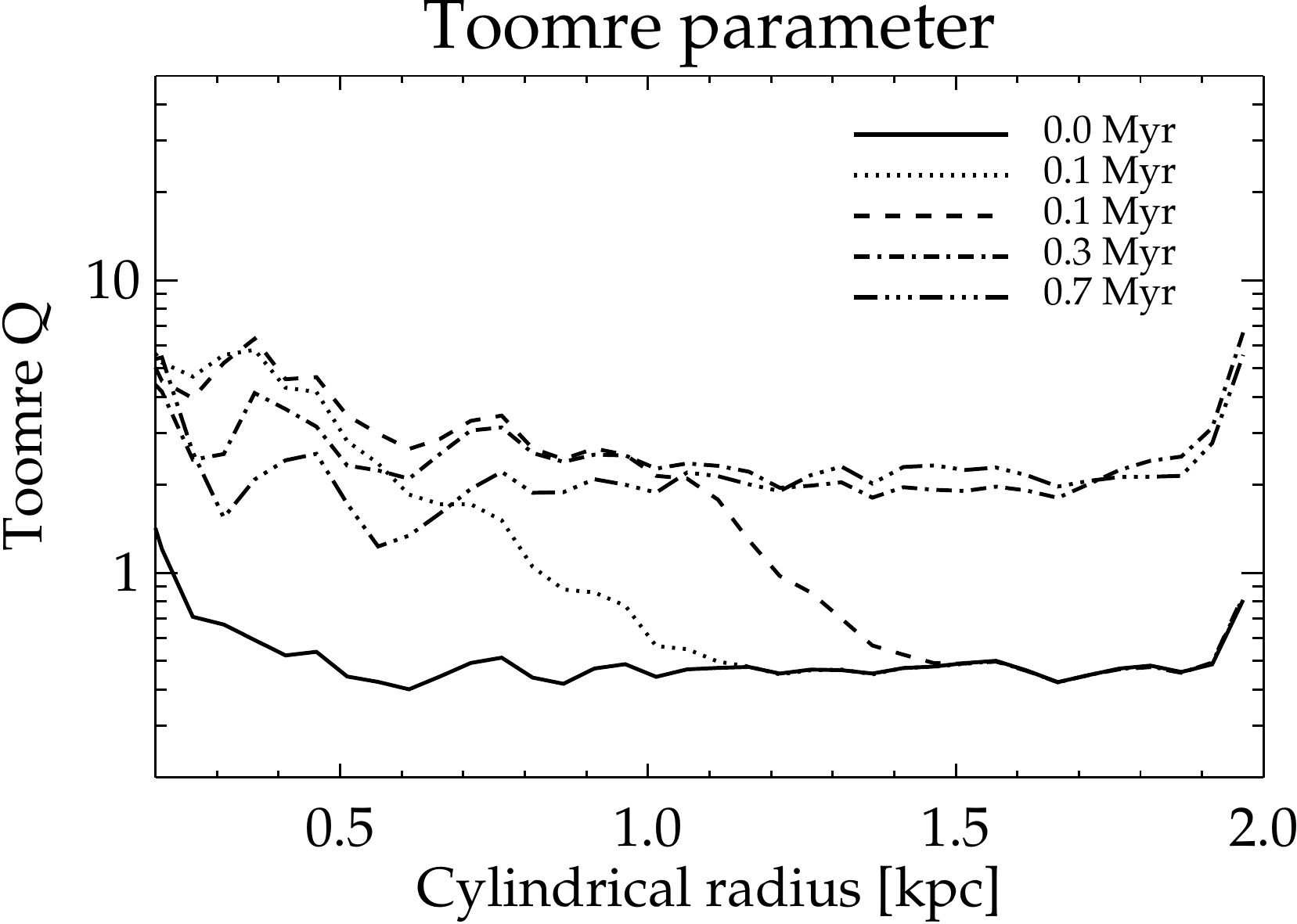}
	\caption{\small The distribution of the Toomre parameter at different times for simulation H. Starting from an initial $Q<1$, the jet driven increase in dispersion enhances the $Q$ to values greater than 1.}
	\label{fig.toomre}
\end{figure}
An often used indicator for ascertaining whether a gas disk in equilibrium will be prone to fragmentation and hence star formation is the Toomre parameter $Q$, defined as $Q \sim \left(\frac{v_\phi}{r}\right) \sigma_v/\left(\pi G \Sigma\right)$. Here $\sigma _v$ is the velocity dispersion and $\Sigma$ the density per unit area.  We have approximated the epicyclic frequency as $\kappa ^2 = \frac{2 \Omega}{r}\frac{d}{dr}\left(r^2 \Omega\right) \sim \langle v_{\phi} \rangle/r $. A gas disk is will be gravitationally stable if $Q >1$ \citep{toomre64a,wang94a}. If we consider the fractal disk in our simulations as a symmetric thick disk on average, the jet induced dispersion is expected to raise the Toomre $Q$, preventing further fragmentation. We present the cylindrically radial profile of $Q$ for simulation H in Fig.~\ref{fig.toomre}. The starting velocity dispersion at the time of injection of the jet in simulation H was selected to be lower than in the other simulations ($\sim 40 \kms$) so that initially, $Q<1$. As the jet interacts with the disk, raising its dispersion,  $Q$ increases by a factor of 2 or more throughout the disk. Hence, on average, the increased velocity dispersion, even on the outer edges of the disk which are not directly affected by the central jet driven outflow, may render the disk stable to fragmentation. 

Thus, to conclude, density enhancement near the jet axis can make the clouds more prone to gravitational collapse  promoting star formation. However, overall, an increase in turbulent kinetic energy may lower the star formation rate. This indicates that both positive and negative feedback may operate in galaxies impacted by relativistic jets. Simulations with higher spatial resolution are needed to better assess the relative importance of these two forms of feedback.

\section{Summary and Discussion}\label{sec.discuss}
In this paper we have studied the interaction of relativistic jets and the energy bubbles they create with dense turbulent gas disks. We have carried out a suite of simulations with different jet powers, densities and jet inclinations. Our primary focus has been in understanding the evolution of the  kinematics, morphology and energetics of the multi-phase ISM under the influence of the relativistic jet and its implications for the evolution of the galaxy properties such as star formation rate. Below we summarize the results and the their implications.

\begin{enumerate}
\item\textbf{Jet driven outflows}
The jets launch an energy bubble that drives shocks into the disk. The jet couples with the ISM driving fast multi-phase outflows. Clouds are dredged up vertically along the axis of the jet with outflows velocities $\sim 500 \kms$. The dense clouds accelerated by the jets form cometary tails as they are ablated by the fast flows. Typically, a cloud is expected to survive a few cloud crushing times \citep{fragile04a,scannapieco18a}, where the cloud crushing time is the time for the incumbent shock to pass the length of the cloud. For a shock of speed $v_{sb}\sim 1000 \kms$, and density contrast of $\chi = 10^4$ between the up-stream and down-stream fields (which would occur for a bubble density of $n_b \sim 0.01 \cc$ and cloud density of $n_c\sim 100 \cc$), the cloud crossing time ($t_{cc}$) is \citep{fragile04a}:
\begin{align}
	v_{sc} &= \left(\frac{n_b}{n_c}\right)^{1/2} v_{sb} \\
	       &= \chi^{-1/2} v_{sb} \\
	t_{cc} &= \frac{L_c}{v_{sc}} \\
	       &= \chi^{1/2} \frac{L_c}{v_{sb}} \\
	       & \sim 1 \mbox{ Myr } \left(\frac{\chi}{10^4}\right)\left( \frac{L_c}{100 \pc}\right) \left(\frac{v_{sb}}{1000 \kms}\right)
\end{align}
		This is comparable to the time scales of the simulation. In our simulations, the gas ejected vertically from the disk, is mostly shock heated to high temperatures, and breaks up to form diffuse cometary tails. The ablated gas being lighter ($n \sim 0.1-10 \cc$) is swept up and attains much higher velocities of $\sim 10000 \kms$. However, we note that to attain numerical convergence of jet-cloud interaction, a much higher resolution is required, with the clouds resolved by at least 100 computational cells \citep{barragan16a}.  As the jet breaks out from the disk, the energy bubble spreads over the disk, driving shocks inward into the disk and compressing them.

\item\textbf{Non-circular motions in the disk}
Besides vertical outflows, the lateral expansion of the jet driven energy bubble also launches radial outflows into the plane of the disk. The jet driven outflows result in a departure of the gas kinematics from the circular rotation initially established (as shown in Fig.~\ref{fig.beta}). The strong radial motions of the gas has two  observational signatures. Firstly, lines of sight probing the disk would observe extreme gas kinematics as highly broadened emission lines. Significant distortion in the galaxy's rotation curve in a position-velocity diagram would also be observed. Such features have been observed in detail for the galaxy IC~5063 \citet{morganti15a,dasyra15a,dasyra16a,oosterloo17a}. Detailed simulations targeted towards explaining the observed kinematics in IC~5063 have been presented in a separate publication \citep{mukherjee18a}. 
		
Secondly, if the radially driven gas cools to form stars, non-circular motions would leave a population of stars with strongly elliptical orbits. Our present work lacks the numerical machinery to address the star formation properties of the gas clumps. However, similar suggestions have been made in recent papers \citet{dugan14a}, raising the possibility that hyper-velocity stars (or gas parcels) with non-circular radial orbits being are a possible signature of AGN activity.

\item\textbf{Turbulence inside the disk}
The jets drive shocks and hydrodynamical flows that penetrate the disk, increasing its turbulence. The turbulence in the shocked gas increases by 4-6 times its initial value (as shown in Fig.~\ref{fig.dispersion}). This increase occurs not just in the central nuclear region where the jet-ISM interaction is expected to be strongest, but all throughout the disk. This implies that the onset of jet activity produces large-scale effects over several kpc. For a disk with an identical initial ISM, jets with power $10^{46} \ergs$ induce velocity dispersions  higher by a factor $\sim 1.5$ than a jet with with power $10^{45} \ergs$.  Lower density clouds are more prone to ablation, resulting in higher velocity dispersion in ISM with lower mean density, for the same jet power. Our simulations support the kinematic observations of jet-ISM interactions in galaxies such as 3C 326 \citep{nesvadba10a,nesvadba11b} and 3C 293 \citep{mahony16}. 

\item\textbf{Inclined jets}
We have carried out simulations of jets inclined to the disk at various angles of inclination. We find that an inclined jet behaves very differently from a jet launched perpendicular to the axis. Being inclined to the disk, the jets encounter a larger column depth of gas with which they interact  before breaking out. The inclined jets decelerate considerably and launch a sub-relativistic wind along the galaxy's minor axis following the path of least resistance. Such a wind is reminiscent of the wide-angle outflow often associated with AGN winds \citep{rupke11a} or nuclear starburst \citep{veilleux05a}. 
		
Similar results of AGN outflows inclined to the disk have  recently been explored in simulations of \citet{dugan17a}. The starting outflow in \citet{dugan17a} was assumed to be non-relativistic \citep[though ultra-fast, akin to][]{wagner13a}. However, as we demonstrate in our simulations, such wide-angle outflow may also arise from an inclined relativistic jet itself. The ensuing bubble evolves spherically, in contrast to the directional outflows seen in powerful radio jets. Often the difference in morphology of the observed winds is used to differentiate between a wind-driven feedback from a jet-driven feedback \citep{rupke11a}. However, our simulations show that even a relativistic jet can give rise to a morphology that that may be otherwise attributed to a quasar driven wind. 

An inclined jet also perturbs the gaseous disk more strongly. The velocity map in Fig,~\ref{fig.p45dir45} shows a gas distribution resembling a wide-angled outflow, compared to an inner outflow followed by compression at outer edges as seen in the case of a jet perpendicular to the disk. The stronger jet-ISM interaction results in an increased velocity dispersion and a higher mean radial velocity with a larger value of the inclination angle. The jet-driven bubble of an inclined jet expands more slowly, than that of a perpendicular jet, as the bubble for the inclined jet is powered by a sub-relativistic wind. The ensuing morphology of the shocked gas and bubbles have strong similarities with some observed systems such as NGC~1052 and NGC~3079.

\item\textbf{Positive vs negative feedback}
\begin{itemize}
\item\textbf{Positive feedback: }
In Sec.~\ref{sec.sfr} we have estimated the impact of the jet on the star formation rate in the galaxy. Assuming a Kennicutt-Schmidt star formation law \citep{kennicutt98a}, we find that regions close to the jet axis may experience an increase in star formation rate resulting from the enhanced density produced by strong radiative shocks. Simulations with $P_{\rm j} = 10^{45} \ergs$ show only a modest increase in star formation (e.g. by $\sim 6\%$ for simulation B), whereas higher power jets ($P_{\rm j} = 10^{46} \ergs$) show a significant overall increase in star formation rate (e.g. by $80\%$ for simulation H). The increase of SFR with higher jet power is due to the enhanced compression of the gas by the stronger shocks. At lower jet powers, the ablation by the outflow wins over shock compression. Similar results were also shown in \citet{zubovas17a}, where it was found that cloud fragmentation rate (a proxy for star formation) peaks at a critical AGN luminosity. Inclined jets, with stronger interactions with the ISM, are also seen to exhibit an increase in star formation. 
		
Jet-induced star formation is a viable mechanism  to explain enhanced star formation in several observed sources with star forming gas clumps located near a radio jet, e.g.,  the Minkowski object \citep{salome15a,lacy17a}, 3C 285 \citep{salome15a}, 4C 41.17 \citep{bicknell00a}. There is also evidence of enhanced star formation rate in galaxies hosting an AGN \citep{zinn13a,bernhard16a}. Several other theoretical works have also explored the possibility of AGN-induced starformation in galaxies, such as \citet{nayakshin12a}, \citet{wagner11a}, \citet{gaibler12a}, \citet{dugan17a} etc., where compression of gas by the AGN driven outflow can trigger star formation, similar to what we find.
		
The enhancement in the star formation rate surface density is highest close to the jet axis, appearing as a circum-nuclear star-forming ring ($\lesssim 1 \kpc$). Similar results have also been presented  by \citet{gaibler12a}. If such positive feedback is indeed at play, a star forming ring with a young stellar population around the jet axis is a possible observational signature for such systems.

Positive feedback can result in the formation of hyper-velocity stars, and more generally  and stars with significant radial orbital perturbations, with kinematics significantly different from the rest of the stellar population (see Fig.~\ref{fig.starpdf}). Recently, star formation ($\sim 15 M_\odot \mbox{yr}^{-1}$) has been detected inside a galactic outflow \citep{maiolino17a} with stellar velocities higher than that of the stars in the centre (by about $\lesssim 100 \kms$). Similar kinematic imprint of hyper-velocity stars resulting from star formation triggered by an AGN driven outflow is likely to be a general feature of star-forming galaxies with a central active super-massive black hole \citep{silk12b,zubovas13a,zubovas17a,wang18a}. If a star-forming galaxy undergoes several episodes of AGN/jet activity \citep[e.g. ][]{rampadarath18a} which also result in positive feedback, such galaxies will have a population of  stars whose formation has been triggered by positive feedback,  with radial velocity  asymmetries \citep[as apparently found in stacked SDSS star-forming galaxies  ][]{cicone16a}. Although supernova-driven outflows can also impact the stellar kinematics \citep[e.g. ][ for low mass galaxies]{badry17a}, the kinematic signature of star formation induced by an AGN driven outflow will be different, in having a strong radial motion imprint  of the newly formed stars. These dynamical signatures will accumulate and be reinforced over successive AGN episodes.

\item\textbf{Negative Feedback:}
In our simulations we do not find any appreciable mass loss due to outflows which would result in significant quenching. However, the increased local velocity dispersion in the ISM induced by the jet driven shocks can provide additional turbulent support preventing formation of stars. Such an effect is not captured by the positive feedback estimates inferred from a simple Kennicutt-Schmidt star formation law that solely depends on the local density \citep[e.g. eq.~\ref{eq.sfr}, as well as in ][]{gaibler12a,bieri16a,dugan17a}. In our simulations we find that turbulence is strongly enhanced ($\sim 400-600 \kms$) to several times the initial values. The dispersion computed here is on scales of the global disk ($\sim 1$ kpc). 

Observational support of turbulence-driven quenching by relativistic jets  has been recently presented for some galaxies, e.g., 3C 326 \citep{nesvadba10a}, a large fraction of the MOHEGs \citep[Molecular Hydrogen Emission Galaxy, ][]{ogle10a}, NGC 1266 \citep{alatalo15a}, etc. Although high-velocity outflows have been detected in several of these galaxies \citep{ogle10a,guillard12a,nesvadba10a}, they do not carry enough mass to unbind the gas reservoir completely. Such a scenario is also true for NGC 1266 with a jet of much lower power \citep{nyland13a}  that does not drive outflows that escape the galaxy \citep{alatalo11a}. This points to a quenching mechanism in which the jet can drive turbulence on scales comparable to the gas extent of the galaxy, as in our simulations.

\end{itemize}

The current simulations represent only a few million years of the lifetime of a galaxy. Such time scales are much smaller than the  dynamical time of the galaxies. Hence, the estimates of positive or negative feedback presented here should be considered more as a singular event associated with one episode of jet activity. Long-term effects of jet-driven feedback on a disk  pressurized by the jet cocoon have also been recently explored by \citet{bieri15a,bieri16a}. \citet{bieri16a}, who keep the disk artificially pressurized for $\sim 400$ Myr, resulting in enhanced fragmentation of the disk and subsequent star formation. However, in a realistic system, the pressure in the energy bubble will decrease with time as a result of its expansion, as shown in Fig.~\ref{fig.meanpres}. Thus it is still uncertain whether there can be sustained positive feedback to significantly affect a galaxy's stellar mass assembly. 
		
		
 We conclude that an onset of jet activity may induce positive feedback in regions directly being impacted by the jet, resulting in localized regions of enhanced star formation during early phases of the jet evolution. Over larger spatial extents, enhanced turbulence from the jet-driven energy bubble may suppress star formation for the duration of jet activity. The above inferences are indirect, as our simulations currently lack the numerical tools to track potential star forming clumps and lack the resolution to model turbulence in the star forming clumps. In future we aim extend the numerical set-up to address these questions in greater detail.

\end{enumerate}

\section{Acknowledgement}
We gratefully acknowledge Nicole Nesvadba, Christoph Federrath, Michael Dopita, Elizabeth Mahony, Sylvain Veilleux and Luisa Ostorero for helpful discussions. We thank the referee for the thorough review and insightful comments. The simulations have been carried out in the supercomputing facilities at NCI and Pawsey, made available to researchers at ANU under the NCMAS and ANU partner share schemes. We thank the timely help from the NCI and Pawsey support staff. DM's visit to JHU was supported by a Balzan grant from New College, Oxford.  AYW has been supported in part by ERC Project No. 267117 (DARK) hosted by Universit\'e Pierre et Marie Curie (UPMC) - Paris 6, PI J.~Silk.

\appendix
\section{Bubble dynamics} \label{append.bub}
In this appendix we discuss the self-similar expansion of a bubble driven by a nuclear wind expanding adiabatically into an ambient medium with a power-law density profile. The bubble has a radius $R_B$, volume $V_B$ and pressure $p_B$ which is assumed to be uniform (as is typical of an over-pressured bubble). The bubble is powered by a nuclear wind with energy flux $F_E$. The bubble is expanding into an ambient medium, whose density varies as a power-law beyond a scale radius $R_c$ (typically the core radius of the galaxy):
\begin{align}
\rho _a &= \rho _c \left(\frac{R}{R_c}\right)^{-\beta}.
\end{align}
Thus the total mass enclosed in a radius $R_B$ (for halo with $\beta \leq 2$) is:
\begin{align}
    \int_0^{R_B} \rho_a 4 \pi R^2 dR &= \int_0^{R_B} \rho _c \left(\frac{R}{R_c}\right)^{-\beta} 4 \pi R^2 dR \nonumber \\
    &=\frac{4}{3-\beta} \pi R_B^3 \rho_c \left(\frac{R}{R_c}\right)^{-\beta} \label{eq.mass}
\end{align}
We consider the bubble radius $R_B$ to be defined by the forward shock. The velocity of the bubble radius is then $v_B=dR_B/dt$
Neglecting the mass-inflow by the wind such that the mass interior to the forward shock is comprised of swept up gas,  the mass and momentum equations can be written as
\begin{align}
\frac{d}{dt} \int \rho _a v_B d^3x &= \int p \hat{r}.dS \label{eq.moment1}\\
\frac{d}{dt}\left(\frac{4}{3-\beta} \pi R_B^3 \rho_c \left(\frac{R_B}{R_c}\right)^{-\beta} \frac{d R_B}{dt} \right) &= 4 \pi R_B^2 p_B. \label{eq.moment}
\end{align}
where we have inserted eq.~\ref{eq.mass} in eq.~\ref{eq.moment1}.
With an energy flux $F_E$ into the bubble, the energy equation can be written as:
\begin{align}
&\frac{d}{dt} \left[\frac{4 \pi}{3} \frac{p_B}{\gamma -1} R_B^3 \right] 
 + 4 \pi R_B^2 p_B \frac{d R_B}{dt}  = F_E  \label{eq.energy}
\end{align}

For a self-similar expansion, the bubble radius is a power-law in time: $R_B=At^\alpha$. Inserting this into eq.~\ref{eq.moment} we find the time evolution of the pressure as:
\begin{equation}
p_B = \frac{\rho_c \alpha}{(3-\beta) R_c^{-\beta}} A^{2-\beta} \left[(4-\beta)\alpha -1\right] t^{(2-\beta)\alpha -2} \label{eq.pres}
\end{equation}
Inserting eq.~\ref{eq.pres} in the energy equation (eq.~\ref{eq.energy}), and assuming the energy flux to be time-independent, we obtain  
\begin{equation}
\alpha = \frac{3}{5-\beta} \label{eq.alpha}
\end{equation}
with the coefficient $A$ given by
\begin{equation}
A=\left[\frac{R_c^{-\beta} (3-\beta)}{6 \pi \rho_c}\frac{(5-\beta)^3}{(7-2\beta)(11-\beta)} F_E \right]^{1/(5-\beta)}
\end{equation}
Thus the expressions for evolution of radius, velocity of the bubble surface and pressure are:
\begin{align}
R_B &= \left[\frac{1}{6 \pi}\frac{(5-\beta)^3(3-\beta)}{(7-2\beta)(11-\beta)} F_E \right]^{1/(5-\beta)} \nonumber \\
& \times R_c^{-\beta/(5-\beta)} \, \rho_c^{-1/(5-\beta)} \, F_E^{1/(5-\beta)} \; t^{3/(5-\beta)} \label{eq.radius2} \\
V_B &= \frac{3}{5-\beta} \left[\frac{1}{6 \pi}\frac{(5-\beta)^3(3-\beta)}{(7-2\beta)(11-\beta)} F_E \right]^{1/(5-\beta)} \nonumber \\
& \times R_c^{\beta/(5-\beta)} \, \rho_c^{-1/(5-\beta)} \, F_E^{1/(5-\beta)} \; t^{-(2-\beta)/(5-\beta)} \label{eq.vel2}\\
p_B &= \frac{3(7-2\beta)}{(3-\beta)(5-\beta)^2} \left[\frac{(5-\beta)^3(3-\beta)}{6\pi (7-2\beta)(11-\beta)}\right]^{(2-\beta)/(5-\beta)} \nonumber \\
&\times R_c^{-3 \beta/(5-\beta)} \, \rho_c^{3/(5-\beta)} \,  F_E^{(2-\beta)/(5-\beta)} \, t^{-(\beta + 4)/(5-\beta)} \label{eq.pres2} \\
p_B &= \frac{3(7-2\beta)}{(3-\beta)(5-\beta)^2} \left[\frac{(3-\beta)(5-\beta)^3}{6 \pi (7-2\beta)(11-\beta)} \right]^{2/3} \nonumber \\
&\times \rho_c^{1/3} R_c^{\beta/3} F^{2/3} R_B^{-(\beta+4)/3} \label{eq.presrad}
\end{align}
For $\beta=0$ the above equations reduce to the expressions derived in \citet{weaver77a} for a uniform external medium.
 The exponent of the time for the evolution of the bubble pressure is similar to that of \citet{kaiser97a}, who also have studied the self-similar evolution of a jet.

Hence for $\beta \sim 1.5-2$, which is typical for the double isothermal potentials used in our simulations, the bubble pressure will thus be $\propto t^{-1.57}$ or $\propto t^{-2}$ respectively. For simulation D, which launches a spherical sub-relativistic wind, the pressure initially ($t < 0.4$ Myr) varies as $\sim t^{-1.21}$, with a later evolution of $\sim t^{-1.54}$, close to the predicted bubble solution. The initial slower decrease occurs because the bubble is trapped inside the disk and suffers radiative losses. Vertical jets show a faster and  non-spherical expansion, resulting in a faster decrease in the bubble pressure (as shown in Fig.~\ref{fig.meanpres}). 

\begin{figure}
	\centering
	\includegraphics[width = 7.cm, keepaspectratio]{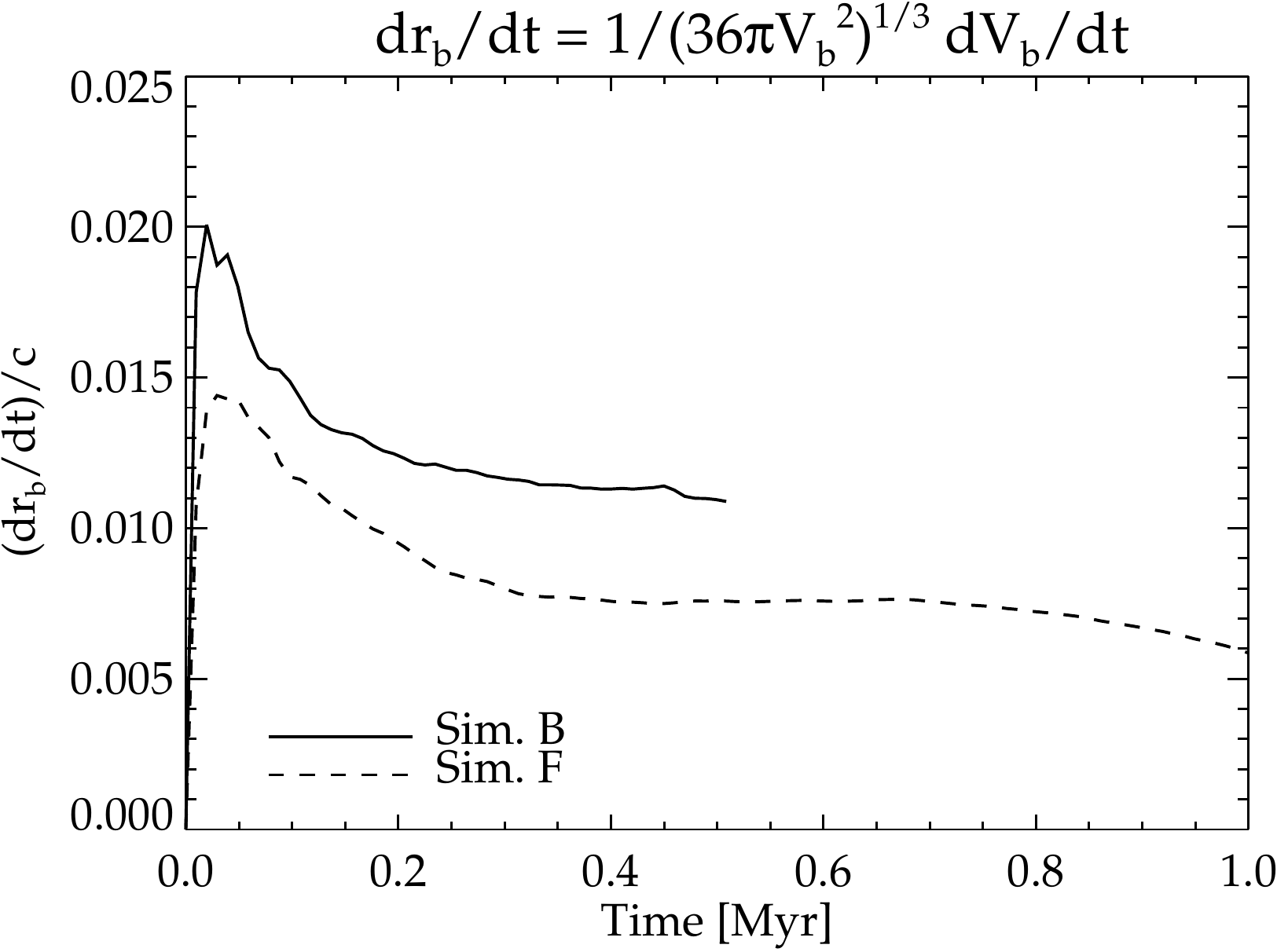}
	\caption{\small Rate of expansion of the bubble for simulation B and F computed as $dr_b/dt = 1/(4 \pi r_b^2) dV_b/dt = 1/(36 \pi V_b^2) dV_b/dt$, where $V_b = (4 \pi r_b^3)/3 $ is the volume of the bubble. }
	\label{fig.machamb}
\end{figure}
The strength of the shock driven into the ambient medium can be approximately estimated by equating the ram pressure of the shock to that of the pressure in the energy bubble.
\begin{align}
	p_{\rm bubble} &= \rho _a v_{\rm shock}^2 \\
	p_{\rm bubble} &\sim 8.95\times10^{-8} - 8.95 \times10^{-9} \mbox{dyne cm}^{-2} \\
	               & (\mbox{for t} \sim 0.1-0.5 \mbox{ Myr respectively, } \mbox{ for Sim. B}) \nonumber \\
	n_a &\sim 0.1 \cc  \mbox{(ambient density in 2-4 kpc)} \\
	v_{\rm shock} &\simeq 3\times10^3 -10^4 \kms \\
	\mathcal{M} &= v_{\rm shock}/\left(\frac{K T_a}{\mu m_a} \right)^{1/2} \simeq 8.1-26
\end{align}
Another approximate estimate of the shock velocity can be obtained from the rate of expansion of the volume of the bubble. For a spherically expanding bubble, the rate of the expansion of the bubble is $dr_b/dt = 1/(4 \pi r_b^2) dV_b/dt = 1/(36 \pi V_b^2) dV_b/dt$, where $V_b = (4 \pi r_b^3)/3 $ is the volume of the bubble. In Fig.~\ref{fig.machamb} we show the  $dr_b/dt$ for simulations B and F, plotted until the bubble reaches the edge of the simulation domain. As can be seen, for simulation B the rate of expansion up to $\sim 0.5$ Myr is $\sim 0.012 c=3.6\times10^3 \kms$, close to the approximate analytical estimates derived above. The rate of expansion of the bubble in simulation F is slower, as explained in Sec.~\ref{sec.inclined}. Estimates of shock velocities obtained from observational modeling of emission lines observed in gas clouds embedded in the cocoon \citep{sutherland93b,bicknell2000a}, yield values comparable to those derived above.

\appendix
\section{Velocity power spectrum}\label{append.powspec}
\begin{figure*}
	\centering
	\includegraphics[width = 5.5cm, keepaspectratio]{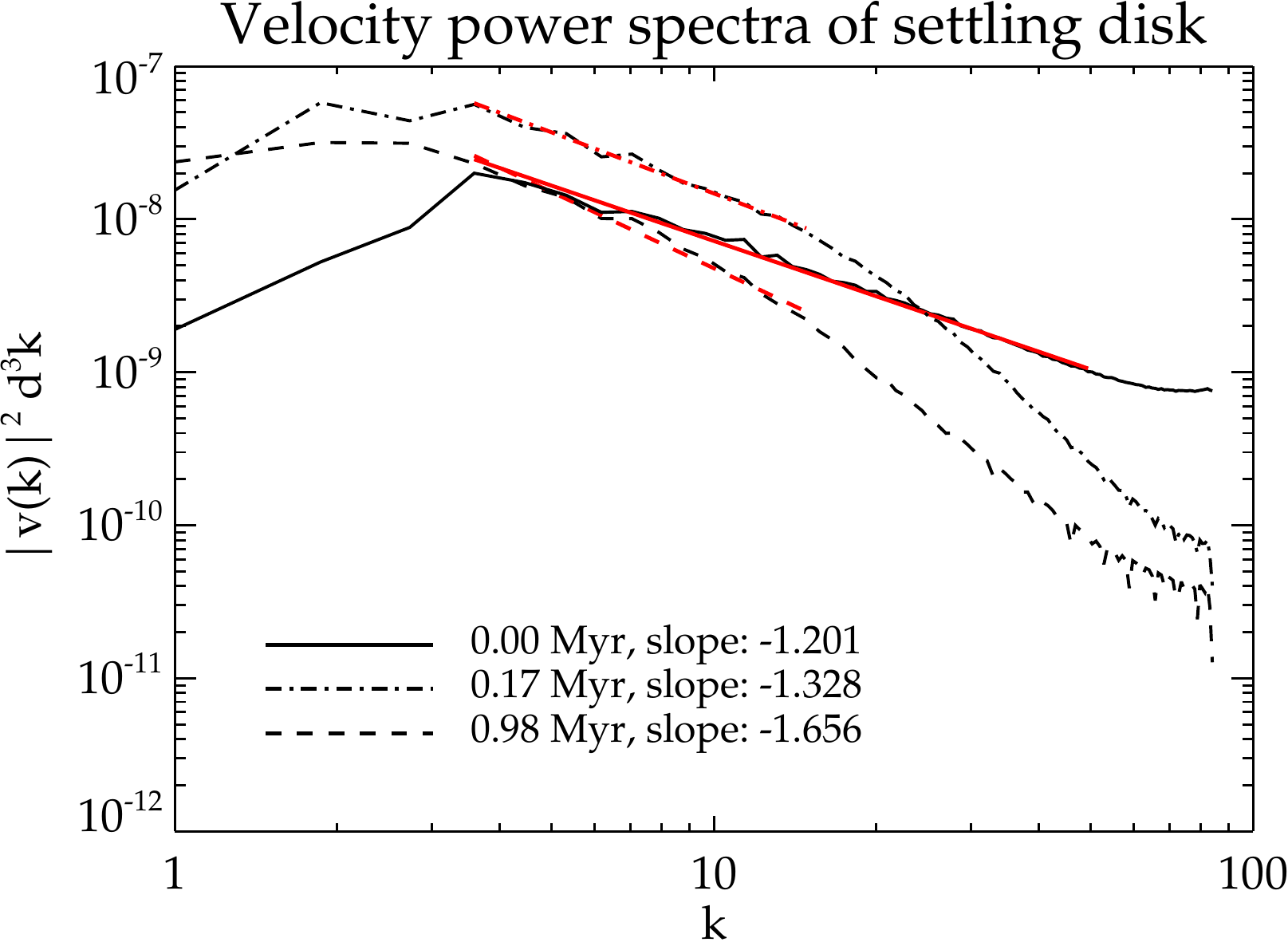}
	\includegraphics[width = 5.5cm, keepaspectratio]{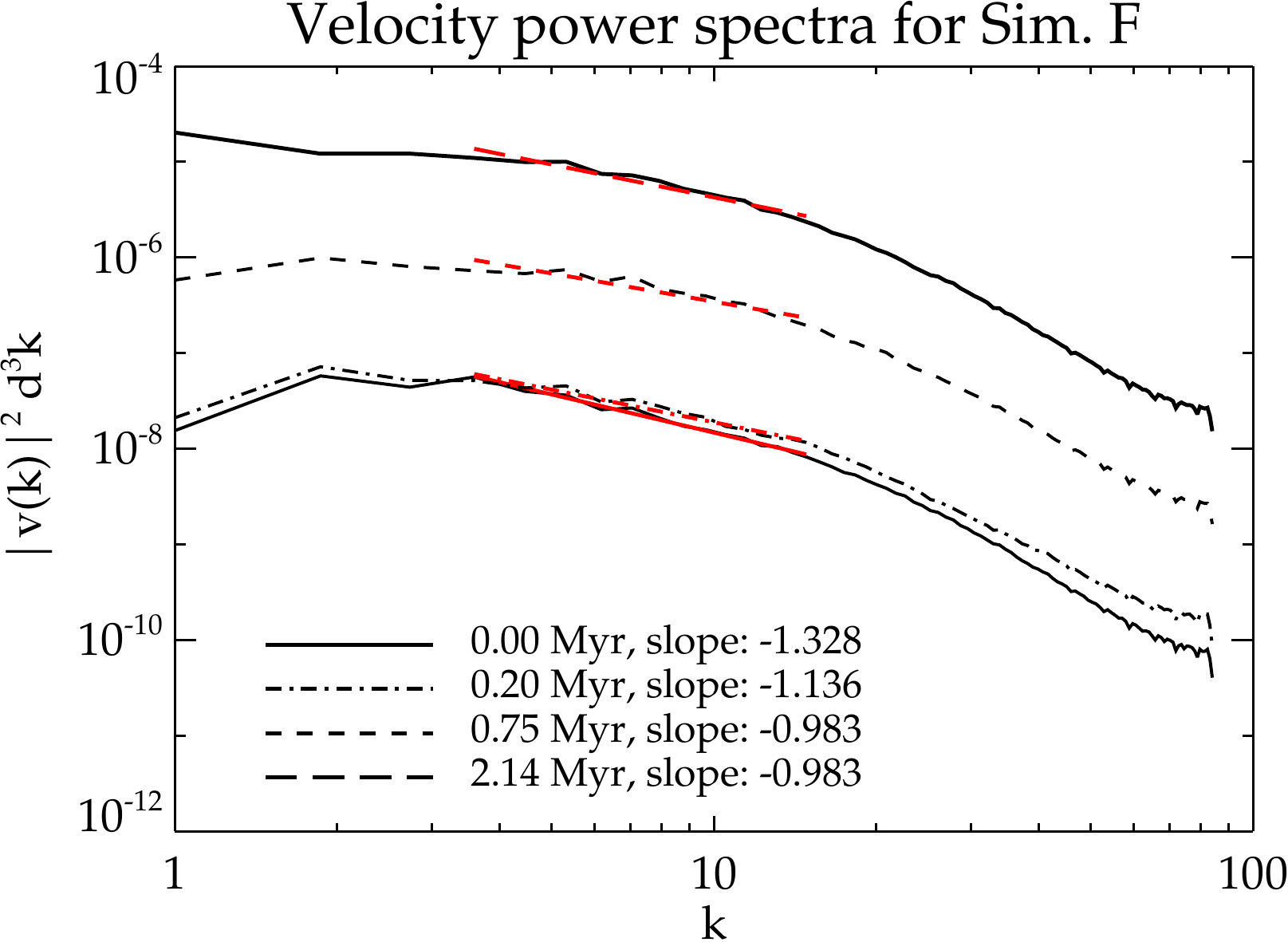}
	\includegraphics[width = 5.5cm, keepaspectratio]{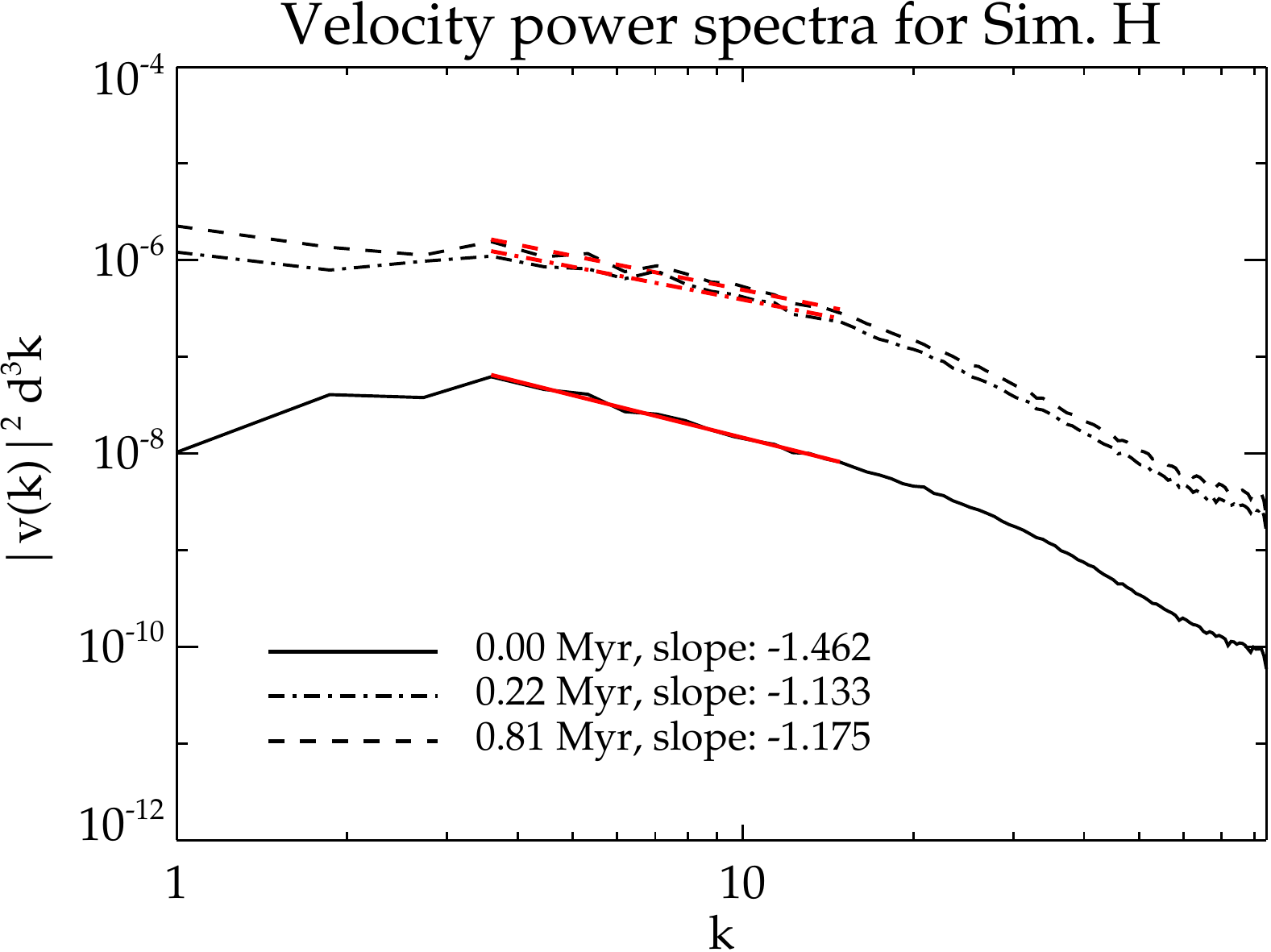}
	\caption{\small Velocity power spectrum for settling disk (left), simulation F (middle) and simulation H (right) at different times.}
	\label{fig.powspec}
\end{figure*}
In fig.powspec we show the evolution of the velocity power spectrum computed for a domain of size $1 \kpc ^3$, centred at coordinates (1,1,0). The domain was chosen to be away from the central axis of the jet which experiences strong outflows, and to be well within the height of the disk, such that a Fourier transform with periodic boundary conditions may be performed. For the settling disk (left panel in Fig.~\ref{fig.powspec}), we note that the slope of the power spectrum at $t=0$ is $\sim k^{-1.2}$. Although the turbulent velocity field is initially created with a slope of $k^{-5/3}$, the apodization with the warm medium when being interpolated on to the PLUTO grid\footnote{During apodization computational cells with $T>3\times10^4$K are replaced with ambient gas. See Sec.~\ref{sec.twophase} for details.} results in departure from a $5/3$ slope. 

As the disk is settled, the slope of the power spectrum converges to a Kolmogorov slope ($\sim k^{1.666}$) in the inertial range of $3<k<15$ (corresponding to physical scales of $\sim 333 \pc - 66 \pc$ respectively). The jet when injected into the settled disk thus initially experiences a settled medium with a Kolmogorov power spectra in the inertial range. With the onset of the jet (middle and right panels of Fig.~\ref{fig.powspec}), the power spectra flattens significantly in the range $3 < k <15$, with a mean slope of $\sim k^{-1}$. This implies that the said range of wavenumbers do not represent the inertial range of turbulent cascade anymore. This is because  the jet and energy bubble forces turbulence on the gas at such scales which are comparable mean cloud sizes ($\sim 100 \pc$) . At higher wavenumbers, the power spectrum may be affected by numerical dissipation and hence the choice of resolution \citep{kritsuk07a}. Higher resolution simulations need to be performed to investigate the nature of the cascade of turbulent energy driven by the jet to smaller length scales.

\def\apj{ApJ}%
\def\mnras{MNRAS}%
\def\aap{A\&A}%
\def\apjl{ApJ}
\def\physrep{PhR}
\def\apjs{ApJS}
\def\pasa{PASA}
\def\pasj{PASJ}
\def\nat{Nature}
\def\memsai{MmSAI}
\def\aj{AJ}%
\def\aaps{A\&AS}%
\def\iaucirc{IAU~Circ.}%
\def\sovast{Soviet~Ast.}%
\def\apss{Ap\&SS}

\bibliographystyle{mnras}
\bibliography{dmrefs}

\end{document}